\documentclass[useAMS,usenatbib]{mn2e}

\let\jnlstyle=\rm
\def\refjnl#1{{\jnlstyle#1}}

\def\apjl{\refjnl{ApJ}}                
\def\apjs{\refjnl{ApJS}}               
\def\ao{\refjnl{Appl.~Opt.}}           
\def\aap{\refjnl{A\&A}}                

\input{psfig.sty}

\usepackage{graphicx}
\usepackage{epsfig}
\usepackage{amssymb}
\usepackage{amsmath}
\usepackage{amsfonts}
\usepackage{txfonts}
\usepackage{multirow}
\usepackage{afterpage}
\usepackage{xcolor}
\usepackage{dsfont}

\definecolor{mypink1}{rgb}{0.858, 0.188, 0.478}
\definecolor{mygreen1}{rgb}{0.258, 0.788, 0.878}
\definecolor{myorange1}{rgb}{0.5, 0.2, 0.2}

\title[KiDS+DES cosmology with peak counts]{KiDS-1000 and DES-Y1 combined: Cosmology from peak count statistics}

\author[J. Harnois-D\'{e}raps \& others.]{Joachim Harnois-D\'{e}raps$^{1}$\thanks{E-mail: joachim.harnois-deraps@ncl.ac.uk}, Sven Heydenreich$^{2}$, Benjamin Giblin$^3$, Nicolas Martinet$^4$, 
\newauthor Tilman Tr\"oster$^{5}$, Marika Asgari$^{1,6,7}$, Pierre Burger$^{8,9,10}$,Tiago Castro$^{11,12,13,14}$, 
\newauthor Klaus Dolag$^{15}$, Catherine Heymans$^{3,16}$,  Hendrik Hildebrandt$^{16}$,  Benjamin Joachimi$^{17}$ \&
\newauthor Angus H. Wright$^{16}$
\\
$^{1}$School of Mathematics, Statistics and Physics, Newcastle University, Herschel Building, NE1 7RU, Newcastle-upon-Tyne, UK\\
$^{2}$Department of Astronomy and Astrophysics, University of California, Santa Cruz, 1156 High Street, Santa Cruz, CA 95064 USA\\
$^{3}$Institute for Astronomy, University of Edinburgh, Blackford Hill, Edinburgh, EH9 3HJ, UK\\
$^{4}$Universit\'e d'Aix-Marseille, CNRS, CNES, LAM, Marseille, France\\
$^{5}$Institute for Particle Physics and Astrophysics, ETH Z\"urich, 8092 Z\"urich, Switzerland\\
$^{6}$E.A Milne Centre, University of Hull, Cottingham Road, Hull, HU6 7RX, UK\\
$^{7}$Excellence for Data Science, AI, and Modelling (DAIM), University of Hull, Cottingham Road, Kingston-upon-Hull, HU6 7RX\\
$^{8}$Waterloo Centre for Astrophysics, University of Waterloo, Waterloo, ON N2L 3G1, Canada\\
$^{9}$Department of Physics and Astronomy, University of Waterloo, Waterloo, ON N2L 3G1, Canada\\
$^{10}$Argelander-Institut f\"ur Astronomie, Auf dem H\"ugel 71, 53121 Bonn, Germany\\
$^{11}$INAF - Osservatorio Astronomico di Trieste, via Tiepolo 11, I-34131 Trieste, Italy\\
$^{12}$INFN - Sezione di Trieste, Via Valerio 2, 34100 Trieste TS, Italy\\
$^{13}$IFPU - Institute for Fundamental Physics of the Universe, via Beirut 2, 34151, Trieste, Italy\\
$^{14}$ ICSC - Centro Nazionale di Ricerca in High Performance Computing, Big Data e Quantum Computing, Via Magnanelli 2, Bologna, Italy\\
$^{15}$Max-Planck-Institut fur Astrophysik, Karl-Schwarzschild Strasse 1, 85748 Garching, Germany\\
$^{16}$Ruhr University Bochum, Faculty of Physics and Astronomy, Astronomical Institute (AIRUB), German Centre for Cosmological Lensing, 44780 Bochum, Germany\\
$^{17}$Department of Physics and Astronomy, University College London, Gower Street, London WC1E 6BT, UK\\
}

\date{Accepted XYZ, 2023. Received XYZ, 2023; in original form XYZ, 2023}

\pubyear{2023}

\begin{document}
\label{firstpage}
\pagerange{\pageref{firstpage}--\pageref{lastpage}}
\maketitle

\begin{abstract}
We analyse the fourth data release of the Kilo Degree Survey (KiDS-1000) and extract cosmological parameter constraints based on the cosmic shear peak count statistics. Peaks are identified in aperture mass maps in which the filter is maximally sensitive to angular scales in the range 2-4 \mbox{arcmin}, probing deep into the non-linear regime of structure formation. We interpret our results with a  simulation-based inference pipeline, sampling over a broad $w$CDM prior volume and marginalising over uncertainties on shape calibration, photometric redshift distribution, intrinsic alignment and baryonic feedback. Our measurements constrain the structure growth parameter and the amplitude of the non-linear intrinsic alignment model to 
$\Sigma_8 \equiv \sigma_8\left[\Omega_{\rm m}/0.3\right]^{0.60}=0.765^{+0.030}_{-0.030}$
and $A_{\rm IA}=  0.71^{+0.42}_{-0.42}$, respectively, in agreement with previous KiDS-1000 results based on two-point shear statistics.  These results are  robust against modelling of the non-linear physics, different scale cuts and selections of tomographic bins. The posterior is also consistent with that from the Dark Energy Survey Year-1 peak count analysis presented in  Harnois-D\'eraps et al (2021), and hence we jointly analyse both surveys with a common pipeline. We obtain $\Sigma_8^{\rm joint} \equiv \sigma_8\left[\Omega_{\rm m}/0.3\right]^{0.57}=0.759^{+0.020}_{-0.017}$, in agreement with the {\it Planck} $w$CDM results. The shear-CMB  tension on this parameter increases to $3.1\sigma$ when forcing $w=-1.0$, and to $4.1\sigma$ if comparing instead with  $S_{8,\Lambda{\rm CDM}}^{\rm joint} = 0.736^{+0.016}_{-0.018}$, one of the tightest constraints to date on this quantity.
 Residual biases in the photometric redshifts of the DES-Y1 data and in the modelling of small scales physics could lower this tension, however it is robust against other systematics. Limits in the accuracy of our emulator prevent us from constraining $\Omega_{\rm m}$.

\end{abstract}




\section{Introduction}
\label{sec:intro}

Cosmic shear cosmology has entered an era of high precision, with recent measurements from  the Kilo Degree Survey\footnote{KiDS:kids.strw.leidenuniv.nl} (KiDS), the Dark Energy Survey\footnote{DES:www.darkenergysurvey.org} (DES) and the Hyper Suprime-Cam Survey\footnote{HSC:www.naoj.org/Projects/HSC} (HSC) reaching a precision of a few percent on parameters central to the standard model of cosmology \citep[e.g.][]{KiDS1000_Asgari, KiDS1000_vdB, KiDS1000_Li, DESY3_Secco, DESY3_Amon, HSCY3_Cl, HSCY3_2PCF}. Based on the detection of weak correlations between the observed shapes of galaxies imparted by the foreground large scale structure, cosmic shear is mostly sensitive to the structure growth parameter $S_8\equiv  \sigma_8 \sqrt{\Omega_{\rm m}/0.3}$, a combination of the matter density  parameter $\Omega_{\rm m}$ and of the amplitude of the linear matter power spectrum smoothed on spheres of $ 8 h^{-1}$Mpc,  labelled as $\sigma_8$ \citep[for lensing reviews, see e.g. ][]{2015RPPh...78h6901K, Mandelbaum18}. These Stage-III lensing surveys have been steadily improving  the data quality and the analysis methods, in preparation for  the next generation of cosmic shear experiments such as the  Rubin observatory\footnote{LSST: www.lsst.org} \citep{LSST-Design},  {\it Euclid}\footnote{{\it Euclid}: www.euclid-ec.org} \citep{RedBook} and the Nancy Grace Roman space telescope\footnote{Roman Space Telescope: roman.gsfc.nasa.gov}  \citep{Roman}. 

Despite the large effort that is being invested by international collaborations in constructing accurate lensing catalogues of hundreds of millions of galaxies, it is not entirely clear  how to best analyse these vast data, striking an optimal compromise between accuracy and precision. To date the shear two-point (2pt)  functions are still  regarded as the baseline summary statistics, having been tested for over a decade and achieving an unmatched level of understanding and control in all aspects of the analysis, including measurements tools \citep[e.g. {\sc TreeCorr} and {\sc NaMaster}, see][]{TreeCorr, NaMaster}, theoretical predictions \citep[e.g.][]{Kilbinger17} and the impact of systematics \citep[see e.g.][]{Mandelbaum18}. The main drawback from these statistics is that they completely disregard the non-Gaussian information that is stored in the non-linear matter field, more precisely in the coupling between the phases of distinct Fourier modes, without which the cosmic web would look like a Gaussian random field. This is obviously sub-optimal, and this waste of  information  will be aggravated  in the upcoming cosmic shear experiments. Accessing this non-Gaussian information is an active field of research: an array of novel weak lensing statistics are being developed specifically  to utilise this complementary small-scale information. These new methods are reaching a level of maturity that makes them competitive at analysing existing cosmic shear data, carefully balancing the precision vs accuracy metric. Recent progress is largely due to the radically improved modelling of the signal, thanks to the increased accuracy of cosmological $N$-body codes and the availability of super-computers \citep[see][for a recent review on $N$-body codes]{AnguloHahn_Nbody}. 
Recent examples of these `beyond-2pt' cosmic shear  data analyses  include the three-point function \citep[][]{Fu2014, DESY3_3pt, KiDS1000_Map3}, peak count statistics  \citep[][]{Kacprzak2016, Martinet18,Shan18,HD21, DESY3_Zuercher, HSCY1_peaks_th, DESY3_Gatti_source_clust, HSCY1_peaks_sims}, density split statistics \citep{Brouwer2018, Gruen2017, KiDS1000_BurgerDSS}, shear clipping \citep{Giblin18}, persistent homology \citep{DESY1_Heydenreich}, moments of convergence maps \citep{VanWaerbeke2013,Gatti20}, cumulative distribution functions \citep{DESY3_CDF},  likelihood-free inference \citep{Jeffrey2021, KiDS1000_SBI, Gatti2023}  or convolutional neural network inference \citep{Fluri2019, KiDS1000_Fluri}. 

At the moment, these alternative methods exhibit a constraining power that is similar to that of two-point functions, which is not surprising given the noise levels of current lensing data, {which make difficult the extraction of information stored in the noisy higher-order moments.} The situation will change drastically with the upcoming surveys, where the cosmic web itself will be detectable with lensing, at which point the non-Gaussian information  will take on a larger proportion of the signal. 

All forecasts are clear about this: joint cosmic shear analyses that combine two-point functions and any complementary probe improve the constraints on cosmological parameters even in presence of systematic uncertainties \citep[e.g.][]{MassiveNu1, Baryonification2, Zuercher2020a, PyneJoachimi,Tidalator, Giblin_PDF, HOWLS_paper1}.  The main difficulty in many of these methods comes from their accrued dependence on numerical simulations, which adds a significant computational overhead to the data analysis compared to those for which an analytical model exists. Typically, simulations are needed for modelling the cosmological signal, for modelling some of the systematics such as baryonic feedback or intrinsic alignments of galaxies, and for the estimation of the covariance matrix (although this is not always necessary,  as demonstrated by the recent likelihood-free inference analyses mentioned above). 

This paper contributes an important step to this effort: we carry out a cosmological analysis based on lensing peak statistics measured from the fourth data release of the Kilo Degree Survey (KiDS-1000 hereafter). We use the exact same data as those used in the two-point function analyses of \citet[][A21 hereafter]{KiDS1000_Asgari}, while ignoring for now other non-lensing KiDS galaxy catalogues designed for galaxy clustering analyses  \citep{KiDS1000_LRGs, KiDS1000_bright}. Our method finds peaks in aperture mass maps with an aperture filter designed for the extraction of small-scale structure, with maximal sensitivity to scales of less than 4 \mbox{arcmin}, as in \citet[][hereafter M18]{Martinet18} and \citet[][HD21]{HD21}. This contrasts with the recent peak count analysis of \citet{DESY3_Zuercher}, in which peaks are extracted from convergence maps with pixel resolution of about 7 \mbox{arcmin}. Both methods have their advantages and downsides, ours strongly focuses on small, non-linear scales, which, as demonstrated in HD21 and \citet {Martinet20}, have a higher potential for complementarity with two-point functions. Finding a posterior that is statistically consistent with that from HD21, we combine both likelihoods and carry out a joint KiDS-1000 + DES DR1 data (DES-Y1 hereafter) peak count  analysis, finding the tightest constraints on $S_8$ to date from peaks alone.

After describing the data and simulations in Sec. \ref{sec:data}, we detail our measurement techniques and analysis pipeline in Sec. \ref{sec:methods}, and we present our mitigation strategy for the key systematic uncertainties in Sec. \ref{sec:systematics}. We finally show our results in Sec. \ref{sec:results} and  discuss our findings  afterwards. Supplementary material is provided in the Appendices, including a thorough discussion of $B$-modes (in Appendix \ref{sec:Bmodes}), supplementary  pipeline validation tests (in Appendix \ref{sec:syst_pipeline})  {and a detailed discussion on goodness-of-fit for noisy covariance matrices (in Appendix \ref{sec:pvalues})}.

\section{Data and Simulations}
\label{sec:data}

We present in this section the survey data and the various simulation suites that are used for the cosmological  analysis.

\subsection{KiDS-1000 data}
\label{subsec:data}
The Kilo Degree Survey \citep{2015MNRAS.454.3500K} is a multi-band photometric galaxy survey explicitly designed for weak lensing cosmology.  Carried out at the European Southern Observatory by the VST-OmegaCAM, we analyse here the public\footnote{KiDS-1000 data:http://kids.strw.leidenuniv.nl/DR4} fourth data release \citep{KiDS1000_Kuijken}. The observation conditions are of exceptional quality, with a mean seeing of 0.7 arcsec in the $r$-band, used for shape measurements. The photometric redshifts are obtained from a combination of nine optical and infrared bands  \citep[$ugriZYJHK_{\rm s}$, see][]{Wright_KV450_SOM}, thanks to the observations of the companion VIKING  survey \citep[VISTA Kilo-degree INfrared Galaxy,][]{VIKING}. The galaxies selected in this analysis exactly match those used in the cosmic shear two-point function analyses of A21 and \citet{KiDS1000_vdB}, covering an effective area of  777.4 deg$^2$. 

The KiDS DR4 data are reduced with the  {\sc theli} \citep{2013MNRAS.433.2545E} and Astro-WISE \citep{astroWISE} pipelines, following which the shear is inferred from lens{\it fit} \citep{2013MNRAS.429.2858M, FenechConti17}. Shear additive and multiplicative biases ($c$- and $m$-corrections) are measured to a high accuracy  \citep{KiDS1000_Giblin}, where it is shown via a series of null tests that known residual systematics in the shear measurement could lead to no more than a 0.1$\sigma$ shift in the structure growth parameter $S_8 \equiv \sigma_8 \sqrt{\Omega_{\rm m}/0.3}$, the composite quantity that is best measured by cosmic shear. Note that strictly speaking, the results from the tests carried out in \citet{KiDS1000_Giblin} are only shown to hold for two-point cosmic shear statistics.

\begin{figure}
\begin{center}
\includegraphics[width=3.3in]{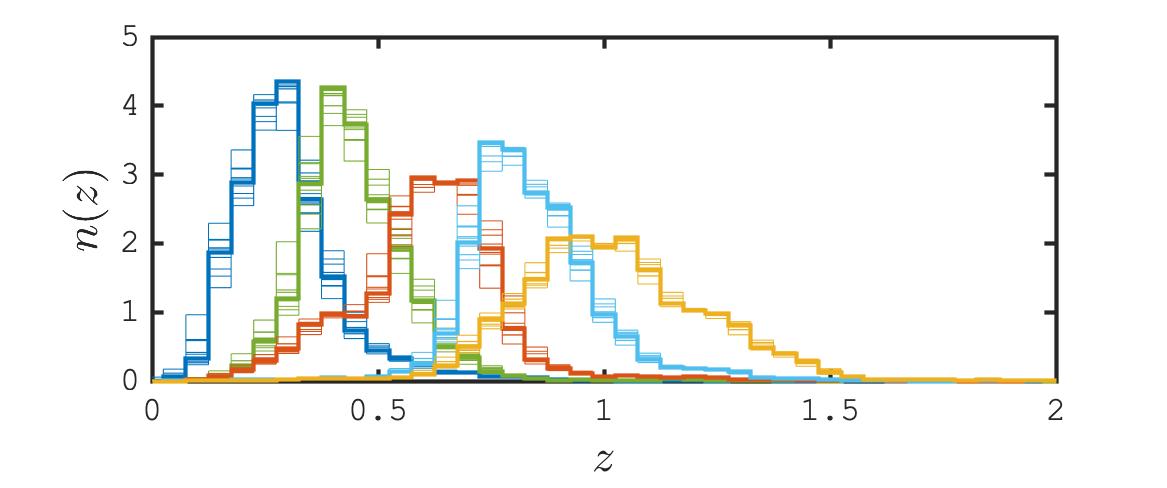}
\caption{Tomographic redshift distribution of the KiDS-1000 data. The thinner lines represent the effect of photometric uncertainty on these distributions, characterised by $n^a(z) \rightarrow n^a(z+\Delta z_a)$, with $\Delta z_a$ sampled 10 times from Gaussian distributions with widths listed in Table \ref{table:survey}. All shifted $n^a(z)$ are then rebinned with the same $z$ bins.}
\label{fig:nz}
\end{center}
\end{figure}

Following A21, we split the full DR4 galaxies in five tomographic bins according to their individual best-fit redshift $z_{\rm B}$ as measured by {\sc bpz} \citep{BPZ},  with  bin edges set to [0.1, 0.3, 0.5, 0.7, 0.9 and 1.2]. The tomographic redshift distributions, $n^a(z)$, are estimated via self-organising maps \citep[SOM, see][]{Wright_KV450_SOM}, which group galaxies based on their nine-band photometric properties and assign redshifts based on similar studies made on spectroscopic samples; galaxies for which no match is found are rejected.  We further reject galaxies for which the SOM redshift catastrophically differs from the initial $z_{\rm B}$, resulting in the so-called `Gold Sample' introduced in  \citet{KiDS1000_redshifts} and used in the subsequent KiDS-1000 cosmic shear analyses mentioned above. As detailed in A21, the means and the error of the SOM redshift distributions are calibrated on KiDS-like mock data constructed from the MICE2 simulations \citep{Fosalba2013, KiDS1000_MICE} and accounted for during the inference stage of our analysis.  The redshift accuracy is excellent due to the nine-band photometry, which helps breaking degeneracies in the galaxy spectral energy distributions: at worst, the difference between the mean redshift and that estimated from the matched spectroscopic sample is  $z_{\rm est} - z_{\rm true} = 0.013 \pm 0.0118$, making this a sub-dominant source of uncertainty in our measurement.  Note that \citet{KiDS1000_redshifts} further show that the SOM redshift distributions are fully consistent with independent estimates based on clustering cross-correlations with spectroscopic reference samples, providing extra robustness to the method. Fig. \ref{fig:nz} shows the redshift distributions estimated in the five tomographic bins, along with the variations on these distributions allowed within our photometric uncertainty.

\begin{table}
   \centering
   \caption{Main properties of the KiDS-1000 data used in the current work. The Gold Sample redshift selection based on $z_{\rm B}$ is identical to that presented in \citet{KiDS1000_redshifts}. The effective number densities are listed in the second column, in gal/\mbox{arcmin}$^2$. The shape noise (per component) listed in the third column reflects the dispersion measured in the observed galaxy shapes, as documented in \citet{KiDS1000_Giblin}, while the fifth column shows the mean shape calibration coefficients.  The redshift bias and errors listed in the fourth column are estimated from the SOM method in \citet{KiDS1000_redshifts}, while the last column shows the additive $c_{1/2}$ terms,  which has an uncertainty of $0.23\times10^{-3}$ \citep{KiDS1000_Giblin}. }
   \tabcolsep=0.11cm
      \begin{tabular}{@{} cccccccc @{}} 
      \hline
      \hline
      tomo &       $n_{\rm eff}$ & $ \sigma_\epsilon$  & $z_{\rm est} - z_{\rm true}$  & $m$ &$(c_1,c_2)\times 10^{3}$\\
       \hline
       bin1     &  0.62 &        0.27 & $0.000 \pm 0.0106$& $-0.009\pm0.019$& (0.295, 0.156) \\
       bin2     &  1.18 &         0.26& $0.002  \pm 0.0113$&$-0.011\pm0.020$& (0.004, 0.621)\\
       bin3     &  1.85 &         0.27&$0.013\pm0.0118$&$-0.015\pm0.017$ & (0.052, 0.728)\\
       bin4     &  1.26 &         0.25& $0.011\pm 0.0087$&$\,\,\,\,0.002\pm0.012$ & (-0.360, 0.948)\\
       bin5     &   1.31&         0.27& $-0.006\pm0.0097$&$\,\,\,\,0.007\pm0.010$ & (-1.363, 1.155)\\
    \hline 
    \hline
    \end{tabular}
    \label{table:survey}
\end{table}

The SOM selection and the shear inference pipelines are both repeated on KiDS-like image simulations  \citep{2019A&A...624A..92K}, 
from which a relation between apparent size, magnitude and the observed galaxy shape is used to calibrate the inferred lens{\it fit} shear\footnote{A KiDS-1000 re-analysis has been presented in \citet{KiDS1000_Li} after correcting for an anisotropic error in the {\it lens}fit likelihood sampler. This error has not been corrected here, but their study shows the correction has a negligible impact on the inferred cosmology.}.
Whereas previous cosmological analyses use a single $m$-calibration factor per tomographic bin, the aperture mass map statistics exploited in 
this paper  are subject to local variations in the noise levels and seeing conditions, and we therefore use the above-mentioned  relation to extract a shear calibration {\it per object}, $m_a$. This is not necessary, but allows to capture possible correlations between the $m$-correction and the lens{\it fit} weights. These are inevitably noisier than the average over the full tomographic bins, but a large fraction of this noise cancels within our aperture mass map calculations as well, while providing optimal estimates of the local 
noise contribution (M18). Let us recall that this calibration corrects for known residual biases such as shape detection biases 
\citep{FenechConti17,2019A&A...624A..92K} or blending of the images of galaxies \citep{2015MNRAS.449..685H}. While we apply the $m$-correction per object, 
the averaged multiplicative biases per redshift bin used in A21 enter our analysis at the inference level in the form of  nuisance parameters over 
which we marginalise. Table \ref{table:survey} summarises the survey properties relevant to our analysis.

\subsection{DES-Y1 data}

The DES-Y1 measurements is based on the public year-1 data release from the Dark Energy Survey Collaboration \citep{DESY1_data}, with source galaxy selections that exactly follow the main cosmic shear results described in  \citet{DESY1_Troxel}. The lensing catalogue consists of 26 million galaxies covering a footprint of 1320 deg$^2$ with a galaxy density of 5.07 gal \mbox{arcmin}$^{-2}$. The per-galaxy shear signal is inferred with the {\sc Metacalibration} method \citep{Metacal}. Every galaxy is assigned to one of the four tomographic bins based on the photometric redshift posteriors estimated from the the {\it griz} flux measurements, as detailed in \citet{DESY1_redshifts}. Following \citet{DESY1_Troxel},  the mean and uncertainty on the shear multiplicative calibration are given by $m_a=0.012 \pm0.023$.

Whereas the original DES-Y1 results estimated the tomographic $n^a(z)$  from a Bayesian photometric redshift analysis calibrated on the COSMOS2015 field \citep{COSMOS15}, the HD21 reanalysis instead opted for  $n^a(z)$ estimates based on a direct reweighted calibration of matched spectroscopic data \citep[][DIR hereafter]{DIR}, following the DES-Y1 reanalyses of \citet{KiDS_DES_Joudaki} and   \citet{Asgari_DES_KiDS_cosebi}. The uncertainty on the DIR mean redshift distributions is $\Delta z_a$ = [0.008, 0.014, 0.011 and 0.009] for redshift bins $a=1...4$, respectively. Both methods have their pros and cons. The calibration with COSMOS is by design based on a complete sample but suffers from imperfect redshifts and sampling variance \citep[see e.g.][]{Alarcon2021}. In contrast, the spec-$z$ samples used for the DIR method have (close to) perfect redshifts but are incomplete and not representative of the source sample, which is alleviated by the reweighting, but often cannot be fully eliminated \citep[see][]{GruenBrimioulle}. Importantly, the DIR $n^a(z)$ favours $S_8$ values that are smaller by 
$\Delta S_8 = 0.03$ compared to the COSMOS-calibrated $n(z)$,  which is a 0.8$\sigma$ shift \citep{KiDS_DES_Joudaki}.

\subsection{Simulations}
\label{subsec:sims}

As mentioned in the introduction, the accuracy of simulation-based inference pipelines fully depends on the quality of the numerical simulations it is calibrated on. The same way 2pt  analyses must carefully understand the scales, cosmologies and redshifts that are well captured by their model, it is critical for our peak count analysis to identify the range of validity of our training simulations. The additional complexity here is that no simulation suite serves all purposes, and therefore we  must carefully investigate, for all of them separately, the accuracy and limits of the measurements and how these impact the peak count statistics. The simulations used in this work are in many aspects identical to those presented in HD21, which we refer to for further details. Specifically: 
\begin{enumerate}
\item the cosmological dependence of the peak count statistics is calibrated on the $w$CDM {\it cosmo}-SLICS $N$-body simulations introduced in \citet{cosmoSLICS}. They sample a wide volume in $S_8$, $\Omega_{\rm m}$, $w_0$ and $h$ with 25 points arranged in a Latin hypercube (plus one $\Lambda$CDM point), each evolved with a pair of $N$-body simulations designed to suppress sample variance in 2pt functions,   then ray-traced in ten light-cones of 100 deg$^2$ ($10000$ deg$^2$ in total area). These form our {\it Cosmology Training Set}, and resolve the non-linear physics to better than 2\% up to $k$-modes of 2.0 $h^{-1}$Mpc, when compared to the Cosmic Emulator \citep{Coyote3}. Smaller scales gradually lose precision, affecting mainly their ability to resolve substructure in most massive objects. The exact impact of this loss on weak lensing peak counts is investigated  in HD21 with a separate set of simulations ran with a much higher force resolution, where it is found that this leads to at most a 1\% loss of the highest peaks, which is largely sub-dominant compared to both baryonic physics and statistical errors. We revisit this in Sec. \ref{sec:systematics} (see also point iv);
\item the covariance matrix that captures the sample variance is estimated from 124 fully independent SLICS $N$-body simulations described in  \citet{SLICS_1}. 
These are evolved from independent initial conditions at a fixed cosmology, and make our {\it Covariance Training Set}. They resolve the same non-linear physics 
as the {\it cosmo}-SLICS, and are shown in \citet{cosmoSLICS} and HD21 to produce marginalised errors on cosmological parameters  that are fully consistent with 
those obtained with an analytical calculation, when analysing 2pt statistics. \citet{KiDS1000_BurgerDSS} further show in the context of density-split statistics 
that a covariance matrix estimated from the SLICS or from a much larger number of log-normal {\sc Flask} mocks \citep{FLASK} produce fully consistent results, 
as expected for these mildly non-linear statistics. We further increase the effective number of covariance mocks by randomly rotating 10 times the shape noise 
components. This works particularly well given that the peak statistics is currently shape-noise dominated \footnote{The average shape noise contribution, computed from the scatter between the 10 noise  realisations for a fixed underlying simulation, takes up about 90\% of the total error budget, 95\% for the auto-bins.}:
while the expectation value of standard 2pt statistics does not depend on the noise (only their variance does), shape noise affects both the signal and 
covariance of map-based statistics \citep[see Appendix D of][]{Homology};
\item for the KiDS-1000 analysis, the impact of galaxy intrinsic alignments is measured from the IA-infused lensing simulations described in \citet{Tidalator}. These are also constructed from the {\it cosmo}-SLICS and therefore resolve the same physical scales. This {\it IA Training Set} assumes a linear coupling between the projected  non-linear tidal field and the intrinsic ellipticity of every galaxy, and is  therefore physically modelling the  non-linear  linear alignment model of \citet{2007NJPh....9..444B} without explicit redshift nor luminosity dependence. It is expected that this effective IA model does not fully capture the alignment signal, and that a more physical model such as the Tidal Alignment And Torquing model \citep{TATT}  or the halo-model of \citet{Fortuna2021} would provide a more accurate description, however current cosmic shear surveys do not have the statistical power to constrain parameters beyond the simpler NLA model \citep{DESY3_Secco}, which is therefore deemed sufficient for the current analysis.  The IA infusion process has been shown in \citet{Tidalator} to accurately reproduce the NLA predictions for the 2pt correlation function down to scales of a few \mbox{arcmin}, beyond which the NLA is expected to fail in a manner that is undetectable in the current data.  \citet{KiDS1000_Map3} further show that these same simulations agree with the IA modelling of three-point shear statistics. The model fails at scales that correspond to high over-densities in our simulations, which contribute to lensing peaks that are excluded from our analysis. We infuse different levels of IA and marginalise over these choices in the end, as described in Sec.  \ref{sec:systematics}. for the DES-Y1 analysis, IA are included with a  non-linear halo-based model, see HD21 for details; 
\item limits in the force resolution of the {\it cosmo}-SLICS are bound to impact the weak lensing statistics in a manner that is not always predictable. We assess this with the SLICS-HR suite \citep{SLICS_1}, a  high-resolution version of the SLICS light-cones recently used in a combined lensing-clustering cosmological analysis \citep{Duncan2022}. The SLICS-HR consist of ten independent $10\times10$ deg$^2$ catalogues that are run at the same cosmology and with the same particle count and volume as the SLICS, but the $N$-body force accuracy has been increased such as to resolve $k$-modes up to 10 $h^{-1}$Mpc. We use these to validate the full inference pipeline in Sec. \ref{subsec:validation}, acting as our {\it Validation Set};
\item the impact of baryon feedback is estimated with the {\it Magneticum} hydrodynamical simulations\footnote{{\it Magneticum} simulations: www.magneticum.org}, forming our {\it Baryons Training Set}. These have been shown to reproduce a number of key observations relevant to weak lensing studies \citep{LensingPDF_baryons}, and notably the feedback on the matter distribution closely matches that of the BAHAMAS  \citep{BAHAMAS}, another suite of hydro simulations with independent prescriptions for their sub-grid physics. The training set consists of ten $10\times10$ deg$^2$ {\it pseudo}-independent light-cones extracted\footnote{The {\it Magneticum} light-cones were built with the public {\sc slicer} code:  https://github.com/TiagoBsCastro/SLICER} from full hydrodynamical simulations, and another 10 light-cones extracted from dark matter-only sister simulations, evolved from the same initial conditions \citep[more details on the used simulations can be found in][]{Martinet21}. There is a large uncertainty on the exact impact of baryonic physics on the matter distribution (and therefore on our lensing statistics), which we account for by linearly scaling the relative baryonic bias with a nuisance parameter, $b_{\rm bary}$, which we marginalise over at the inference stage\footnote{Note that this parameter is not to be confused with  $A_{\rm bary}$ used in A21 (see their Table 2), which specifically relates to one of the free parameters entering their {\sc HMCode} halo model.}.
\item  different $N$-body codes and ray-tracing methods, even at fixed cosmology, will have a residual impact on the peaks statistics \citep{Hilbert_LensSimAccuracy}. We explore these numerical systematics with the public full-sky weak lensing simulations from \citet[][T17 hereafter]{HSCmocks}\footnote{T17: http://cosmo.phys.hirosaki-u.ac.jp/takahasi/allsky\_raytracing/}, post-processed into KiDS-1000 mock  data (North and South patches) as in \citet{KiDS1000_Map3}. The T17 simulations follow the non-linear evolution of 2048$^3$ particles in a series of nested cosmological volumes with side length starting at $L$ = 450$h^{-1}$Mpc at low redshift, then increasing at higher redshifts. These result in 108 pseudo-independent full-sky lensing maps, seven of which are used in this work, with flat $\Lambda$CDM cosmological parameters set to $(\Omega_{\rm m}, \Omega_{\rm b}, \sigma_8, h, n_{\rm s}, w_0) = $(0.279, 0.046, 0.82, 0.7, 0.97, -1.0).
\end{enumerate}

\subsubsection{Assembling mock surveys}
\label{subsubsec:mosaic}

\begin{figure}
\begin{center}
\includegraphics[width=3.3in]{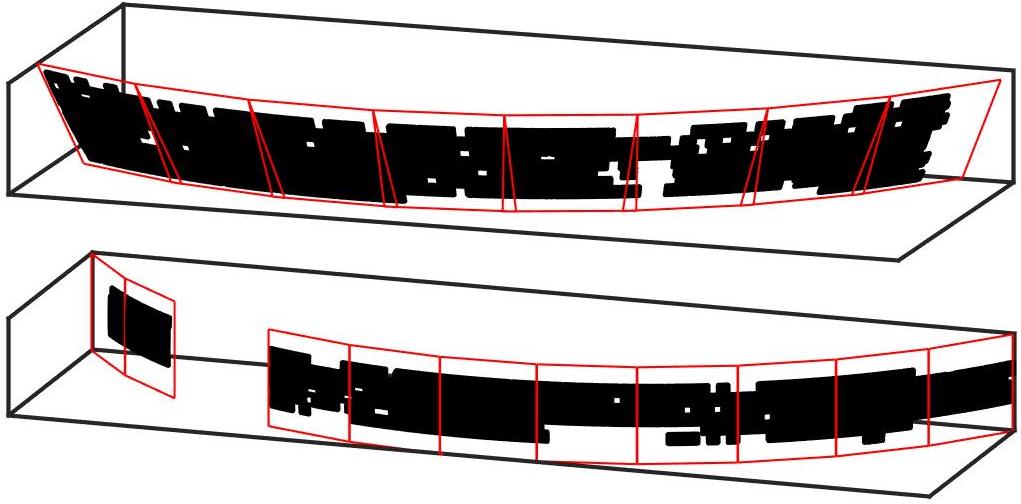}
\caption{Tiling strategy adopted to pave the full KiDS-1000 data (black) with flat-sky $10\times10$ deg$^2$ simulations (red squares). 
Some of the tiles slightly overlap due to the sky curvature, in which case the data is split at the mean Dec in overlapping regions. }
\label{fig:KiDS1000_tiles}
\end{center}
\end{figure}

Most of these simulations have been introduced in HD21, in \citet{DESY1_Heydenreich}, in  \citet{KiDS1000_BurgerDSS} and in the references listed in the previous section;  we encourage the interested reader to consult these for a more complete technical description. To summarise some of the key properties, all of the above-mentioned simulations are organised in light-cones of 100 deg$^2$ each, populated with galaxy samples that match the tomographic $n^a(z)$ distributions, number densities and shape noise levels of the KiDS-1000 Gold Sample and DES-Y1  data. Except for the IA-infused simulations, the galaxy positions, the amplitude of their ellipticities $|\boldsymbol{\epsilon}_{\rm data}|$ and their multiplicative shear calibration factors $m_a$ are exactly reproduced in each of the mock survey realisations (i.e. in the {\it Cosmology, Covariance, Validation} and {\it Baryon Training Sets}). To achieve this, the KiDS-1000 data are split into 18 tiles that each fit within 100 deg$^2$ regions, as depicted in Fig. \ref{fig:KiDS1000_tiles}, and the shear and convergence from every simulation is repeated across them, interpolated at the local galaxy positions. These tiles are analysed separately and combined only at the level of the summary statistics, ensuring that  cross-tile correlations that exist in the data but not in the simulations are explicitly ignored. This effect is minor for localised non-Gaussian probes such as peak statistics, but is critical for e.g. shear 2pt functions. The shear and convergence are interpolated from the underlying simulations at the position of every galaxy,  infused with the $m_a$ from the data, then combined with the (randomly rotated) observed ellipticity following:
\begin{eqnarray}
\boldsymbol \epsilon_{\rm mock} = \frac{{\boldsymbol \epsilon}_{\rm data}^{\rm rand} + {\boldsymbol g}}{1 + {\boldsymbol \epsilon}_{\rm data}^{\rm rand}{\boldsymbol g}^*} \, . 
\label{eq:eps_obs}
\end{eqnarray}
In the above expressions, bold-font symbols are spin-2 complex quantities and $\boldsymbol g$ is the $m$-biased simulated reduced shear. As described in HD21, this involves rotating each tile at the equator, which preserves the relative positions of galaxies but modifies their ellipticities, defined with respect to the North pole.

We repeat this construction for all light-cones of the  {\it Cosmology Training Set}, the {\it Covariance Training Set}, the {\it Baryon Training Set} and the  {\it Validation Set}. Additionally, the  uncertainty in the photometric redshifts is forward-modelled with a further 10 full survey realisations computed at the fiducial cosmology, in which the $n(z)$ is shifted by small amounts (details provided in Sec. \ref{sec:systematics}). In total, this results in  414 simulated mosaic surveys that we analyse in preparation for the inference stage, with the majority (260) contributing to the  {\it Cosmology Training Set}. Each mock further contains 10 random rotations of $\boldsymbol{\epsilon}_{\rm data}$, to improve convergence of the signal\footnote{For peak statistics, removing the shape noise from simulated data changes both the mean of the signal and the covariance, whereas for shear 2PCF, the mean of the signal is unchanged. For this reason, the best way to achieve convergence on the mean peaks signal is by averaging over multiple noise realisations.}.

The {\it IA Training Set}  are treated slightly differently, since for these the positions of the mock galaxies must be sampled from  the simulated over-density maps  or halo catalogues, which do not correlate with the positions in the data \citep[see][for more details]{Tidalator}.  The mosaic survey tiling is therefore not possible, so we use instead 5 light-cones per IA model and explore  four alignment strengths  in KiDS, and one model in DES. Although these represent a lesser total area than the real data, their sole purpose  is to capture the relative impact of IA on the signal, computed from ratios in which the sample variance cancels by design. We use $4\times5\times100$ deg$^2$ of training data, which is enough to capture this.

The simulations are free of additive biases by construction, however \citet{KiDS1000_Giblin} measures residual additive terms $c_{1/2}$ in the KiDS-1000 cosmic shear catalogues, caused by the shape measurement method itself. These are reported in Table \ref{table:survey} and subtracted from the observed ellipticities when analysing the real data.  We follow \citet{DESY1_Troxel} by not accounting for any low-level additive terms in the DES Y1 catalogue. 
%
%
The multiplicative biases are not easily removed from the data, hence we instead infuse the mocks with the $m_a$ terms per object, and treat thereafter data and simulations on equal footings.


\section{Methods}
\label{sec:methods}

\subsection{Aperture mass map statistics}
\label{subsec:peaks}

There exists a number of methods for identifying and  counting lensing peaks, including finding maxima  on convergence maps \citep{MassiveNu1}, on wavelet-transformed maps \citep{Ajani2020} or on aperture mass maps \citep{Schneider1996}. We here opted for the aperture mass maps for the following reasons: as  argued  in M18, this statistics is immune to masking-induced biases and strong $B$-mode leakage common to  methods based on reconstruction of  convergence maps, plus it benefits from a local estimation of the shape and Poisson noise, yielding more accurate signal-to-noise maps. 

Specifically, we cover each of the 18 tiles with a two-dimensional grid with a pixel size of 0.59 \mbox{arcmin}. We next reconstruct  the mass inside an aperture filter $Q$ centred on each pixel, at position $\boldsymbol \theta$ on the sky, from the sum of all tangential ellipticities  $\epsilon_{a,{\rm t}}$  contained therein as:
 \begin{eqnarray}
M_{\rm ap}(\boldsymbol \theta) = \frac{1}{n_{\rm gal}(\boldsymbol \theta) \sum_a w_a (1 + m_a)}\sum_a w_a \epsilon_{a,{\rm t}}({\boldsymbol \theta}, {\boldsymbol \theta}_a) Q(|{\boldsymbol \theta} - {\boldsymbol \theta}_a|, \theta_{\rm ap}, x_c) \, .
\label{eq:Map}
\end{eqnarray}
The tangential ellipticity about $\boldsymbol \theta$ is computed as $\epsilon_{a,{\rm t}}({\boldsymbol \theta}, {\boldsymbol \theta}_a)= -[\epsilon_1({\boldsymbol \theta}_a)\ {\rm cos}(2\phi({\boldsymbol \theta}, {\boldsymbol \theta}_a))+\epsilon_2({\boldsymbol \theta}_a)\ {\rm sin}(2\phi({\boldsymbol \theta}, {\boldsymbol \theta}_a))]$,  where $\boldsymbol \theta_a$ is the position of galaxy $a$ and  $\phi({\boldsymbol \theta}, {\boldsymbol \theta}_a)$ is the angle between both coordinates. The sum runs over all galaxies in the aperture, and $n_{\rm gal}(\boldsymbol \theta) $ is the local galaxy density  in the filter when centred at $\boldsymbol \theta$. 
As in M18 and HD21, our filter $Q(\theta, \theta_{\rm ap}, x_c)$, abridged to $Q(\theta)$, matches that of  \citet{Schirmer2007}, which is optimised for efficiently detecting NFW haloes:
 \begin{eqnarray}
Q(x) = \frac{{\rm tanh}(x/x_c)}{x/x_c} \big[1 + {\rm exp}(6 - 150x) + {\rm exp}(- 47 + 50x)\big]^{-1}.
\label{eq:Q}
\end{eqnarray}
In the above expression,  we use the standard value of  $x_c = 0.15$, while $x=\theta/\theta_{\rm ap}$, with $\theta$  the distance to the filter centre. We additionally use the same filter size, set to $\theta_{\rm ap}=12.5$ \mbox{arcmin}, which is shown in M18 to better detect the cosmological signal over other filter sizes in KiDS data.  We compute Eq. (\ref{eq:Map}) at every pixel location to construct  our signal map. The variance about this map is computed  at every pixel location with:
 \begin{eqnarray}
\sigma^2_{\rm ap}(\boldsymbol \theta) = \frac{1}{2 n_{\rm gal}^2(\boldsymbol \theta)  \left[ \sum_a w_a \right]^2} \sum_a w_a^2 |{\boldsymbol \epsilon}_{a}|^2 Q^2(|{\boldsymbol \theta} - {\boldsymbol \theta}_a|) \, ,
\label{eq:MapNoise}
\end{eqnarray}
where again the sum runs over all galaxies in the filter. The $m$-calibration estimated from the image simulations of \citet{2019A&A...624A..92K} is meant to correct the inferred shear, not the ellipticity, which explains why it appears in the denominator of Eq. (\ref{eq:Map})  but not in that of Eq. (\ref{eq:MapNoise}), which describes the noise map. Finally, we  take the ratio between Eq. (\ref{eq:Map}) and the square root of Eq. (\ref{eq:MapNoise}) at every pixel location to construct our signal-to-noise maps, $\mathcal{S}/\mathcal{N}(\boldsymbol \theta) \equiv M_{\rm ap}(\boldsymbol \theta)  / \sqrt{\sigma_{\rm ap}^2(\boldsymbol \theta)}$, from which we identify peaks as local maxima and record their $\mathcal{S}/\mathcal{N}$-values. We repeat this process for the 10 realisations of random rotations and report the average, except for the {\it Covariance Training Set}, for which we do not take the average; instead, each noise realisation leads to an estimate of the covariance matrix, of which we take the average in the end.

As detailed in HD21, masking is dealt with naturally in aperture mass statistics, and no special treatment needs to be enforced as long as data and simulations are masked and analysed the same way. This is achieved by fixing galaxy positions in the simulations to that of the observed data, which ensures the impact of the mask is identical. In our case, we decided nevertheless to act upon masked pixels. These are identified from the galaxy catalogues as regions with an aperture galaxy density that is either critically low or null, then removed from the final  $\mathcal{S}/\mathcal{N}(\boldsymbol \theta)$ maps. 

It has been shown that some additional information can be extracted by combining the peak count statistics measured from multiple filter sizes \citep[e.g.][]{DESY3_Zuercher, Giblin_PDF}, however M18 shows that this gain is mild for Stage III surveys. We therefore opted for a single-scale analysis here, but intend to revisit this in the future. 

\subsection{Tomography and selection}

Tomographic decomposition of the lensing data allows us to probe the redshift evolution of the large scale structures, which is largely driven by $\Omega_{\rm m}$ and $w_0$ via their impact on the growth of perturbations. A direct consequence of the improved sensitivity to these is a gain in precision in $S_8$, arising from degeneracy breaking.  This decomposition is different for the KiDS and DES surveys, which we detail here. 

\subsubsection{KiDS-1000}

\begin{figure*}
\begin{center}
\includegraphics[width=7.0in]{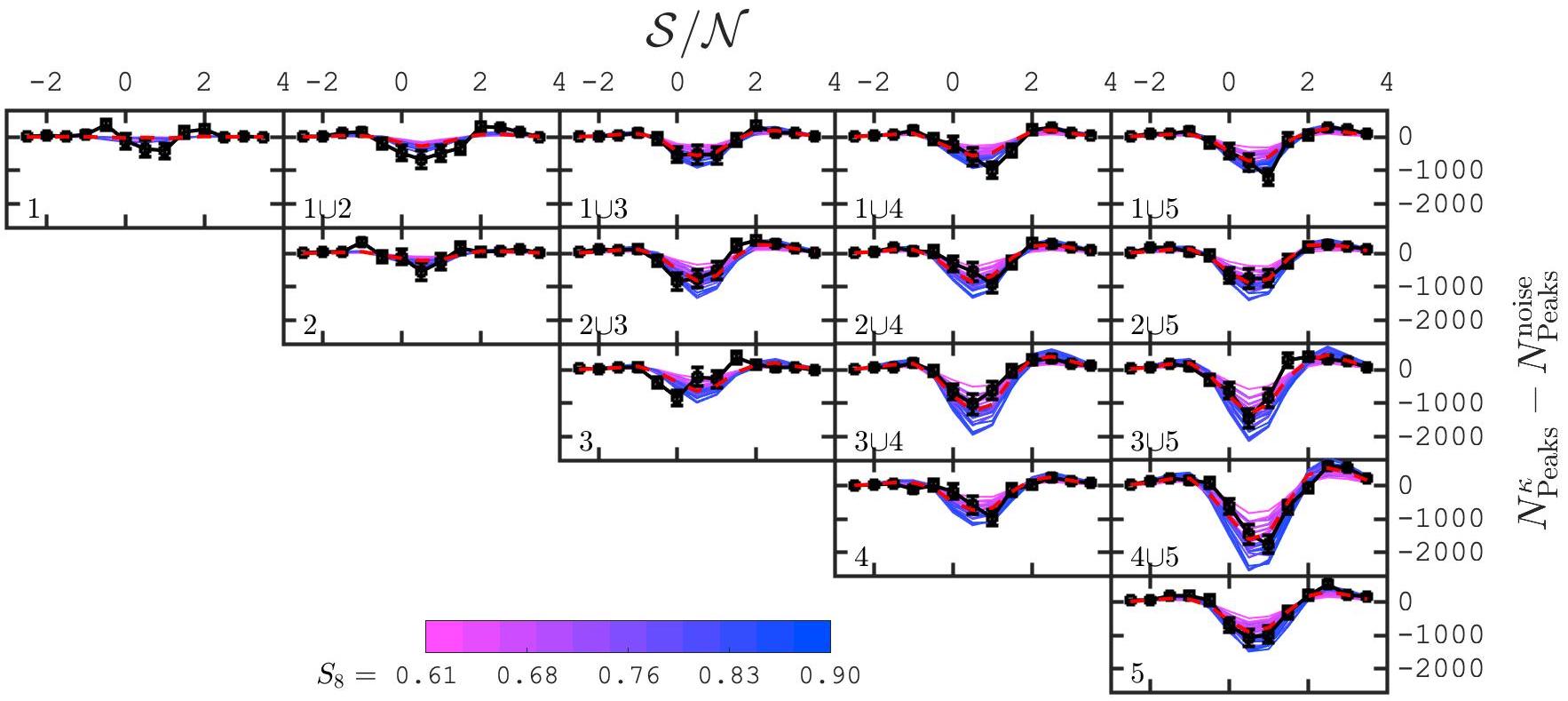}
\includegraphics[width=7.4in]{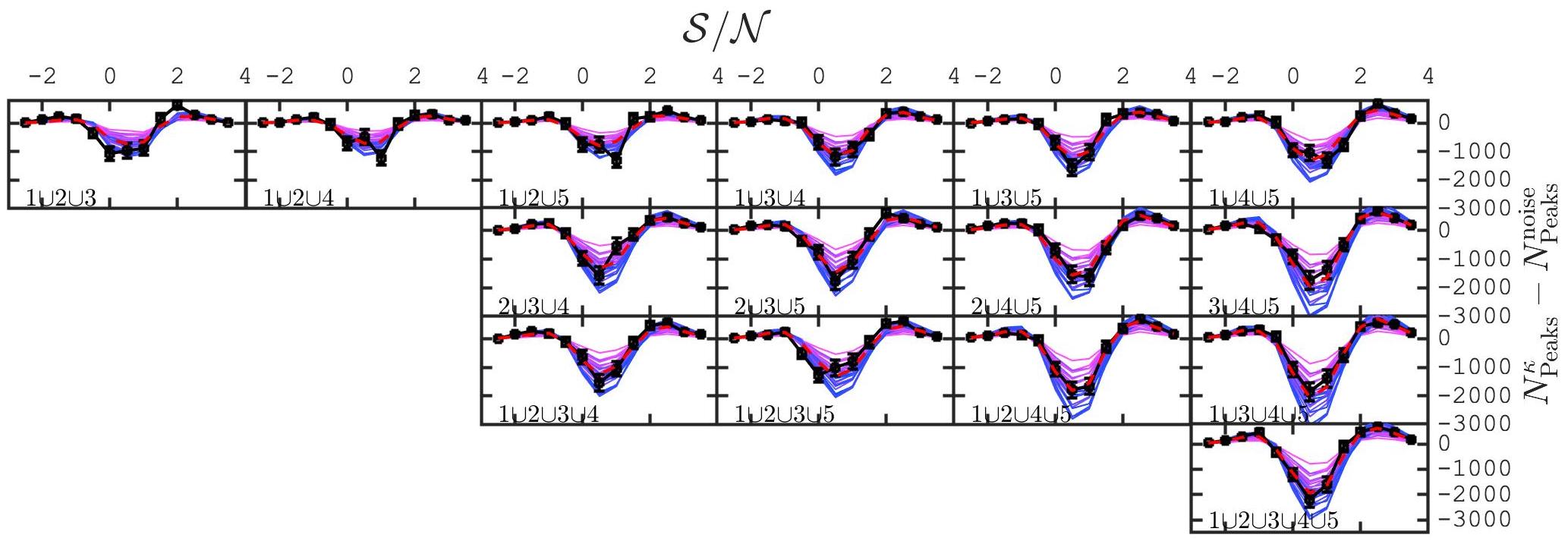}
\caption{Tomographic weak lensing peak function $N_{\rm peaks}^{\kappa}(\mathcal{S}/\mathcal{N})$ measured in the KiDS-1000 data (black squares) and in the {\it Cosmology Training Set} simulations,  colour-coded by their  $S_8$ value.
The pure noise signal $N_{\rm peaks}^{\rm noise}$ has been removed to better highlight the variations with respect to cosmology. The panels show the results from different combinations of tomographic bins, in which the red dashed lines represents the best-fit model inferred from our fiducial analysis, see Sec. \ref{sec:results}.}
\label{fig:N_peaks_data}
\end{center}
\end{figure*}

From the five  KiDS tomographic bins,  we include both the auto- and the cross-redshift measurements, as first defined in \citet{Martinet20}. To be specific, peaks  are identified from the individual tomographic galaxy catalogues (the `auto' redshift bins 1, 2, 3, 4, 5),  from every possible combination of bin pairs (1$\cup$2, 1$\cup$3, 1$\cup$4, 1$\cup$5... 4$\cup$5), triplets (1$\cup$2$\cup$3, 1$\cup$2$\cup$4, 1$\cup$2$\cup$5...), quadruplets (1$\cup$2$\cup$3$\cup$4, 1$\cup$2$\cup$4$\cup$5...) and quintets (i.e. no tomography).  As shown in  HD21,  \citet{DESY3_Zuercher} and \citet{DESY1_Heydenreich}, these `cross-tomographic' catalogues contain a significant amount of additional information that is not contained within the `auto' case. The tomographic peak function is presented in Fig. \ref{fig:N_peaks_data}, showing in the different panels the 30 different redshift bin combinations. For each case we overlay the predictions from the {\it Cosmology Training Set} in colour with the data measurements in black; the error bars are obtained from the {\it Covariance Training Set}.  A similar measurement is presented in Fig \ref{fig:N_peaks_Magneticum}, where the data is replaced by the mean over our {\it Baryons Training Set}
In these figures, we have subtracted the peak function  measured from pure shape noise fields, $N^{\rm noise}_{\rm peaks}$, to better highlight the cosmological dependence of the signal.  

In all cases, we measure the peak function in $\mathcal{S}/\mathcal{N}$ bins of width 0.5 in the range [-2.5, 4.0], for a total of 13 bins per sub-panel and 390 elements in total. The motivation behind this initial choice of range is driven by a number of requirements, notably that of having a large number of peaks per bin to ensure the data is Gaussian-distributed (with our selection, every bin has at least 200 objects, while bins outside this range have far fewer objects).  Additionally, our  analysis has strict requirements on the modelling precision and on the level of contamination by residual systematic effects, resulting in this bin selection being in fact ``aggressive". We expand on this in Sec. \ref{sec:systematics}, where we argue that instead the range $[-1.0 < \mathcal{S}/\mathcal{N} < 3.0]$ is a better choice with lower modelling errors, forming a `clean' data vector of $7\times30=210$ elements in total that is used for the main cosmological analysis. 

 \subsubsection{DES-Y1}

Following HD21, our DES peak count analysis includes the auto- and cross-redshift measurements up to pairs of tomographic bins, for a total of 10 bin combinations. The peak function is  measured in 12 $\mathcal{S}/\mathcal{N}$ bins in the range $[0.0 < \mathcal{S}/\mathcal{N} < 4.0]$, forming a data vector with 120 elements. Although these details differ compared to the KiDS-1000 case described above, it is shown in HD21  to be accurate and competitive.


\begin{figure}
\begin{center}
\includegraphics[width=3.5in]{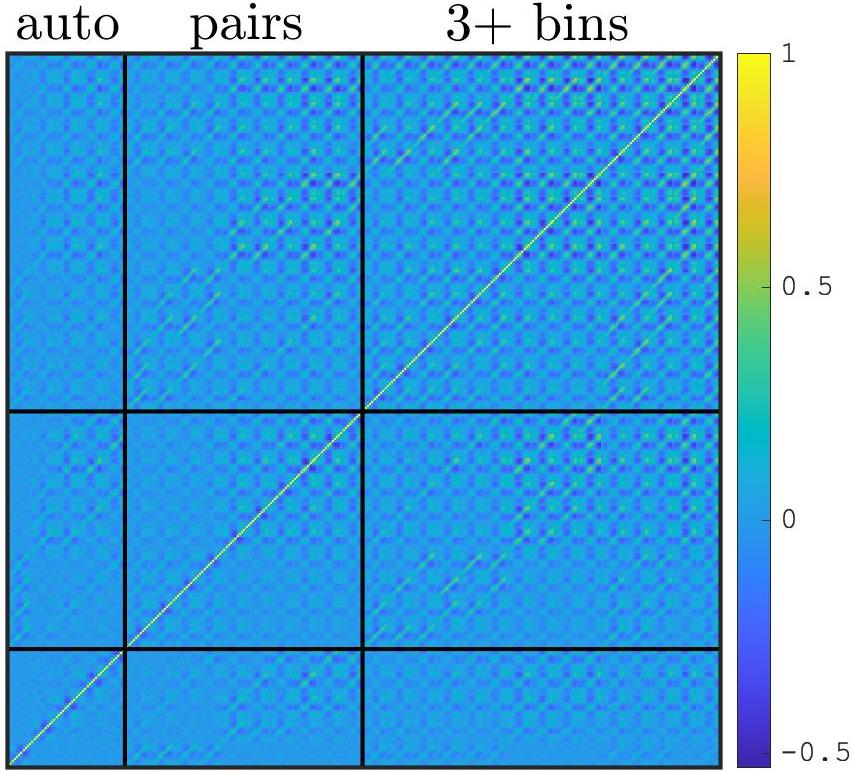}
\caption{This figure highlights the correlations between the different elements of the KiDS-1000 data vector. From left to right, the 30 blocks show the correlation coefficients for the different redshift bin combinations, starting with singlets (i.e. auto-bins), pairs, triplets, quadruplet and the no-tomographic case, with redshift increasing towards the right and the top of the figure.}
\label{fig:CovMatrix}
\end{center}
\end{figure}

\subsection{Analysis pipelines}
\label{subsec:pipeline_overview}

\begin{table}
   \centering
   \caption{Priors used in the  KiDS likelihood sampling. The ranges for the four cosmological parameters  are determined by the cosmo-SLICS simulations, while the prescription from sampling the nuisance parameters describing the photometric redshifts $\Delta z_a$ and  intrinsic alignments  $A_{\rm IA}$ are taken from \citet{KiDS1000_Joachimi}. In particular, the  redshift parameters  are correlated and drawn from  a multivariate Gaussian distribution with means $\boldsymbol{\mu}$ taken from Table \ref{table:survey} (fourth column) and a  covariance matrix $\mathcal{C}_z$ described in Sec. \ref{subsec:photoz}. The shear calibration parameters $\Delta m_a$ are sampled from Gaussian priors centred on zero with a standard deviation $(\mu,\sigma)$ estimated in \citet{KiDS1000_Giblin}. The baryonic feedback parameter $b_{\rm bary}$ is used to scale the effect measured in the {\it Baryon Training Set}.}
   \tabcolsep=0.11cm
      \begin{tabular}{@{} cccccccc @{}} 
      \hline
      \hline
       Parameter       &  range & prior \\
       \hline
       Cosmology\\ 
       $\Omega_{\rm m}$ &[0.1, 0.55] & Flat\\
       $S_8$ & [0.6, 0.9] & Flat\\
       $h$ & [0.6,  0.82] & Flat\\
       $w_0$ & [-2.0, -0.5]& Flat\\
       \hline
       Nuisance\\
       $\Delta z_a\times10^2$ & [-10, 10] & $\mathcal{G}(\boldsymbol{\mu}, \mathcal{C}_z)$\\                    
       $\Delta m_1\times10^2$ & [-10, 10]& $\mathcal{G}(0.0,  1.9)$ \\  
       $\Delta m_2\times10^2$ & [-10, 10]& $\mathcal{G}(0.0,  2.0)$ \\  
       $\Delta m_3\times10^2$ & [-10, 10]& $\mathcal{G}(0.0,  1.7)$ \\  
       $\Delta m_4\times10^2$ & [-10, 10]& $\mathcal{G}(0.0,  1.2)$ \\  
       $\Delta m_5\times10^2$ & [-10, 10]& $\mathcal{G}(0.0,  1.0)$ \\  
        \hline
       Astrophysics\\
       $A_{\rm IA}$ & [-5, 5]&  Flat\\                    
        $b_{\rm bary}$ & [0, 2] & Flat\\
     \hline
    \hline 
    \end{tabular}
    \label{table:priors}
\end{table}

\begin{table}
   \centering
   \caption{Priors used for sampling the nuisance parameters in the DES-Y1 peak statistics analysis. The sampling of photometric redshifts $\Delta z_a$ and  shear bias $\Delta m_a$  nuisance parameters follows  the original cosmic shear paper by \citet{DESY1_Troxel}.  The baryonic feedback parameter $b_{\rm bary}$ is the same as in the KiDS-1000 likelihood, however there is no IA parameter here.}
   \tabcolsep=0.11cm
      \begin{tabular}{@{} cccccccc @{}} 
      \hline
      \hline
       Parameter       &  range & prior \\
       \hline
       $\Delta z_1\times10^2$ & [-10, 10]& $\mathcal{G}(0.1,  1.6)$ \\  
       $\Delta z_2\times10^2$ & [-10, 10]& $\mathcal{G}(1.9,  1.3)$ \\  
       $\Delta z_3\times10^2$ & [-10, 10]& $\mathcal{G}(0.9,  1.1)$ \\  
       $\Delta z_4\times10^2$ & [-10, 10]& $\mathcal{G}(1.8,  2.2)$ \\  
       $\Delta m_a\times10^2$ & [-10, 10]& $\mathcal{G}(1.2,  2.3)$ \\  
        \hline
       Astrophysics\\
        $b_{\rm bary}$ & [0, 2] & Flat\\
     \hline
    \hline 
    \end{tabular}
    \label{table:priorsDES}
\end{table}

Our cosmological inference pipeline heavily builds on the methods presented in HD21 and \citet{DESY1_Heydenreich}, which  we briefly overview here. First, we model the peak function by training a Gaussian Process Regression\footnote{We use the GPR toolkit provided by {\sc SciKitLearn} \citep{scikit}.} emulator (GPR) on the  measurements  from the {\it Cosmology Training Set}, after averaging over the 10 noise realisations. The GPR can subsequently produce $N^{\kappa}_{\rm peaks}$ predictions within a fraction of a second everywhere inside the parameter volume covered by the {\it cosmo}-SLICS. This therefore determines the prior ranges over $\Omega_{\rm m}, S_8, w_0$ and $h$, which we report in Table \ref{table:priors}.

Second, we must estimate the covariance matrix, which captures the correlation between the elements of our data vector, central to the error propagation.
As mentioned before,  the {\it Covariance Training Set} consists of 124 full survey realisations, each duplicated 10 times with a distinct shape noise realisation, producing 1240 {\it pseudo}-independent data vectors from which our covariance matrix C is extracted\footnote{In practice, we follow HD21 and estimate C from the average over 10 matrices, each computed from one of the noise realisation.}. Since shape noise is added at the galaxy level, cross-redshift bins are correlated with the auto-bins. We show in Fig. \ref{fig:CovMatrix} the cross-correlation coefficient matrix, defined as $C_{ij}/\sqrt{C_{ii} C_{jj}}$, which better highlights the correlations between the negative and positive peaks in each of the tomographic block. Also visible is the significant amount of correlation (and anti-correlation) present in the off-diagonal component.  This matrix contains at most $390^2$ elements and is thus invertible \citep[since 390 $< 1240$, see][]{Hartlap2007}, a criteria that is also naturally satisfied by the `clean' KiDS-1000 data vector, which contains only 210 entries,   and by the DES-Y1 data vector, which contains 120. 

Having our model and covariance matrix, we are now in a position to evaluate the likelihood $\mathcal{L}$ of the model $\boldsymbol{x}(\boldsymbol{\pi})$ with parameters $\boldsymbol{\pi}$, given the data vector $\boldsymbol{d}$.
We use the  \citet{SellentinHeavens} $t$-distribution likelihood, which is well suited for nearly Gaussian data vectors with simulation-based covariance matrices \footnote{Using instead a multivariate Gaussian likelihood along with a Hartlap factor is less accurate, see \citet{SellentinHeavens} for a full discussion.}. It is constructed as: 
\begin{eqnarray}
{\rm ln}\mathcal{L}(\boldsymbol{\pi}|\boldsymbol{d}) =   \frac{N_{\rm sims} }{2}  {\rm ln}\bigg[1 + \chi^2 / (N_{\rm sims} - 1)\bigg] + {\rm const},\mbox{\hspace{5mm} with} 
\label{eq:like}
\end{eqnarray}
\begin{eqnarray}
\label{eq:T2def}
\chi^2 = [\boldsymbol{x}(\boldsymbol{\pi}) - \boldsymbol{d}]^{\rm T}{\rm C}^{-1} [\boldsymbol{x}(\boldsymbol{\pi}) - \boldsymbol{d}] .
\end{eqnarray}
In the above, $N_{\rm sims}=1240$ is the number of realisations used to evaluate the covariance matrix C. The model depends on the four cosmological parameters $\Omega_{\rm m}, S_8, w_0$ and $h$, and on a set of  12 (9) astrophysical and nuisance parameters for KiDS (DES), which characterise the dependence of our signal on the systematic effects mentioned previously. This is an excellent approximation to the more general likelihood suggested by \citet{PercivalLike} in our case. 
Finally, the posteriors are sampled  both by the nested sampling algorithm {\sc MultiNest} \citep{Multinest} and by {\sc Nautilus} \citep{Nautilus}, implemented within {\sc CosmoSIS} \citep{cosmoSIS}.  While the latter sampler is more robust  \citep{Nautilus}, the former has been more widely used in the literature and is therefore useful to make fair comparisons with previous analyses.  We report from these the mean and 68\% credible intervals computed from the one-dimensional projected posteriors\footnote{We refer to these intervals as $1\sigma$ regions, even though strictly speaking this notation should only apply to Gaussian posteriors.}, as well as the maximum {\it a posteriori} for some of our key results.

Since our likelihood function differs from the widely used multi-variate Gaussian, the goodness-of-fit evaluation must be adapted accordingly. For Gaussian likelihoods, the $\chi^2_{\rm best-fit}$, estimated at the best-fit parameters, is to a very good approximation sampling an underlying  $\chi^2_\nu$ probability distribution, which depends only on the number of degrees of freedom $\nu$ -- this is only an approximation however, because of informative priors, non-linear modelling and correlated error bars \citep[see e.g.][]{KiDS1000_Joachimi}. A good fit will have a $\chi^2_{\rm best-fit}$ close to the maximum of the $\chi^2_\nu$ probability distribution, while a bad fit will land far in the tail, leading to a probability to exceed (PTE) that is smaller than our  acceptance  threshold, set to 0.01.
For our Student-$t$ distribution likelihood, we still assess the goodness-of-fit with  PTE values, however the  $\chi^2_\nu$ curve needs to be modified (see Appendix \ref{sec:pvalues} for details).


A few differences exist between the KiDS-1000 and DES-Y1 likelihoods which  are worth highlighting here, as these influence  the construction of our joint pipeline. First, the original DES-Y1 peak count analysis samples  $\sigma_8$ instead of $S_8$; the latter is a better option as it exactly covers the training volume and is therefore adopted for our DES-Y1 re-analysis.

Second, the treatment of intrinsic alignments are simpler in the DES-Y1 analysis: the IA contribution is estimated from the alignments of dark matter haloes, which are assumed to fully correlate with the alignment  of central galaxies. This non-linear prescription provides a single IA model that is then added to the predictions, without marginalisation. As discussed in Sec. \ref{subsec:syst_results}, not marginalising over the IA  in the KiDS analysis  slightly underestimates the total error. This is likely less important in the DES-Y1 likelihood since the statistical error is larger.

This also connects with the third difference, which is that in the baseline DES-Y1 measurement, only the auto-tomographic redshift bins are included, in an attempt to avoid possible residual contamination from unmodelled IA in the cross-redshift bins. This turns out to be an over-conservative data cut. Indeed, the recent  DES-Y1 persistent homology cosmic shear analysis from \citet{DESY1_Heydenreich} reveals that the constraints on $S_8$ are negligibly affected by these IA terms: they show that a full tomographic analysis including all cross-tomographic combinations shift the parameter by at most $0.3\sigma$ towards higher $S_8$ values, even when the inferred $A_{\rm IA}$ is as large as unity. Although their analysis is based on the different statistics (they use persistent homology instead of peak count), their results should hold here too, given that peaks are a subset of their data vectors. Therefore residual IA cannot play an important role in the DES-Y1 peak count analysis, justifying our choice to include the cross-redshift bins here (up-to-pairs, but not the triplets nor the quadruplets since these are not fully modelled yet for the DES-Y1 data).

A fourth difference in the likelihood concerns the  treatment of the baryonic feedback: in HD21 the peak statistics are measured in the {\it Magneticum} simulations to ensure  that the selected elements from the data vectors are immune to unmodelled baryonic mechanisms, but no marginalisation is included. This can potentially lead to a slightly over-optimistic precision on the DES likelihood compared to the KiDS-1000 likelihood, which includes marginalisation over the $b_{\rm bary}$ parameter. We therefore decided to include in our joint analysis the same marginalisation machinery for both the KiDS-1000 and the DES-Y1 pipelines. Moreover, we use a unique $b_{\rm bary}$ parameter to infuse baryonic feedback into both surveys, since these physical processes describe physics that affect the foreground matter distribution independently of survey-specific source selection. In total, the combined-survey analysis marginalises over nine redshift bias parameters, nine shear bias parameters, one IA  and one baryon parameter.  The sampling strategy of the  DES-related parameters are listed in Table \ref{table:priorsDES}. 
Finally, given the absence of  overlap between two survey footprints and the compatibility of the priors, the two likelihoods can be directly added at each evaluation point, without needing to consider cross-survey covariance. 

Before running the analysis on the KiDS and DES data, we validated our pipelines on simulated data, as presented in the next section, and made no further modifications to thereafter. This method is not as strong as adopting a full blinding strategy at the catalogue level, however this avenue was not available anymore since many authors were already unblinded, having worked on previous cosmic shear analyses with the same data. In these conditions, our validation strategy is an excellent option to protect ourselves against confirmation bias. 


\section{Systematic Uncertainties}
\label{sec:systematics}

As the amount of high quality lensing data keeps increasing, the statistical precision reaches unprecedented high levels, and consequently understanding and controlling the residual systematics in every segment of the data analyses has become one of the primary objectives and focus of development in the field of beyond-2pt statistics. We investigate here a number of such systematic effects that have been identified in the literature and mentioned earlier, including residual uncertainties related to interpolation in the modelling, shear calibration,  photometric redshifts, astrophysics (intrinsic alignments and baryonic feedback),  simulation-based covariance matrix, non-linear physics,  source-lens coupling and likelihood sampling strategies. Some of the systematics that are ignored in the current work are those related to the effect of source blending, depth variations, PSF leakage, or the cosmology dependence of the IA signal. These will likely become important in the future, but can be safely omitted  in current Stage-III lensing surveys \citep[see HD21,][]{DESY3_Zuercher}. Amongst those that we investigate here, many are shown to be sub-dominant or heavily suppressed by our range of $\mathcal{S}/\mathcal{N}$, while others are forward-modelled with nuisance parameters that are marginalised over in the likelihood analysis. 

\subsection{Modelling}

\begin{figure}
\begin{center}
\includegraphics[width=3.3in]{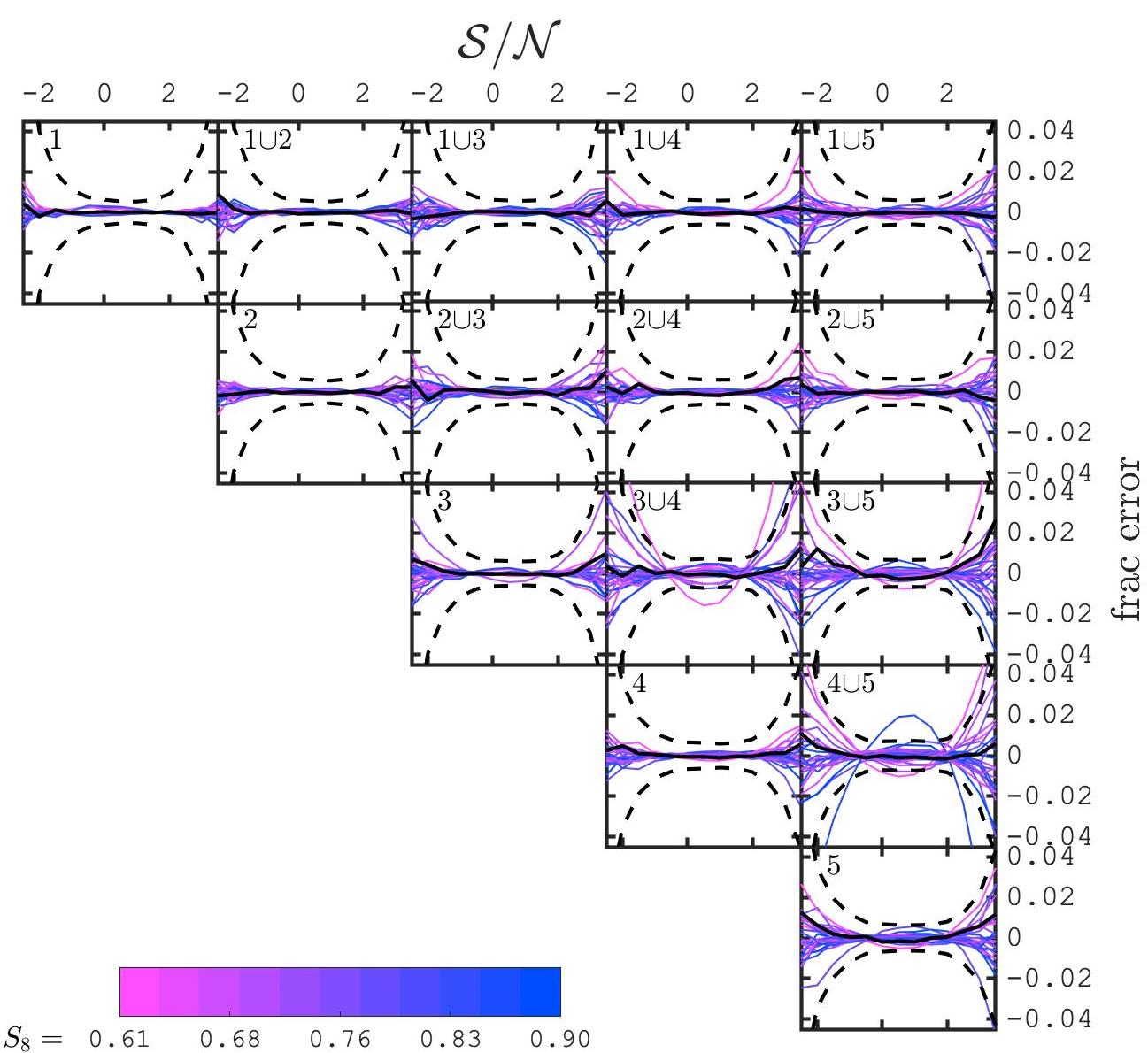}
\caption{Accuracy of the KiDS-1000 GPR emulator, computed with a leave-one-out cross-validation test. The results are colour-coded with the $S_8$ value of the removed training point, and compared with the statistical precision on the measurement of the peak function (shown with the black dashed lines). The black solid line indicates  the $\Lambda$CDM node, and the different panels  show the auto- and cross-redshift (up to pairs) measurements; the other 15 tomographic combinations show a similar precision  and hence are not shown. The outliers seen in a few panels are of extreme $S_8$ values and as such required the test emulator to extrapolate; this does not occur with the full emulator, and should therefore not be considered when estimating the interpolation error. }
\label{fig:GPR_CV}
\end{center}
\end{figure}

The accuracy of the peak function modelling has two aspects to be considered:  we must understand both how well the cosmology scaling is captured by the emulator, and whether any elements of our data vector are affected by either resolution limits of our  simulations or the choice of gravity solver.  Regarding the first aspect,  the cosmo-SLICS have been shown to match in precision the  commonly  used {\sc CosmicEmu} \citep{Coyote3}, and to even outperform the latter in terms of range, benefiting from more training nodes \citep{cosmoSLICS}. The GPR interpolation uncertainty  is fully propagated in the likelihood, and is quantified with a leave-one-out cross-validation test, in which the emulator is trained on all but one node, producing predictions at that node that are then compared with the measurement. As shown in Fig. \ref{fig:GPR_CV}, we cycle over all 26 nodes in this way and estimate an upper limit on the error (since the actual GPR has all the nodes). The accuracy degrades towards large positive or negative $\mathcal{S}/\mathcal{N}$ values, but most of the cross-validation lines lie well within the statistical precision on the data, shown with the dashed black lines. There are a handful of exceptions with poorer accuracy, attributed to removing extreme values of $S_8$ from the training set and therefore effectively demanding the GPR to extrapolate. With these edge nodes included, the full emulator has no such outliers. In fact, as argued in HD21, the most reliable estimate of the emulator's precision is evaluated by removing the fiducial cosmology and training on the others, which is shown as the thick black lines in the figure and always well within the statistical precision.  For the KiDS modelling, the interpolation error is mostly at the 1\% error over the range $-1.0 < \mathcal{S}/\mathcal{N} < 3.0$ (our  `clean'  range), and is otherwise always under 10\%. Similarly, the DES interpolation error is everywhere under 2\% (see HD21),  the difference coming from the choice of $\mathcal{S}/\mathcal{N}$ cuts. This will likely be a limiting factor for future data analysis with sub-percent accuracy requirement on the  modelling, and will  be addressed by increasing the number of  nodes  in the next generation of the {\it Cosmology Training Set}. At the moment however, the interpolation error is low enough for our analysis. We nevertheless include it in our error budget by averaging over the square of the residuals (after the outliers have been removed):
\begin{eqnarray}
{\rm Cov}_{\rm interp} = {\rm diag}\left<  \left( N_{\rm peaks}^{\rm GPR} -  N_{\rm peaks}^{\rm sim}\right)^2\right> \, ,
\label{eq:Cov_GPR}
\end{eqnarray}
 and adding this to our statistical covariance. We could have instead used the errors directly provided by the Gaussian Process Emulator, however \citet{Homology} have shown that the two methods yield  posteriors with negligible differences. 
This contribution, although small, helps with the goodness-of-fit in the data analysis. 

The second aspect, concerning the training sets themselves, is discussed below in the section on $N$-body resolution.

\subsection{Shape calibration}


Table \ref{table:survey} shows the average multiplicative correction factors $m_a$ that must be applied to the observed galaxy shapes in order to correct for a combination of residual  PSF leakage, blending and measurement noise, as assessed from  \citet{KiDS1000_Giblin}. While in A21 the uncertainty on the shape calibration is absorbed directly in the analytical covariance matrix, our simulation-based method works instead  at the level of the data vector, as for all  other nuisance parameters.  The $M_{\rm ap}$ estimator itself is unbiased (see Eq. \ref{eq:Map}), however we must propagate forward the uncertainty on the $m_a$ calibration. The impact of potentially mis-calibrated shape measurements is estimated  by infusing a non-corrected global term $m_a \rightarrow m_a + \Delta m_a$ directly in the simulations and measuring the effect on the different elements of the peak function $N_{\rm peaks}^{\kappa}$. As we show later, this systematic effect is completely subdominant compared to the others due to the tight priors on $\Delta m_a$ (reported in Table \ref{table:priors}), and hence it is sufficient to model its impact with a reduced accuracy. In HD21 the estimation is based on a linear regression (i.e. $\partial N_{\rm peaks} / \partial \Delta m$ per data element) that is fit through 10 values of $\Delta m_a$. We use here only two points, at $\pm1\sigma$, which is sufficient given the small values of $\Delta m_a$. The measured $\partial N_{\rm peaks} / \partial \Delta m$ is  further discussed in Appendix \ref{sec:syst_pipeline}, and is used to modify the data vector for any value of $\Delta m_a$ (see Eq. \ref{eq:syst}) sampled in the likelihood. For cross-redshift tomographic bins, we use the mean shift, e.g. $\Delta m_a^{1\cup2}=(\Delta m_a^1 + \Delta m_a^2)/2$,   which is consistent with what is currently done for all shear two-point function analyses. We could instead use an $n_{\rm gal}$-weighted mean to compute the $\Delta m_a$ shift in cross-redshift tomographic bin, however this should have a negligible effect given the tight priors on these parameters, and we therefore leave this for the future. 

\subsection{Photometric redshifts}
\label{subsec:photoz}

The KiDS-1000 uncertainty on the redshift distributions has been fully quantified in \citet{KiDS1000_redshifts}, where it is shown that the  mean of the $n(z)$ is captured to a high accuracy, varying by no more than  
0.014 at the $1\sigma$ level\footnote{The full shape of the $n(z)$ is less accurate than its mean, and which consequences we leave for future work. }. The posteriors on the mean of the redshift distributions are used as priors on nuisance parameters in the current work, summarised in Table \ref{table:priors}.
In this case however, the five redshift bias parameters  $\Delta z_a$ must be drawn from a  correlated distribution. This is achieved in a two-step operation where we first draw five uncorrelated numbers from the priors, then rotate into the correlated space using a Cholesky decomposition of the redshift covariance matrix:
\begin{gather*}
C_{z} \times 10^{5}= 
\begin{bmatrix} 
11.20  & 2.600   &1.562  & 0.056   &0.622\\
2.600  & 12.78   &4.081 &-1.692 &-0.2140\\
1.562  & 4.081   &13.81 &-1.139  & 0.525\\
0.056  &-1.692  &-1.139  & 7.551   &3.054\\
0.622  &-0.2140   &0.525 & 3.054  &9.496
\end{bmatrix} \, ,
\end{gather*}
which results in a correlated sampling of these five nuisance parameters  \citep[see A21,][for more details]{KiDS1000_redshifts}.
We produced a dedicated set of {\it Redshift Training Set} simulations in which the $n(z)$ are shifted, but which are otherwise identical to the {\it Cosmology Training Set} at the fiducial cosmology. 
Following HD21, we measure the peak function on full mock surveys with 10 shifts, each with a slightly different value of $\Delta z_a$ sampled from the prior, 
then extract a linear fit per data element and estimate $\partial N_{\rm peaks} /  \partial \Delta z_a$. This derivative is used to forward model redshift uncertainties on our data vector for arbitrary  $\Delta z_a$ values. Again, we use the mean shift when considering cross-redshift bins, and the DES-Y1 $\partial N_{\rm peaks} /  \partial \Delta z_a$ measurements from HD21.

\subsection{Astrophysics}

Cosmic shear measurements  are strongly affected by IA and baryon feedback. Using the {\it IA} and the {\it Baryons Training Sets} described in Sec. \ref{subsec:sims}, we estimate in a similar way  $\partial N_{\rm peaks}/  \partial A_{\rm IA}$ and  $\partial N_{\rm peaks} /\partial  b_{\rm bary}$, where $A_{\rm IA}$ and $b_{\rm bary}$ are free parameters that control the levels of IA and baryon contamination, respectively.  The IA derivative is obtained by linear fitting the peak function's response to changes in $A_{\rm IA}$,  measured from the {\it IA Training Set} infused with $A_{\rm IA}$ = 2.0,1.0, 0.0, $-$1.0 and $-$2.0. Since IA is currently not well constrained and the NLA parameterisation is an effective model, we adopt a wide top-hat prior over the range [$-$5.0 ; 5.0], as argued in \citet{KiDS1000_Joachimi}.  This extrapolates our fit to larger  $A_{\rm IA}$  values, which can in principle become inaccurate, however in the end the $3\sigma$ region of our posterior is fully contained within the training range (see Sec. \ref{sec:results}). Similarly, the baryon derivative is measured  from the {\it Baryons Training Set}, which we use to infuse a baryonic correction whose strength is controlled by the parameter $b_{\rm bary}$. The case  $b_{\rm bary}=0.0$ corresponds to a dark matter-only universe, while $b_{\rm bary}=1.0$ corresponds to the case where the feedback processes  is exactly described by the {\it Magneticum} physics.  There is a large uncertainty on the amplitude of this baryon correction, hence we scale the measured baryonic correction with  a free parameter $b_{\rm bary}$. Since the  {\it Magneticum} suites are already a strong model \citep[see][for a comparison with other hydrodynamical simulations]{Martinet21}, we sample the range $b_{\rm bary} \in [0.0,  2.0]$, thereby spanning a variety of realistic models (albeit imposing a fixed  shape for the relative signal).   As seen later,   low $b_{\rm bary}$ values are not well constrained by the data while larger values are strongly disfavoured, hence we do not extend the prior limit beyond 2.0.

\subsection{Implementation of forward-modelled systematics}

Four sources of systematics are forward-modelled in our pipeline. Following  \citet{DESY1_Heydenreich}, we  construct systematics-infused data vector as:
\begin{flalign}
&N_{\rm peaks}^{\rm syst}({\boldsymbol \pi}, \Delta m_a,\Delta z_a, A_{\rm IA},b_{\rm bary}) = & \nonumber \\
&\hspace{1.3cm}N_{\rm peaks}^{\rm GPR}({\boldsymbol \pi})+  \left[\partial N_{\rm peaks} /\partial \Delta m_a\right] \Delta m_a  +\left[\partial N_{\rm peaks} /\partial \Delta z_a \right]\Delta z_a \, \, ... & \nonumber \\
&\hspace{2cm} + \left[\partial N_{\rm peaks}  /\partial A_{\rm IA}\right] A_{\rm IA} +  \left[\partial N_{\rm peaks} /\partial  b_{\rm bary}\right]  b_{\rm bary} \, ,&
\label{eq:syst}
\end{flalign} 
where the  twelve parameters ($\Delta m_a,\Delta z_a, A_{\rm IA},b_{\rm bary}$) are sampled from the priors described in Table \ref{table:priors}. We marginalise over these nuisance parameters when inferring the values of the cosmological parameters. Equation (\ref{eq:syst}) assumes that these different systematics are independent of cosmology and from each other,  which we know is not entirely true. It has been shown that the cosmology dependence of the baryon feedback is a second order effect \citep{BAHAMAS}, supporting our simplified approach, however  the intrinsic alignments couple to the tidal field that is in itself cosmology dependent. The shear calibration and redshift errors are independent of cosmology {\it a priori}, however the derivatives of the peak function with respect to $\Delta m_a$ and $\Delta z_a$ are not (see HD21), a secondary effect we neglect here. Moreover, it has been shown that the photometric and shape calibration errors are sometimes correlated \citep{DESY3_MacCrann}. Although these approximation will become important in Stage-IV surveys, the current level of statistical precision allows us to relax the  modelling of these effects without hurting our results. We illustrate this point  in Section \ref{subsec:syst_results} by running inference MCMC chains in which the modelling of some or all of these systematic effects are switched off: the minor impact this has on the inference validates this approach. We also assume here that these systematic effects have a linear dependence on the nuisance parameter, which is probably not entirely true, but has been shown to be good enough for Stage-III lensing data in \citet[][see their figure 7]{DESY1_Heydenreich}.

 \subsection{Other sources of systematics}

In addition to the main systematic effects described in the last section, we consider here other known sources of errors that could potentially impact our results. 

\subsubsection*{$N$-body resolution}
\label{subsubsec:Nbody}
Being completely simulation-based, our analysis  relies on the quality of the underlying training samples. As mentioned already in Sec. \ref{subsec:sims}, the {\it Cosmology Training Set} has been shown to closely reproduce the non-linear clustering of the Cosmic Emulator \citep{Coyote3}, which is based on a completely independent $N$-body code. This agreement between different gravity solvers is key to assert the accuracy of the non-linear solution to structure formation \citep[see, e.g.][for a comparison between different $N$-body solvers]{EuclidEmulator}, and the convergence of the solution must be assessed via a comparison with calculations carried out with a higher force/mass resolution simulations. As shown in HD21, known limits in the mass resolution of the {\it cosmo}-SLICS used for the  peak function emulation mainly affect high peaks. More precisely,  $N^{\kappa}_{\rm peaks}(\mathcal{S}/\mathcal{N} > 4.0)$ is systematically under-predicted by tens of  percent, while the $\mathcal{S}/\mathcal{N} = 4.0$ count is affected by no more than 5\%. This is in fact one of the main justification for our initial choice of upper $\mathcal{S}/\mathcal{N}$ limit. 

The KiDS-1000 data are deeper than DES-Y1, and hence the sensitivity to such non-linear effects could be accrued here. We verify this by running our cosmological inference on the peak count statistics measured from the SLICS-HR, in which the increased force resolution results in a slightly larger number of large positive and negative peaks. Details are presented in Appendix \ref{sec:syst_pipeline}, but in short our data selection and marginalisation scheme almost completely protects us  against this, yielding no noticeable shifts on $\Omega_{\rm m}$ nor  $S_8$. 
As in HD21, we nevertheless compute a multiplicative factor from the ratio between the SLICS-HR and the mean of the SLICS and optionally apply it on our model predictions during the likelihood sampling. The overall effect is smaller than the baryon and IA corrections, hence marginalising over these latter two significantly washes out the impact of inaccurately-modelled non-linear physics under question here. In the future we intend to look into multi-fidelity emulators as in \citet{Bird_Multifidelity}. The T17 and {\it Magneticum} simulations were produced with a different $N$-body solver, and we show later that their reduced spatial resolution can affect quite significantly the peak count statistics. We treat this as a further uncertainty on the small scale physics and optionally add their scatter to the theoretical error in the covariance matrix, similar to Eq. (\ref{eq:Cov_GPR}). In two-point statistics analysis, this would be equivalent to including a theoretical error in the covariance matrix to account for difference between the $P(k)$ predictions provided by {\sc HaloFIT} \citep{Takahashi2012}, {\sc HMCode} \citep{2016MNRAS.459.1468M} or the {\sc baccoEMU} \citep{BACCOEmu}, which can have a significant impact on the results \citep{DESY3_BACCO}.

\subsubsection*{Ray-tracing approximations}
Our ray-tracing method  in itself contains approximations and algorithmic components that are bound to affect to some level the lensing statistics. Most importantly, the finite thickness of the mass sheets and the randomisation process between them destroys correlations along the line of sight; in particular it can slice large galaxy clusters in two, and no structures larger  than 257.5 $h^{-1}$Mpc can exist along the line-of-sight in our light-cones (except for the T17 mocks, which we discuss below). This suppresses some  of the large-scale power, as documented in  \citet[][see their Appendix B]{HSCmocks}. However, smaller structures, such as those probed by the peak statistics, are left completely unaffected by this, which is why no forward modelling is needed here. This has been measured specifically for peak statistics in \citet{LensingHyperparameters} where it was found to play a subdominant role even for Stage-IV surveys. Of course, full on-the-fly light-cones such as the `Onion Universe' methods \citep{Onion} avoid these problems, which we will consider for future analyses.  The T17 simulations have thinner mass shells of 150.0 $h^{-1}$Mpc, but they are constructed such that the structures are preserved in groups of three shells, thus yielding a coherence length of 450.0 $h^{-1}$Mpc, further suppressing this residual systematic effect.

Another source of error comes from the fact that  our simulations assume the Born approximation in the flat-sky limit, which introduces small inaccuracies at high-$\ell$ and low redshift, respectively \citep{Hilbert_LensSimAccuracy}. However, these are affecting the signal at a level much smaller than the statistical accuracy of  our lensing data, and are not expected to matter here.

\subsubsection*{Covariance matrix}
Estimation of the covariance matrix is one of the main computational challenges for non-Gaussian weak lensing probes, as it requires a large number of simulations with a resolution  that is high enough to capture the non-linear physics being measured. Resorting to approximate methods such a {\sc Flask} \citep{FLASK} and ICE-COLA \citep{ICE-COLA} can significantly lower the computational cost of creating such mocks, but at the price of a reduced precision on the physics under investigation. We instead opted for mocks produced by a full $N$-body suite, our {\it Covariance Training Set}, and are therefore only limited by the number of mocks and their box size. To test the convergence of our covariance matrix with respect to $N_{\rm sims}$, we run an inference analysis in which we increase the number of pseudo-independent realisations to 2120 (and adjusted the likelihood $N_{\rm sims}$ parameter accordingly), and find an excellent match to the posterior, with only the tail of the distribution being slightly modified. We could also have opted for a data compression such as in \citet{DESY3_Zuercher} but that is not necessary given our results have converged, and our choice of likelihood accounts for the noise in the covariance matrix. 

The simulation box size could also affect our results, however it has been shown in \citet{cosmoSLICS} that the SLICS contains about 75\% of the ``super-sample covariance" term (SSC), when applied to 2pt statistics, yielding constraints on cosmological parameters that are highly accurate. Although this has not been demonstrated to date, peak count statistics are thought to be even less affected by the SSC, given that the covariance is close to being Poissonian, not Gaussian.  As such, it scales with the number of peaks measured, which is independent of the survey window. In addition, as mentioned earlier,  \citet{KiDS1000_BurgerDSS} finds for the density-split statistics an excellent agreement between  the SLICS covariance and that from full sky  log-normal {\sc Flask} mocks (which contain an incomplete contribution from the trispectrum term but the full SSC), supporting our claim that the partly missing SSC must have a minimal influence on our error budget. This is also consistent with the recent findings from \citet{SSC_Linke} according to which the SSC term affects only the Fourier space estimators, whereas covariance matrix measured from intra-survey real-space statistics such as the $M_{\rm ap}$ are unbiased.

\subsubsection*{Source-lens coupling and blending}

An important difference between real and mock galaxies is that those in the data are clustered, which leads to a number of effects that are systematically absent from the calibration sample. For example, the quality of the shape measurements is lowered in regions of high density due to blending and obscuration. More importantly, the uncertainty in photometric redshifts is particularly severe in such areas, which often results in cluster members being wrongly assigned a higher redshift. This subsequently creates a small population of apparently high-redshift outliers that carry an unexpectedly weak shear component, thus diluting the overall lensing signal.   Correcting for  this can be partially achieved with `boost factors', however it was shown in HD21 and  \citet{DESY3_Zuercher} that even though the excess clustering around high peaks is indeed measured in the data, the impact this has on the inferred cosmology  can be safely ignored. It was also shown in \citet{DESY3_Gatti_source_clust} that source clustering had a minimal effect on the peak count statistics, supporting our choice to neglect this here.

\subsubsection*{Sampling the likelihood}
Our likelihood sampling strategy, described in Sec. \ref{subsec:pipeline_overview}, assumes a flat prior for the four main cosmological parameters and the two astrophysical parameters ($A_{\rm IA}$ and $b_{\rm bary}$), and Gaussian priors for the parameters associated with photometric redshifts and shape calibration. This is not strictly speaking a non-informative approach, 
however the prior edges about the key measured parameters are sufficiently broad to have negligible impact on the posterior.  Since it is found in \citet{Lemos} that {\sc MultiNest} tends to yield slightly over-precise constraints, we use the {\sc Nautilus} sampler for our fiducial results, but report both.

We also note that our cosmology sampling strategy is different from the other KiDS-1000 cosmic shear analyses, mainly due to the volume where our  emulator is valid. For example, A21 sample uniformly the parameters $S_8$, $\omega_{\rm c}\equiv \Omega_{\rm c} h^2$, $\omega_{\rm b}\equiv \Omega_{\rm b} h^2$, $h$ and $n_{\rm s}$. This choice  is designed to avoid regions of parameter space that are strongly disfavoured by external data, and it was shown in \citet{KiDS1000_Joachimi} that while it disfavoured high $\Omega_{\rm m}$ values already in prior space,  the resulting $S_8$ prior space is  highly uninformative. We could have taken a similar approach, however our emulator is much quicker, and hence it is more natural to sample the full training space, ensuring a wide sampling of $\Omega_{\rm m}$, $S_8$ and $w_0$.
 
 Another aspect that currently limits our sampling strategy is the fact that we hold the value of many parameters fixed, notably $\Omega_{\rm b}$ and $n_{\rm s}$. In contrast, the DES-Y3 peak count  analysis of \citet{DESY3_Zuercher} use derivatives to marginalise over variation in these parameters, following the approach we adopt for IA and baryons. Neglecting to account for these has a small effect on current data sets (the DES-Y3 joint peaks+power spectrum analysis finds to be of about 0.13$\sigma$), which are thus ignored here.
 

\subsubsection*{$M_{\times}$ modes}
\label{subsec:null}

The observed weak lensing signal can generally be decomposed into a combination of $E$- and $B$-modes, the latter of which can be estimated for any measurement by rotating all galaxies by 45 degrees; therefore, for the aperture mass map statistics, it is often referred to as $M_{\times}({ \boldsymbol \theta})$. The cosmic shear signal being a pure $E$-mode generator to first order, measurements of $B$-modes are therefore routinely used to assess the presence of residual systematics in lensing data \citep[see, e.g. ][for a recent application to peak statistics]{DESY3_Zuercher}. Whereas the two-point function $B$-mode signal is zero in absence of systematics, the construction of aperture mass maps on a grid inevitably injects non-zero $M_{\times}$-modes  due to  the missing contribution from sub-pixel scales \citep{2006A&A...457...15K}. This can be important: for a small-angle cut-off scale of 10 arcsec and an aperture of $\theta$=2.0 \mbox{arcmin},   $B$-modes measured this way can reach about 10 percent the size of the $E$-mode $M_{\rm ap}^2$ signal. This effect is accentuated for larger cut-off scales and smaller opening angles $\theta$.  Given our pixel scale of 35 arcsec, we do expect non-zero $M_{\times}$-modes to be introduced by our aperture map making, which we fully quantify in   Appendix \ref{sec:Bmodes}. We show therein that the level of contamination is consistent with noise, that there is no evidence for residual systematics in the data from this measurement, hence that our cosmological analysis is clean of $B$-modes.

\subsection{Peak count to cosmology pipeline validation}
\label{subsec:validation}

\begin{figure*}
\begin{center}
\includegraphics[width=5.5in]{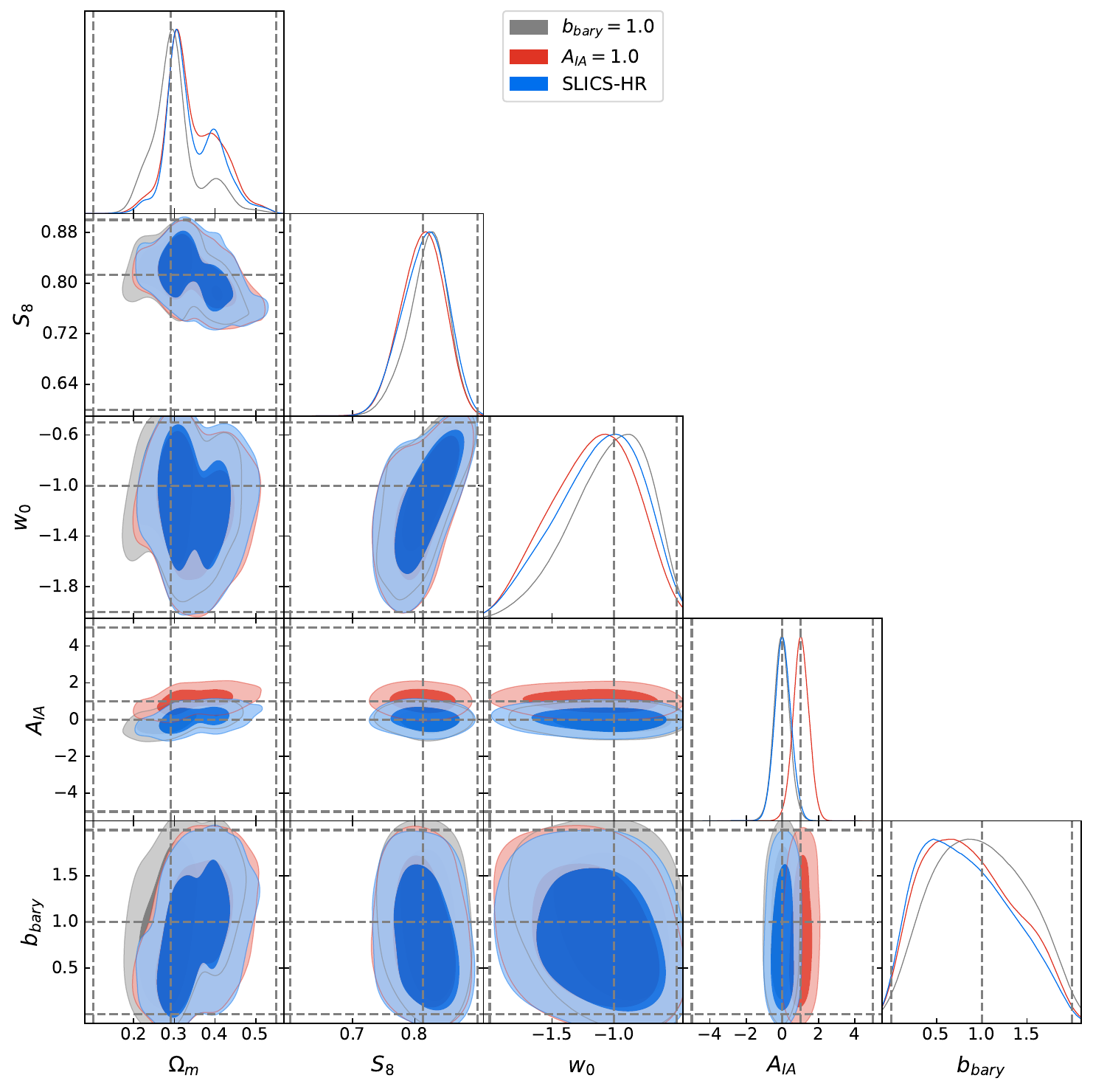}
\caption{Full inference analysis on the {\it Validation Set} (SLICS-HR) peak count data (blue) with {\sc Multinest}, optionally infused with intrinsic alignments (red) or baryon feedback (grey). We marginalise over these two effects plus shape calibration and photometric uncertainty. 
The priors are shown by the dashed grey lines at the edge of the panels, while the cross-hairs show the input truths. }
\label{fig:cosmo_syst_infusion}
\end{center}
\end{figure*}

We test our KiDS-1000 cosmology inference pipeline by analysing simulated data of known cosmology, infused with a controlled amount of residual systematics.
In order to avoid confirmation bias, these tests are carried out with the {\it Validation Set}, which have not been used in the cosmology training nor for the covariance estimation, 
with an  $N$-body  force resolution that is  higher than the other simulations used in this work\footnote{We have further verified that the $w$CDM cosmology is correctly inferred when analysing data from the {\it Cosmology Training Set} but these tests are easier to satisfy since the data is used for training the emulator. We discuss these in greater details in Appendix \ref{sec:syst_pipeline}.}.  In addition, we use the forward-modelling approach presented in Sec. \ref{sec:systematics} to infuse the simulated data vectors with either intrinsic alignments (assuming $A_{\rm IA} = 1.0$) or baryonic feedback (with  $b_{\rm bary}=1.0$). Fig. \ref{fig:cosmo_syst_infusion} shows the results for these three analysis cases. The maxima of the projected posterior distributions are all centred on the input truth, except for the $b_{\rm bary}$ parameter, which are away from zero even in the no-baryon cases. This is a projection effect similar to those discussed in  \citet{KiDS1000_Joachimi}, \citet{LensingProjection} and \citet{DESY3-KiDS1000}, and we have verified that reducing the lower prior limit to $b_{\rm bary}=-2.0$ pushed both the red and blue maxima towards the ground truth.

  In this test, there is a secondary solution for $\Omega_{\rm m}\sim0.4$ that is unexpected, and not observed in other peak count analyses \citep{Martinet18, DESY3_Zuercher, HSCY1_peaks_sims}. As detailed in Appendix \ref{sec:syst_pipeline}, this feature persists when analysing data from the {\it Cosmology Training Set} at the fiducial cosmology,  from the {\it Baryons Training Set} and from the T17 mocks, but can vanish at other cosmologies. This is caused first by the poor sensitivity of the current lensing data to $\Omega_{\rm m}$, as also seen in the large   $\Omega_{\rm m}$ scatter reported in A21 between different two-point functions,  but more importantly by limits in our GPR emulator, whose residual inaccuracy mostly affects this parameter. Inferring $\Omega_{\rm m}$ from single mock survey realisations (as opposed to a mean over several light-cones or shape noise realisations) yields posteriors drawn from either one of these peaks, resulting in occasional strong biases on this cosmological parameter. We thoroughly verify that only $\Omega_{\rm m}$ is affected by this, and therefore do not report its value in our main analyses. See  Appendix \ref{sec:syst_pipeline} for full details.

However, the secondary  $\Omega_{\rm m}$ solution corresponds to an $S_8$ posterior that is slightly lower than the main solution, which means that if a particular realisation of the data prefers this region, it will on average have an $S_8$ value  about 0.03 lower, which is of the size of our statistical precision. Conversely, realisations that prefer lower $\Omega_{\rm m}$ tend to have $S_8$ values that are 0.02 higher than the input truth. We further observe that  this is not always the case: some individual mock survey realisations from the {\it Covariance Training Set} have a best fit $\Omega_{\rm m} \sim 0.45$, yet their $S_8$ is unbiased compared to the input truth. Given that this $0.02-0.03$ shift is about a $1\sigma$ shift, this potentially dominates the systematic error budget on $S_8$, which we therefore must report as $\sigma$(syst)$=^{+0.03}_{-0.02}$.

 This additional systematics error take its roots from the tilt in the $[S_8 - \Omega_{\rm m}]$ posterior, which indicates residual correlation between these two parameters. We can suppress this tilt, and hence the additional error, by replacing $S_8$ with $\Sigma_8^\alpha \equiv \sigma_8 [\Omega_{\rm m}/0.3]^\alpha$, where $\alpha$ is the parameter that best fits the $[\Omega_{\rm m} - \sigma_8]$ degeneracy. According to this metric, $\Sigma_8^\alpha$ is the most robustly measured quantity from peak statistics, with no need for a standalone $\sigma$(syst) term, in this case a significant advantage. With the validation data, we find $\Sigma^{\alpha}_8 = 0.824^{+0.033}_{-0.033}$, with $\alpha=0.582$, in excellent agreement with the input truth of 0.811 with the same $\alpha$. We report the measurements of both $S_8$ and $\Sigma^{\alpha}_8$ in this paper, but while emphasise is on the former to better compare with previous measurements from the literature, the latter is more robust and has interesting properties which we highlight as well, notably on increasing the agreement with previous KiDS-1000 measurements and lowering the tension with external probes.

Back to Fig. \ref{fig:cosmo_syst_infusion}, we observe that the posteriors on $w_0$ and $b_{\rm bary}$ are wide and significantly overlap with the prior limits,  and we thus expect to be unable  to place meaningful constraints on these parameters  with our main KiDS-1000 analysis alone. We observe a degeneracy in the $[S_8 - w_0]$ plane here, however we show in Appendix \ref{sec:syst_pipeline} that it is not always seen when analysing other cosmologies,  making it impossible to draw physically meaningful conclusions about this. Only the $[S_8 - A_{\rm IA}]$ plane  is well constrained with the current KiDS-1000 peak count analysis: we achieve a 4.4\% precision measurement on $S_8$, with $S_8^{\rm SLICS-HR}=0.816^{+0.039}_{-0.033}$ (truth is 0.813), and a precision of $\sigma_{A_{\rm IA}}=0.45$ on $A_{\rm IA}$,  sampling the likelihood with {\sc Multinest}. 

 The DES-Y1 pipeline  validation is presented in HD21, while that for the joint KiDS-DES  is presented in Appendix \ref{sec:syst_pipeline}, showing again an  excellent agreement between the inferred cosmology and the input truth.

\section{Results: KiDS-1000}
\label{sec:results}


\begin{figure}
\begin{center}
\includegraphics[width=3.3in]{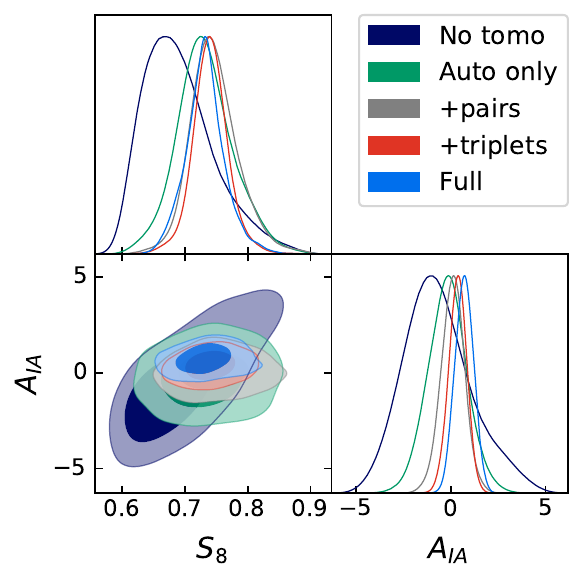}
\caption{KiDS-1000 constraints on the two best-measured parameters from  peaks count statistics, for different selection of redshift bins.  Tomographic analyses all break the $S_8 - A_{\rm IA}$ degeneracy.}
\label{fig:cosmo_wCDM}
\end{center}
\end{figure}

We present in this section the results from our cosmological  inference analyses, beginning with the fiducial KiDS-1000 pipeline, then reporting on the importance of various selection cuts and systematic effects. 
For reasons explained in Sec. \ref{subsec:validation}, we report only the constraints on $S_8$ and $A_{\rm IA}$; results are summarised in Table \ref{table:cosmo_pipeline_test} and further condensed in Fig. \ref{fig:S8_constraints}.

From our fiducial full tomographic KiDS-1000  analysis of the measurements presented in Fig. \ref{fig:N_peaks_data}, we obtain:
\begin{eqnarray}
S_8^{\rm KiDS} =  0.733^{+0.032+0.020 \,(\rm syst)}_{-0.032-0.030 \,(\rm syst)}\, ,  \hspace{1cm} A_{\rm IA} =   0.71^{+0.49}_{-0.49} \, , 
\end{eqnarray}
\begin{eqnarray}
\Sigma_8^{\rm KiDS} =  0.765^{+0.030}_{-0.030}\,,   \hspace{1cm} \alpha = 0.600 \, , 
\end{eqnarray}
after marginalising over three cosmological parameters ($\Omega_{\rm m}$, $w_0$ and $h$) and 11 nuisance parameters ($5\times\Delta m_a$, $5\times\Delta z_a$, $b_{\rm bary}$). Unless explicitly mentioned, all quoted parameter constraints correspond to the mean $\pm1\sigma$ region of the marginalised posterior, not to be confused with the point of maximum likelihood in the higher-dimensional space. This is therefore a 3.9\% measurement of the structure growth parameter $\Sigma_8$. The best-fit model is shown with the red line in Fig. \ref{fig:N_peaks_data}. 
The joint constraints on two of these parameters are shown in Fig. \ref{fig:cosmo_wCDM}, along with results from different selections of  tomographic bins. Importantly, the strong $S_8 - A_{\rm IA}$ degeneracy seen in the no-tomographic case (the tilted dark purple contour) is lifted by tomographic decomposition, which capture the different redshift dependence of the cosmological and IA signals. Indeed, in the no-tomographic case only, and under the NLA framework, large $S_8$ values can be hidden by large tidal alignments, both fitting equally well the same data. However, as seen by the coloured histograms in Fig. 3, the cosmological signal in all tomographic bins is affected by changes in $S_8$, while IA mostly modifies the parts of the data vector that include the lowest tomographic bins. This difference allows one to break the $[S_8-A_{\rm IA}]$ degeneracy,  an important verification of our IA modelling. This result would be slightly different had we included redshift evolution of the IA signal, but this effect will be subdominant given the size of our statistical error bars. This will clearly need to be investigated with upcoming data sets.

Back to Fig. \ref{fig:cosmo_wCDM}, all tomographic additions contribute to further tightening the constraints, once again demonstrating the power of using cross-redshift bins in non-Gaussian statistics. We also observe that all cases shown in Fig. \ref{fig:cosmo_wCDM} are consistent, providing statistical robustness to our measurement.

At our best-fit parameters the measurement yields a $\chi^2$ of 250, which  reduces to   $\chi^2_{\rm red}$ = 1.22 after dividing by $\nu = (220 - 4.5) = 205.5$ degrees of freedom. Note that although we use six  unconstrained parameters\footnote{We do not count as free parameters those nuisance parameters for which we impose a tight prior.} in our likelihood evaluation (the four cosmological parameters plus $A_{\rm IA}$ and $b_{\rm bary}$), it was shown in \citet{KiDS1000_Joachimi} that an effective number of $\nu = 4.5$ free parameters better describes the weak lensing data given the existing correlations and degeneracies, results which we have used here\footnote{It is not guaranteed that the exact same effective number of degrees of freedom applies here, given that the likelihood is not sampled over the same volume. We have checked that our goodness-of-fit is robust over choices for this quantity, with PTE varying between 0.48 and 0.37 over the range $2 < \nu < 7$.}. Our PTE for this measurement is 0.43, which is well above our threshold of 0.01,  using the non-$\chi^2$ distribution described in Appendix \ref{sec:pvalues}. 
It is worth noting that the KiDS-1000 shear two-point correlation functions and band power analyses had a lower goodness-of-fit, with  PTE=0.034 and 0.013, respectively. 

\begin{figure}
\begin{center}
\includegraphics[width=3.3in]{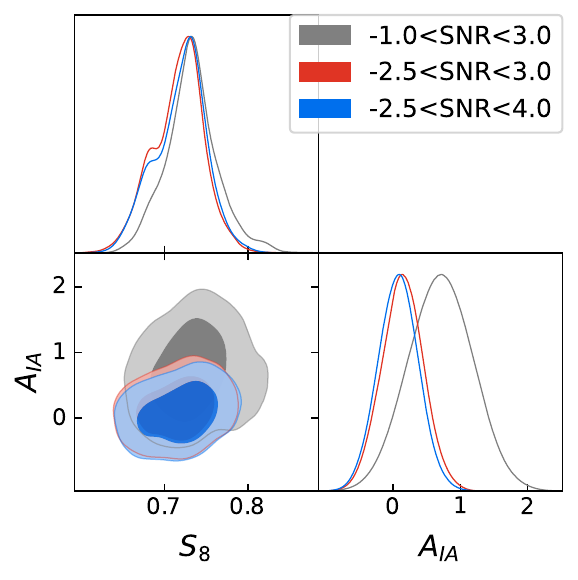}
\caption{Effect of $\mathcal{S}/\mathcal{N}$ cuts on the KiDS-1000 constraints.   Large peaks ($\mathcal{S}/\mathcal{N}$ $>$ 3) slightly increase the statistical precision on $S_8$, as seen by comparing the red and blue contours (see values in Table \ref{table:cosmo_pipeline_test}).
Negative peaks ($\mathcal{S}/\mathcal{N}$ $<$ -1), included in the red but excluded from the grey contours, help in breaking the $[S_8-A_{\rm IA}]$ degeneracy. The grey and blue contours correspond to the `clean' and `aggressive' cases, respectively.}
\label{fig:cosmo_SNRcuts}
\end{center}
\end{figure}

In many previous analyses, sampling and marginalisation over $w_0$ is often excluded, being considered an extension to the vanilla $\Lambda$CDM scenario.  In the present case, fixing $w_0$ to $-1.0$ when sampling the likelihood\footnote{This still uses the same $w$CDM emulator, but only varying the other three cosmological parameters.} results in minor changes to the reported $(S_8, \Sigma_8, A_{\rm IA})$ constraints, leading to  $0.732^{+0.029}_{-0.029}$,  $0.767^{+0.026}_{-0.026} $ and  $0.73^{+0.48}_{-0.48}$. Interestingly, 
we find that the impact of opening up the $w_0$ dimension is far lower than for the two-point statistics, where \citet{KiDS1000_Troester} finds a degradation by a factor of a few on the $S_8$ constraints (compare their figures 1 and 6). Different degeneracy-breaking directions are likely causing this difference, which is promising for upcoming measurements of $w_0$ with alternative  statistics \citep[see][for a Stage-IV lensing forecast on the dark energy parameter with peak statistics]{Martinet21}. 

One of the key questions to be explored by beyond-2pt statistics concerns the exact origin of the non-Gaussian cosmological information. Large peaks are often associated with massive galaxy clusters, which are known to be highly sensitive to the dark energy equation-of-state parameters for instance, however the wide  projection effect and the  fact that baryons, IA and non-linear physics maximally affect these large $\mathcal{S}/\mathcal{N}$ peaks \citep{Martinet21, Tidalator} complicate the picture. To (partly) answer this question, we investigate the constraining power contained in the highest ($\mathcal{S}/\mathcal{N}$$>$3) and lowest ($\mathcal{S}/\mathcal{N}$$<$0) bins by removing these sequentially from the `aggressive' data vector ($-2.5 \le \mathcal{S}/\mathcal{N}\le4.0$) in the likelihood. The results are shown in Fig. \ref{fig:cosmo_SNRcuts}, where we observe that the negative $\mathcal{S}/\mathcal{N}$ peaks significantly help break the $[S_8 - A_{\rm IA}]$ degeneracy, while the highest peaks help in tightening the $S_8$ constraints. 
 In an analysis that ignored the role of IA, M20 found that the amount of information about $S_8$ that is contained in negative peaks is quite small, however here we find that they actually play a key role once IA are forward-modelled.  

\begin{table}
   \centering
       \caption{Summary of our cosmological inference analyses. Posteriors on $\Omega_{\rm m}$, $h$ and $w_0$ are prior-limited, so their constraints are not reported here.   Unless explicitly specified in the first column, the KiDS-1000 measurements are based on the `clean' data vector, i.e. $-1.0 < \mathcal{S}/\mathcal{N}$$\le3.0$.  The last column presents the {\it maximum a posteriori} (MAP) values. Validation of the inference pipelines on mock data are presented in Appendix \ref{sec:syst_pipeline}. 
       }
   \tabcolsep=0.11cm
      \begin{tabular}{@{} lcccccc @{}} 
      \hline
      \hline
   \multicolumn{5}{c}{Peak count analysis of KiDS-1000}\\		
   \hline
           		&  \multicolumn{2}{c}{{\sc Nautilus}} &  \multicolumn{2}{c}{{\sc Multinest}} & MAP\\		
          		&            $S_8$    			&      	$A_{\rm IA}$			& $S_8$   		 &      $A_{\rm IA}$ &$S_8$\\
  Fiducial		 & ${ 0.733^{+0.032}_{-0.032}}$	&	${  0.71^{+0.49}_{-0.49}}$ & ${ 0.733^{+0.021}_{-0.027}}$&	${  0.74^{+0.43}_{-0.43}}$ & 0.726\\
  $\Lambda$CDM & $0.732^{+0.029}_{-0.029}$&$0.73^{+0.48}_{-0.48}$&${ 0.729^{+0.026}_{-0.026}}$&	${  0.73^{+0.43}_{-0.43}}$ & 0.718\\
  Auto-only 	& $0.734^{+0.050}_{-0.050}$ &$-0.2^{+1.0}_{-1.0}$&${ 0.732^{+0.039}_{-0.048} }$ & 	$ { -0.2^{+1.0}_{-1.0}}$ & 0.725	\\
  Up to pairs 	& $0.750^{+0.036}_{-0.050}   $ &$0.10^{+0.69}_{-0.69}$ &$ { 0.742^{+0.032}_{-0.043}}  $&	$ { 0.11^{+0.63}_{-0.63}}$ & 0.742\\
  Up to triplets 	& $0.740^{+0.035}_{-0.035}$  &$0.35^{+0.53}_{-0.53}$&${ 0.740^{+0.029}_{-0.029}}$ & $	 { 0.37^{+0.48}_{-0.48}} $ 	& 0.738\\
  No tomo 		& $0.695^{+0.033}_{-0.087}$ &$-0.6^{+1.7}_{-2.2}$&$ { 0.690^{+0.038}_{-0.068} } $&	$  { -0.7^{+1.5}_{-2.0} }$ & 0.682	\\
   \hline
  -2.5$<\!\mathcal{S}/\mathcal{N}\!\le$4.0& $0.720^{+0.036}_{-0.026}$&$0.08^{+0.30}_{-0.30}$&$  { 0.717^{+0.031}_{-0.022}} $& $ { 0.07^{+0.27}_{-0.27}} $& 0.728 \\
  -2.5$<\!\mathcal{S}/\mathcal{N}\!\le$3.0&$0.717^{+0.031}_{-0.031}$ &$0.14^{+0.31}_{-0.31}$&$  0.713^{+0.023}_{-0.023} $ & $  0.22^{+0.27}_{-0.27}$ & 0.712\\ 
  $ \,$  0.0$<\!\mathcal{S}/\mathcal{N}\!\le$4.0 & $0.739^{+0.031}_{-0.026}$ &$0.77^{+0.47}_{-0.47}$&$0.744^{+0.023}_{-0.023}$   &$  0.80^{+0.38}_{-0.38} $ & 0.734\\
\hline
 No IA & $0.726^{+0.024}_{-0.042} $& $-$ &$ { 0.720^{+0.021}_{-0.031} } $&$ - $  & 0.725\\
No baryons &$0.732^{+0.032}_{-0.032} $&$0.71^{+0.49}_{-0.49}$& ${ 0.725^{+0.022}_{-0.027}} $ &$  { 0.70^{+0.43}_{-0.43}}$ & 0.725\\
 No syst & $0.729^{+0.024}_{-0.056}$& $-$ &$  { 0.723^{+0.022}_{-0.048}}$ & $-$ $-$ & 0.709\\
 No GPR error & $-$& $-$& $0.732^{+0.019}_{-0.025}$&$0.71^{+0.40}_{-0.40}              $ & 0.708\\
 $N$-body error & $-$ &$-$ & $0.725^{+0.027}_{-0.027}            $ &  $0.70^{+0.43}_{-0.43}$ & 0.717\\
\hline
 No bin1 & $0.734^{+0.043}_{-0.037} $ &$0.10^{+0.74}_{-0.74} $&$ { 0.735^{+0.035}_{-0.035}  } $&$   { 0.10^{+0.70}_{-0.10}    }$& 0.727 \\
 No bin2 & $0.740^{+0.042}_{-0.048}$&$-0.78^{+0.72}_{-0.72}$&$  { 0.740^{+0.040}_{-0.040}  }  $&$  { -0.76^{+0.70}_{-0.70}     }$ & 0.738 \\
 No bin3 & $0.775^{+0.049}_{-0.055}$&$0.97^{+0.63}_{-0.63}$&$  {  0.777^{+0.048}_{-0.048}  } $&$  {  0.95^{+0.60}_{-0.60}    }$ & 0.836 \\
 No bin4 & $0.701^{+0.037}_{-0.037}$&$0.41^{+0.68}_{-0.68}$&$  { 0.702^{+0.029}_{-0.034}    }$&$  {  0.45^{+0.59}_{-0.59}    }$ & 0.659 \\
 No bin5 & $0.720^{+0.036}_{-0.028} $&$0.53^{+0.61}_{-0.61}$&$  { 0.723^{+0.030}_{-0.025}    }$&$  {  0.53^{+0.57}_{-0.57}    }$ & 0.716 \\
  \hline
  \hline
     \multicolumn{5}{c}{Peak count analysis of DES-Y1}\\		
\hline
 DH21 & --&--&$0.737^{+0.027}_{-0.031}$ & $-$ &$-$\\
This work  &$0.743^{+0.036}_{-0.036}$ &$-$&$0.742^{+0.030}_{-0.034}$ &  $-$ & 0.712\\
\hline
  \hline
 \multicolumn{5}{c}{Joint peak count analysis}\\
 \hline		
           		&  \multicolumn{2}{c}{{\sc Nautilus}} &  \multicolumn{2}{c}{{\sc Multinest}}\\		
 Fiducial &  $ { 0.732^{+0.020}_{-0.020}}   $ & ${ 0.82^{+0.47}_{-0.47}}$ &$ { 0.732^{+0.012}_{-0.010}}   $ & ${ 0.82^{+0.33}_{-0.33}}$ & 0.745\\
$\Lambda$CDM &  $ 0.736^{+0.016}_{-0.018}  $ & $0.81^{+0.46}_{-0.46}$ & $ { 0.736^{+0.012}_{-0.015}}   $ & ${ 0.79^{+0.40}_{-0.40}}$ & 0.732\\
No baryons&  $0.728^{+0.020}_{-0.016}   $&$0.82^{+0.46}_{-0.46}$&$ { 0.725^{+0.018}_{-0.014} }$ & $ { 0.83^{+0.39}_{-0.39} }$ & 0.726 \\
No IA& $0.726^{+0.020}_{-0.016}$ &-- &$ { 0.729^{+0.015}_{-0.015} }$ & -- & 0.721\\
   \hline
    \hline
    \end{tabular}
    \label{table:cosmo_pipeline_test}
\end{table}

\subsection{Internal consistency}
\label{subsec:consistency}

\begin{figure}
\begin{center}
\includegraphics[width=3.3in]{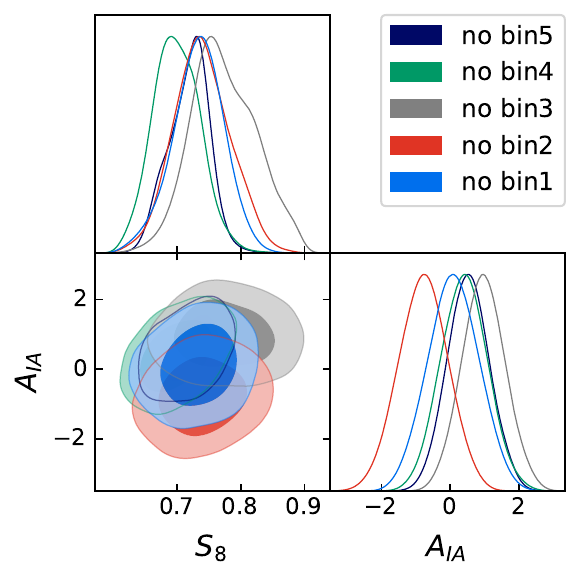}
\caption{Internal consistency: effect of removing tomographic data from the KiDS-1000 analysis.}
\label{fig:zbincuts}
\end{center}
\end{figure}

\begin{figure}
\begin{center}
\includegraphics[width=3.6in]{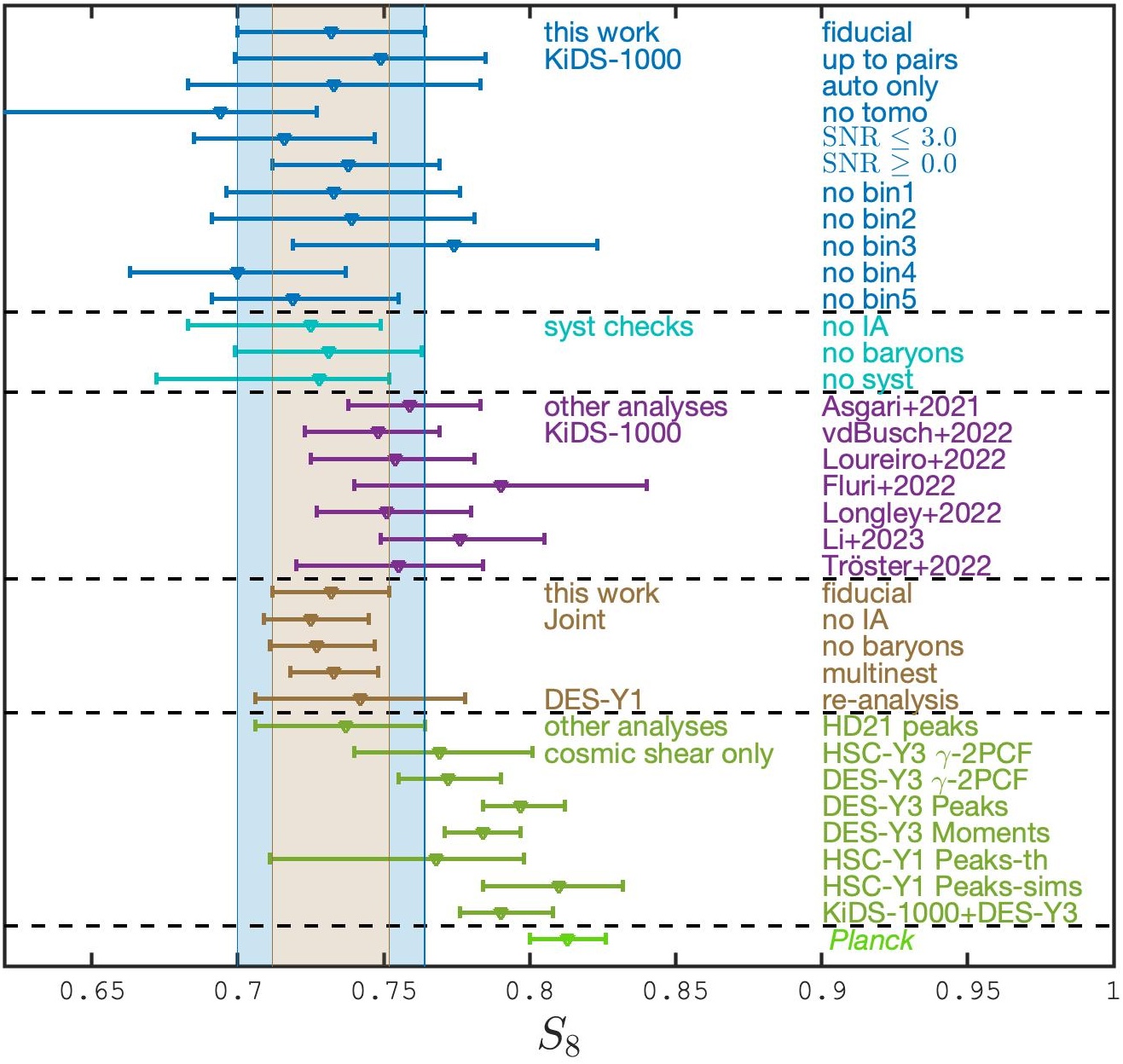}
\caption{Summary of $S_8$ constraints from this work, from recent cosmic shear data analyses and from {\it Planck}. This figure shows the projected 1$\sigma$ errors.}
\label{fig:S8_constraints}
\end{center}
\end{figure}

It has been found in previous cosmic shear analyses \citep[e.g. A21, ][]{ HSCY1_2PCF, DESY3_Amon} that internal consistency tests can help differentiate residual systematics from statistical fluctuations.
We therefore stress-test our results by removing data from tomographic bins one at a time before proceeding to the inference. For example, we consider results obtained from an analysis where exactly no data from bin1 (i.e. 1, 1$\cup$2...1$\cup$2, 1$\cup$2$\cup$3 ... 1$\cup$2$\cup$3$\cup$4$\cup$5) is used, then  no data from bin2, and so on. The results are shown in Fig. \ref{fig:zbincuts}, where we observe that all cases are self-consistent, in agreement with the full selection. Note that the $S_8$ shifts per-bin are not expected to match exactly  those measured with other lensing probes due to different responses of the cosmic shear estimators to noise in the data. For example,  A21 found that removing the fifth tomographic bin maximally degrades the precision on $S_8$, confirming the large amount of information on this parameter carried by high redshift bins in shear two-point functions. In contrast, we find here that removing the third redshift bin has the worst impact on the precision. The third bin  has the greatest number density of galaxies, hence better captures the information in peak statistics, whose mean value is affected by the noise level. The constraints on $A_{\rm IA}$ fluctuate about the fiducial results by less than $2\sigma$, while those on $S_8$ agree within $1\sigma$, as expected.

\subsection{Impact of systematics}
\label{subsec:syst_results}

We present in this section additional variations with respect to the fiducial analysis, designed to better understand our results and assess their robustness to residual systematics. 
We first investigate the impact of IA on the uncertainty by fixing $A_{\rm IA}$ to 0.72, the best-fit value in the fiducial analysis. Doing so, the error bars on $S_8$ shrink by  less than  10\%, while the mean value is not affected, by construction. Setting instead the IA parameter to 0.0, we can estimate the bias on the inferred cosmology if IA are completely neglected. We measure in this case $S_8 ^{\rm no-IA} = 0.725^{+0.024}_{-0.042} $, a 0.22$\sigma$ shift from the fiducial results. Intrinsic alignments are therefore a modest part of the error budget, suggesting that peak count analyses where IA are not modelled or held fixed \citep[e.g. M18, HD21,][]{HSCY1_peaks_sims} likely yield both biased low and slightly optimistic constraints for $S_8$.

We next carry out a similar study this time removing the modelling of baryons, fixing the associated nuisance parameter to $b_{\rm bary}=0.0$. As reported in Table \ref{table:cosmo_pipeline_test},   the measurements are mostly unchanged. As shown in M21, any non-zero residual feedback tends to lower the number of high $\mathcal{S}/\mathcal{N}$ peaks in all tomographic bins, which, when confronted to fixed data, must be compensated with an increased value of inferred $S_8$. Therefore, removing the baryon modelling goes the other way and reduces the inferred $S_8$.  This is not clearly  seen with the {\sc Nautilus} chains, but the {\sc Multinest} runs shows this shift with $0.2\sigma$ significance.

Then, removing modelling of all systematics (photo-$z$, shape calibration, IA and baryons) results in $S_8$ values half way between the no-baryon and no-IA cases, but the error bars are the  larger. This suggests that marginalisation over these systematics helps in finding the true maximal likelihood, which is not at $b_{\rm bary} = A_{\rm IA} = 0$. Indeed, the error on $S_8$ becomes smaller than the fiducial case if $A_{\rm IA}$ and $b_{\rm bary}$ are fixed to their best-fit value of 0.72 and 0.5, respectively, leading to $S_8^{\rm syst-fixed} = 0.728\pm 0.030$.

 The contribution to the error budget coming from the GPR interpolation uncertainty (Eq. \ref{eq:Cov_GPR}) can be estimated from an MCMC run where the covariance matrix excludes this term, and we observe that the error on $S_8$ is reduced by just under 10\%. Similarly, adding an error on small scale non-linear physics estimated from the scatter between the cosmo-SLICS, {\it Magneticum} dark matter-only and the T17 simulations (see Sec. \ref{subsubsec:Nbody}), can degrade the error on $S_8$ by 12\%. This is an upper limit on the degradation, given that the {\it Cosmology Training Set} has better resolution than these, hence the real uncertainty is certainly smaller. We do not include this latter error in the fiducial analysis here, because it is not accurately estimated, and instead report an upper bound on the effect.

Finally, we compared our fiducial {\sc Nautilus} results with those from the {\sc Multinest} nested sampler and recover negligible biases in the inferred parameters, but with smaller error bars ($S_8 = 0.733\pm0.032$ versus ${ 0.733^{+0.021}_{-0.027}}$ for {\sc Multinest}).  This is consistent with previous findings \citep{Lemos} and justifies our choice of {\sc Nautilus} as our main sampler. We nevertheless report results from both samplers to ease comparison with previous results.

\subsection{Comparison with previous KiDS-1000 results}
\label{subsec:previous_results}

The $S_8$ measurement presented here is not the first carried out from KiDS-1000. Previous analyses include the measurements of A21 and \citet{KiDS1000_vdB}, the latter of which used an upgraded photometric calibration compared to the former, followed by that of \citet{KiDS1000_Li} based on upgraded shear measurements. \citet{KiDS1000_Loureiro} carried out a  {\it pseudo}-$C_{\ell}$ analysis, \citet{KiDS1000_Fluri} used instead a convolutional neural network, while \citet{Longley2023} re-analysed the data within the LSST-DESC pipeline.  We report these results as the purple symbols in Fig. \ref{fig:S8_constraints}, where we see that they all seem to prefer slightly higher values of $S_8$ compared to our own measurements, albeit not by a significant amount.
Given the important differences in the analysis pipelines between these efforts, it is reassuring to recover $<1\sigma$ agreements. The constraints from the  $w$CDM band power analysis from \citet{KiDS1000_Troester} are reported in Fig. \ref{fig:cosmo_comp} and are broadly consistent with our peak statistics constraints, even though peaks are clearly more constraining on $S_8$ ($0.732\pm0.032$ for peaks vs $0.742\pm 0.047$ for band power), due to the reduced degeneracy in the [$S_8 - w_0$] plane.  It is worth noting that both statistics provide similar constraints on the $A_{\rm IA}$ parameter ($\sigma_{A_{\rm IA}} = 0.42$ for peaks, compared to $\sigma_{A_{\rm IA}} = 0.36$ for band power), which is reassuring given that both use the same NLA approach. This error is significantly reduced ($\sigma_{A_{\rm IA}} = 0.30$) when considering the more aggressive  data selection ( $-2.5 \le  \mathcal{S}/\mathcal{N} \le 4.0$), but since the associated goodness-of-fit  is poor, the results are not straight-forward to interpret. We nevertheless expect tighter constraints on $A_{\rm IA}$ to be achievable coming from non-Gaussian probes. 

 We finally remark that our constraints on $\Sigma_8^{\alpha}$ aligns remarkably well with the Band Power measurements presented in A21 (they found $\Sigma_8^{\alpha} =0.765^{+0.018}_{-0.024}$ with $\alpha=0.58$, compared to our measurement of $\Sigma_8^{\alpha} =0.765^{+0.030}_{-0.030}$ with $\alpha=0.60$).

\section{Joint analysis with DES-Y1}
\label{sec:joint_DESY1}

\begin{figure}
\begin{center}
\includegraphics[width=3.3in]{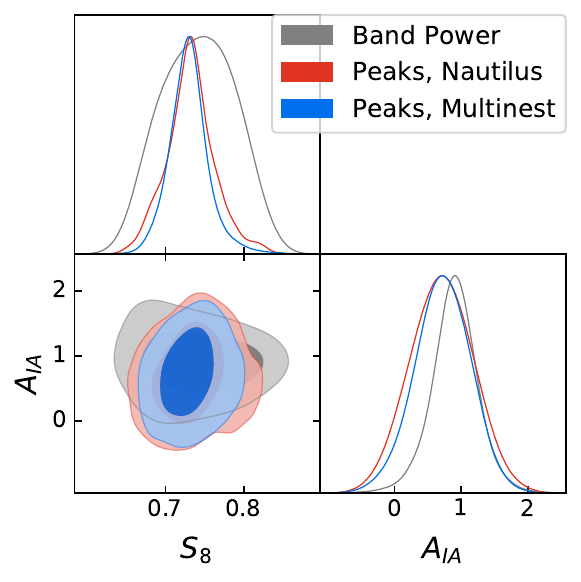}
\includegraphics[width=3.5in]{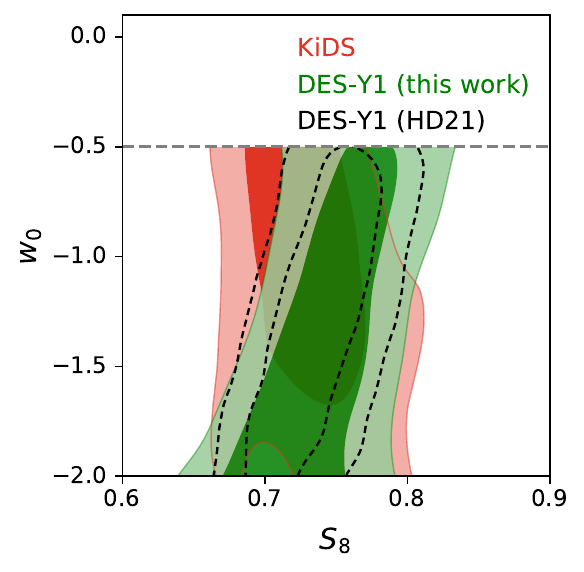}
\caption{Comparison with the previous cosmic shear results. The top part shows a comparison with the  KiDS-1000  band power analysis from \citet[][based on {\sc Multinest}]{KiDS1000_Troester},  while the bottom part 
compares the KiDS results with the  DES-Y1 peak count analysis from HD21 (black) along with current re-analysis (green) and detailed in Sec. \ref{sec:joint_DESY1}.  Note that the posteriors obtained from the {\sc Nautilus} sampler are typically wider and more accurate than those from {\sc Multinest} used in HD21. The excellent agreement seen in this figure warrants the joint survey analysis.  
}
\label{fig:cosmo_comp}
\end{center}
\end{figure}

The posterior obtained from the KiDS-1000 peak count analysis is fully consistent with that from the peak count analysis of the Dark Energy Year 1 (DES-Y1) presented in HD21. In particular,  the latter finds  
$S_8^{\rm HD21} =  0.737^{+0.027}_{-0.031}$, which significantly overlaps with our $S_8^{\rm KiDS}$ $1\sigma$ results. Other parameters less well measured such as $\Omega_{\rm m}$ and $w_0$ are also largely overlapping at the $1\sigma$ level   (see the lower part of Fig. \ref{fig:cosmo_comp}), which means  the intersection between the two likelihood hyper-volumes  must be large enough to safely combine the two data sets. 
Furthermore, both measurements are based on the similar analysis pipeline and, in particular, exploit the same simulations to model the cosmology dependence, thereby suppressing the risk of 
mis-interpreting the joint data due to non-uniform modelling of the signal. 

\subsection{Results: DES-Y1 re-analysis}

 As detailed in Sec. \ref{subsec:pipeline_overview}, there are differences between our DES-Y1 pipeline and that presented in HD21, including the $S_8$ sampling, the treatment of baryons, the inclusion of the emulator uncertainty on the covariance and the choice of sampler.  The results from these re-analyses are presented in the lower panel of Fig. \ref{fig:cosmo_comp} (in green and black). The difference induced on these contours are small, but the goodness-of-fit improvement is important, with a  PTE of 0.53 (using the same PTE estimator as HD21, we obtain 0.25, which is still a massive improvement compared to their PTE=0.005.



We remark that our joint pipeline contains a slight inconsistency: we include IA with the NLA model  in the KiDS-1000 data (with marginalisation over $A_{\rm IA}$) and with the non-linear halo-based IA model for the DES-Y1 data (without marginalisation, but with an on/off switch instead).  We verify the impact of this feature by analysing the likelihood with the DES IA model turned on and off and report on the difference, which is sub-dominant ($\Delta S_8 = 0.002$). We also compare the results from turning off the modelling of baryons, and from  replacing the $w$CDM by a $\Lambda$CDM analysis,  finding in all cases results consistent with the fiducial analysis. The re-analysis presented in this work has slightly larger error bars compared to that of HD21, due to the marginalisation over baryons,  and to the fact that  {\sc Nautilus} yields constraints slightly larger compared to {\sc Multinest}, as summarised in Table \ref{table:cosmo_pipeline_test}.
 Notably, we infer:
\begin{eqnarray}
S_8^{\rm DES} =  0.743^{+0.036}_{-0.036}\, , \,\,\, \Sigma_8^{\rm DES} =  0.762^{+0.036}_{-0.036}\,  , \,\,\,  {\rm with }\, \alpha = 0.559 \, .
\end{eqnarray}

\begin{figure}
\begin{center}
\includegraphics[width=3.3in]{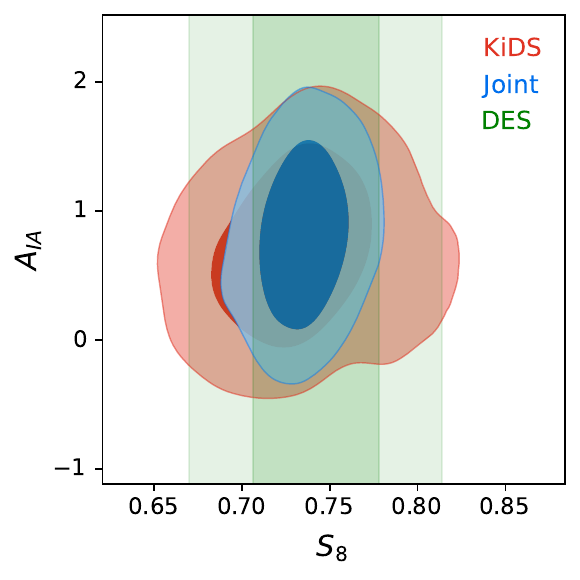}
\includegraphics[width=3.5in]{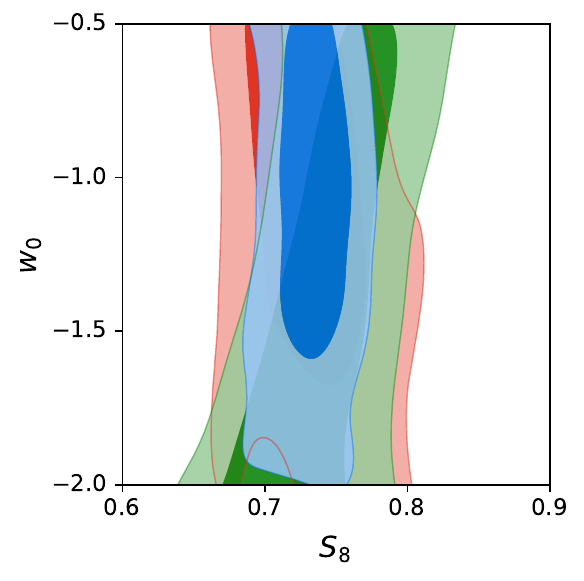}
\caption{Joint peak count analysis of the KiDS-1000 and DES-Y1 data. In the upper panel, the green bands indicate the 1$\sigma$ and 2$\sigma$ confidence intervals from the  DES-Y1 re-analysis presented in this paper. The $A_{\rm IA}$ parameter shown here describes only the IA signal in the KiDS likelihood, since the IA is fixed to a non-linear model in the DES likelihood (see main text for details).} 
\label{fig:cosmo_KiDS-DES}
\end{center}
\end{figure}
 

\subsection{Results: joint KiDS + DES}

We present in this section the results from our joint KiDS-1000 + DES-Y1 peak count analysis. 
Sampling the joint likelihood with our fiducial setup,   we achieve improved constraints on $S_8$ and $\Sigma_8^\alpha$ with:
\begin{eqnarray}
S_8^{\rm joint, {\it w}CDM} = 0.732^{+0.020}_{-0.020}\, , \,\,\,  \Sigma_8^{\rm joint, {\it w}CDM} = 0.759^{+0.020}_{-0.017} 
\end{eqnarray}
and
\begin{eqnarray}
{S_8^{\rm joint, \Lambda CDM} = 0.735^{+0.016}_{-0.018}} \, , \,\,\,  \Sigma_8^{\rm joint, \Lambda CDM} = 0.762^{+0.017}_{-0.017}
\end{eqnarray}
computed with $\alpha = 0.572$ in both cases. These are the tightest results obtained from non-Gaussian cosmic shear statistics to date, comparable to the recent joint $\Lambda$CDM analysis of the KiDS-1000 and DES-Y3 data \citep{DESY3-KiDS1000}, which measured $S_8^{\rm NLA} = 0.792^{+0.016}_{-0.013}$. The two-dimensional posterior is shown in Fig. \ref{fig:cosmo_KiDS-DES} (in blue) and compared to the fiducial KiDS-1000 (red) and DES-Y1 (green) peak statistics constraints. Recall that the $A_{\rm IA}$ parameter affects only the KiDS likelihood since, as explained in the previous section, the DES likelihood assumes instead a fixed halo-based IA model with no free parameter. We should therefore use caution when interpreting this parameter. The reported value is close to the point of maximum likelihood (${S_8^{\rm ML} = 0.728}$), and the size of the error bars on $S_8$ is consistent with our expectation: for example, we read from Table \ref{table:cosmo_pipeline_test} that the $\Lambda$CDM KiDS-1000 analysis has a mean error of $\sigma_{S_8}=0.029$. Scaling this precision by the square root of the area, we naively predict a joint survey error of around 0.018, and obtain 0.017.  The error would be slightly larger had we included as well a marginalisation of the IA in the DES-Y1 part of this analysis, possibly explaining this slight difference. At the joint best-fit cosmology, the PTE values for the KiDS and DES pipelines are basically unchanged, while the joint analysis has a $\chi^2_{\rm red} = 1.15$ and a PTE of 0.96, all satisfying our goodness-of-fit criteria.

If we restrict the joint analysis to $w_0=-1.0$, the $S_8$ values are minimally affected while the uncertainty is reduced,  as expected from lowering the dimensionality of the likelihood. 
Alternatively, turning on the IA modelling in the DES likelihood only yields a 0.2$\sigma$ downward shift, also expected whenever IA modelling is added. The smallness of this shift is once again showing that the intrinsic alignment do not significantly impact the peak count statistics as measured in the DES-Y1 data. In comparison, setting to zero the IA model in both KiDS and DES results in  ${S_8^{\rm joint, no-IA} = 0.725^{+0.020}_{-0.016} }$. Holding fixed the baryonic feedback parameter to $b_{\rm bary}=0.0$   has similar consequences on this joint analysis, shifting the best fit value to $S_8^{\rm joint, no-bary} =0.727^{+0.020}_{-0.016}$, a 0.25$\sigma$ shift compared to the fiducial case. All these values are summarised in Table \ref{table:cosmo_pipeline_test} and in Fig. \ref{fig:S8_constraints} (with the brown symbols).

The dark energy equation-of-state is constrained from this joint analysis, with 
\begin{eqnarray}
 w_0^{\rm joint} = -1.12^{+0.42}_{-0.31} \, ,
\end{eqnarray}
which is the first measurement of this quantity from peak statistics, and arguably one of the best from cosmic shear-only data analyses.  The upper limit is close to the prior edge on $w_0$, which might lead to a slight under-estimation of the error on this side. However, this measurement is robust against the choice of sampler ($w_0 = -1.09^{+0.29}_{-0.29}$ for {\sc Multinest}), against baryon modelling ($w_0 = -1.05^{+0.51}_{-0.22}$ setting $b_{\rm bary}=0.0$), IA ($w_0 = -1.13^{+0.44}_{-0.33}$ setting $A_{\rm IA}=0.0$) and scale cuts ($w_0=-0.958^{+0.45}_{-0.093}$ when including the aggressive $\mathcal{S}/\mathcal{N}$ cut in the KiDS-1000 data vector).  As shown in \citet{Martinet20}, aperture-mass maps statistics are highly sensitive to dark energy and these results seem to be showing exactly that. Previously,  the shear two-point function measurement from \citet{DESY1_Troxel} on DES-Y1 achieved $w_0 = -0.77^{+0.30}_{-0.37}$ when varying the baryonic feedback model,  using the  {\sc Multinest} sampler. The GCNN analysis of \citet{KiDS1000_Fluri} was also able to set constraints on dark energy, with $w_0=-0.93^{+0.32}_{-0.29}$, although they recognise that their results are affected by the prior boundary on the low side, just like ours is on the high side\footnote{We review the A21 definition that constraints are uninformed by the prior when the posterior drops below 0.135 of its maximum at the edges of a uniform prior volume. In the case of $w_0$ we find that the posterior is slightly above this threshold (0.156) at the upper edge. As this is at the borderline of the  A21 criteria, we therefore caution that the error on this side might be slightly  under-estimated.}. Similarly, HD21 found $w_0 > -1.5$, also prior-dominated on one side. Other cosmic shear measurements of $w_0$  involve additional data \citep{DESY3_extensions, KiDS1000_Troester}, making this an unfair comparison.

 It is worth mentioning that all peak count analyses based on the {\it cosmo}-SLICS yield $S_8$ constraints that are lower than the 2PCFs fiducial analyses. This could be pointing to limitations in the training set, but is quite speculative at this stage given that the $\Sigma_8$ values align well. Further investigations and novel simulation suites would be required to ascertain this, which we post-pone for future work.

\subsection{Tension with {\it Planck}}

\begin{figure}
\begin{center}
\includegraphics[width=3.5in]{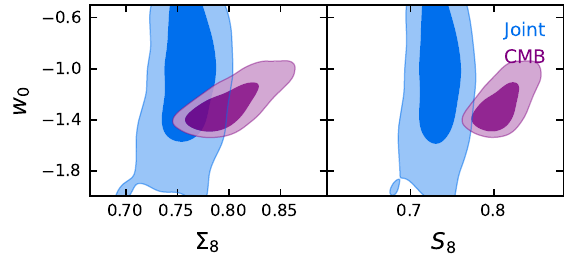}
\caption{ Joint constraint on $S_8$, $w_0$ and $\Sigma_8^\alpha$, where $\alpha = 0.572$, comparing here the combined cosmic shear surveys with the CMB results. The  tension is lower on  $\Sigma_8$ than on $S_8$.}
\label{fig:tension_S8_Sigma8}
\end{center}
\end{figure}

The $S_8$ tension between recent CMB anisotropy and weak lensing data analyses is drawing a lot of attention, as it could point towards new physics or hidden systematics \citep[see, e.g.][for a review]{S8TensionReview}.
The {\it Planck} mission reports $S_8^{\it Planck} = 0.830\pm0.013$ \citep{PlanckLegacyCosmo}, which is higher than many lensing results \citep[see][and references therein]{LensingIsLow2023}.
The tension $\tau$ can be evaluated with a number of metrics, and we use here a relatively simple one used in A21, which compares the difference in the mean with the combined variances, ${\rm var}[S_8]^{\it Planck},{\rm var}[S_8]^{\rm peaks} $ :

\begin{eqnarray}
\tau = \frac{S_8^{\it Planck} - S_8^{\rm peaks}}{\sqrt{{\rm var}[S_8]^{\it Planck} +{\rm var}[S_8]^{\rm peaks}   }}
\end{eqnarray}
With the fiducial setup shown in Table \ref{table:cosmo_pipeline_test}, and using the above definition, our results from the KiDS-1000  peak count analysis are in 
$\tau = 2.0\sigma$ tension with the {\it Planck}  nominal constraints on $S_8$ from their $w$CDM analysis.  \citet{KiDS1000_Troester}  finds a similarly low tension with the same KiDS-1000 data in a $w$CDM analysis, either using this simple tension metric or a more sophisticated method based on the full shape of the posteriors. Similarly, we evaluate our joint KiDS-DES analysis to be in $\tau = 2.7\sigma$ tension with {\it Planck}, an increase that is driven by the decrease in error bars. Using instead the MAP values (listed in Table \ref{table:cosmo_pipeline_test}) lowers the KiDS  tension to 1.7$\sigma$, and the joint-survey  to 1.65.

The tension reaches $\tau = 4.1\sigma$ in our joint-survey $\Lambda$CDM analysis, which could be pointing to a resolution of the $S_8$ tension that includes modification to the dark energy equation of state, as suggested by \citet{KiDS1000_Troester} and by the recent results from the Dark Energy Spectroscopic Instrument \citep{DESI_cosmo}, however such statement cannot be definitive until better $w$ measurements are obtained from cosmic shear data.

Note that this tension is not only seen in weak lensing, but also in other  late-time probes, e.g. data involving galaxy clustering, as recently summarised in  \citet[][see their figure 7]{Quaia_Alonso2023} and reviewed with greater details in  \citet{CosmologyIntertwinedReview}. The current work aligns with the existing trend, without providing an obvious solution. Again, large unaccounted contributions from IA and baryons could push the inferred $S_8$ value towards {\it Planck},  but our analysis prefers lower values: in particular, we measure $b_{\rm bary} < 1.05$ at 95\% CL in the KiDS-1000 analysis, and $< 0.82$ in the joint analysis, excluding baryonic feedback models that are stronger than the {\it Magneticum}. Also, the redshift estimation methods used to analyse the DES-Y1 data are suboptimal compared to recent developments \citep[see e.g.][]{KiDS1000_redshifts}, potentially causing biases of up to 0.03 in the inferred $S_8$ value. The observed tension with {\it Planck} would be different if that bias was real.

Interestingly, the tension on  $\Sigma_8$ is reduced to  $\tau = 0.72\sigma$ with the KiDS-1000 data, and to  $\tau = 1.33\sigma$ with the joint data, in both cases evaluated at the $\alpha$ value preferred by the cosmic shear measurements\footnote{The {\it Planck} $w$CDM measurements of $\Sigma_8$ are of $0.793^{+0.019}_{-0.031}$ and $0.797^{+0.018}_{-0.028}$ for the KiDS-1000 and joint-survey $\alpha$ parameters, respectively.}. This is in part caused by a degradation of the CMB constraints along this quantity, which completely relaxes the tension, and in part by a different projection angle of the high-dimensional posterior, as seen in Fig. \ref{fig:tension_S8_Sigma8}. At the same time, when holding $w$ fixed, the tension on $\Sigma_8$ is again increased reaching $2.3\sigma$ for KiDS, slightly lower than the $3\sigma$ reported in A21, and  $3.1\sigma$ for our joint analysis. To summarise, the tension with {\it Planck} on $S_8$ is lowered in $w$CDM compared to $\Lambda$CDM, and is further lowered when considering the more robust $\Sigma_8$ parameter instead of $S_8$. This is discussed in more details in Appendix \ref{sec:syst_pipeline}.

A more in-depth analysis of this tension requires a robust determination of the inferred $\Omega_{\rm m}$ parameter. We defer this improvement to future work, which will benefit from a denser training set, yielding a more accurate and robust emulator.

\section{Conclusions}
\label{sec:conclusion}

We report in this paper a 4.4\% measurement of $S_8$ from the tomographic peak count statistics measured from the KiDS-1000 data. Our simulation-based inference method exploits the non-linear features extracted from aperture mass statistics, sensitive to scales as small as 2.0 \mbox{arcmin}. We model the cosmological dependence with simulated $w$CDM weak lensing light-cones, we estimate the covariance matrix numerically, and forward model the effect of intrinsic alignments, baryonic feedback, photometric redshift error and galaxy shape mis-calibrations. We find a value of $S_8^{\rm KiDS} =  0.733^{+0.032}_{-0.032}$, which aligns well with  previous KiDS-1000 measurements.
We show that our results are robust to residual systematics and that, of these, intrinsic alignment of galaxies plays the most important role, shifting the best-fit $S_8$ value by 0.22$\sigma$ if left unmodelled.  We also show that the most robustly measured parameter in our analysis is $\Sigma_8^{\rm KiDS}\equiv \sigma_8 \left( \Omega_{\rm m}/0.3\right)^\alpha =  0.765^{+0.030}_{-0.030}$, with $\alpha = 0.60$, in excellent agreement with previous KiDS-1000 analysis.

The inferred posterior distribution is consistent with the peak count measurement carried out on the DES-Y1 data using a similar analysis pipeline (HD21), allowing us to jointly analyse the two data sets, which yields $S_8^{\rm joint} =  0.732^{+0.020}_{-0.020}$, one of the tightest constraints on this parameter from lensing data alone. The combined data sets have enough statistical precision to allow the first measurement of the dark energy equation-of-state parameter from non-Gaussian statistics: $w_0^{\rm joint}=-1.12^{+0.42}_{-0.31}$, in agreement with the $\Lambda$CDM scenario, and robust to variations in the analysis choices. 

Our best-fit $S_8^{\rm joint}$ is also in statistical agreement with all previous KiDS-1000 analyses and with the HSC-Y3 and DES-Y3 $\gamma$-2PCF results, but lower than the DES-Y3 measurements from peaks and moments, and in $2.7\sigma$ tension with {\it Planck}.  This joint-survey tension increases to 4$\sigma$ in our $\Lambda$CDM analysis, but lowers to 2.3 and 3.1$\sigma$ when considering instead the $\Sigma_8$ parameter,  for the KiDS-only and joint survey analyses, respectively.  

Our pipeline has been thoroughly tested, however we recognise it is incomplete. As detailed in Sec. \ref{sec:systematics}, we hold fixed a number of cosmological parameters, which likely affect our results, including $\Omega_{\rm b}, n_{\rm s}$ and $m_{\nu}$. We also consider a single baryonic feedback model (although we allow its amplitude to vary), knowing that other hydrodynamical simulations would provide slightly different responses. Furthermore, we model IA with the redshift-independent NLA model, which we know is an incomplete effective model, and we neglect source clustering, as it was shown to be completely subdominant.  Finally, we have identified a systematic effect in our Gaussian process emulator that is caused by the relatively small number of training nodes,  preventing us from extracting meaningful information about $\Omega_{\rm m}$, however all other parameters are unaffected by this. This limits our ability to further study in higher-dimensional space the potential $S_8$ tension with the CMB.  Addressing these will therefore be the object of future work. It will also be informative to compare our results to other non-Gaussian probes of cosmic shear, and possibly combine the methods to further reduce the uncertainty on $S_8$ and $w_0$.

\section*{Acknowledgements}

We would like to thank Dr. Eric Tittley from the Royal Observatory of Edinburgh for his continuous help and support with the Cuillin cluster on which the bulk of the computations were carried on, as well as Elena Sellentin for discussions on the use of Hotelling's $T^2$ statistics.  

JHD acknowledges support from an STFC Ernest Rutherford Fellowship (project reference ST/S004858/1). SH is supported by the U.D Department of Energy, Office of Science, Office of High Energy Physics under Award Number DE-SC0019301. BG acknowledges support from a UK STFC Consolidated Grant and from the Royal Society of Edinburgh through the Saltire Early Career Fellowship (ref. number 1914). NM acknowledges support from the French Agence Nationale de la Recherche (PISCO project / ANR-22-CE31-0004-01). TT acknowledges funding from the Swiss National Science Foundation under the Ambizione project PZ00P2\_193352. KD is supported by the grant agreements ANR-21-CE31-0019 / 490702358 from the French Agence Nationale de la Recherche / Deutsche Forschungsgemeinschaft (DFG, German Research Foundation) for the LOCALIZATION project. TC is supported by the INFN INDARK PD51 grant, by the Fondazione ICSC National Recovery and Resilience Plan (PNRR) Project ID CN-00000013 ``Italian Research Center on High-Performance Computing, Big Data and Quantum Computing'' funded by MUR Missione 4 Componente 2 Investimento 1.4: ``Potenziamento strutture di ricerca e creazione di ``campioni nazionali'' di R$\&$S (M4C2-19)'' -- Next Generation EU (NGEU), and by the FARE MIUR grant `ClustersXEuclid'. HH is supported by a DFG Heisenberg grant (Hi 1495/5-1), the DFG Collaborative Research Center SFB1491, as well as an ERC Consolidator Grant (No. 770935). 
BJ acknowledges support by the ERC-selected UKRI Frontier Research Grant EP/Y03015X/1 and by STFC Consolidated Grant ST/V000780/1. AHW is supported by the Deutsches Zentrum f\"ur Luft- und Raumfahrt (DLR), made possible by the Bundesministerium f\"ur Wirtschaft und Klimaschutz, and acknowledges funding from the German Science Foundation DFG, via the Collaborative Research Center SFB1491 "Cosmic Interacting Matters - From Source to Signal". 
\\
\\
The KiDS-1000 results in this paper are based on data products from observations made with ESO Telescopes at the La Silla Paranal Observatory under programme IDs 177.A-3016, 177.A-3017 and 177.A-3018, and on data products produced by Target/OmegaCEN, INAF-OACN, INAF-OAPD and the KiDS production team, on behalf of the KiDS consortium. This project also used public archival data from the Dark Energy Survey (DES). Funding for the DES Projects has been provided by the U.S. Department of Energy, the U.S. National Science Foundation, the Ministry of Science and Education of Spain, the Science and Technology FacilitiesCouncil of the United Kingdom, the Higher Education Funding Council for England, the National Center for Supercomputing Applications at the University of Illinois at Urbana-Champaign, the Kavli Institute of Cosmological Physics at the University of Chicago, the Center for Cosmology and Astro-Particle Physics at the Ohio State University, the Mitchell Institute for Fundamental Physics and Astronomy at Texas A\&M University, Financiadora de Estudos e Projetos, Funda{\c c}{\~a}o Carlos Chagas Filho de Amparo {\`a} Pesquisa do Estado do Rio de Janeiro, Conselho Nacional de Desenvolvimento Cient{\'i}fico e Tecnol{\'o}gico and the Minist{\'e}rio da Ci{\^e}ncia, Tecnologia e Inova{\c c}{\~a}o, the Deutsche Forschungsgemeinschaft, and the Collaborating Institutions in the Dark Energy Survey. The Collaborating Institutions are Argonne National Laboratory, the University of California at Santa Cruz, the University of Cambridge, Centro de Investigaciones Energ{\'e}ticas, Medioambientales y Tecnol{\'o}gicas-Madrid, the University of Chicago, University College London, the DES-Brazil Consortium, the University of Edinburgh, the Eidgen{\"o}ssische Technische Hochschule (ETH) Z{\"u}rich,  Fermi National Accelerator Laboratory, the University of Illinois at Urbana-Champaign, the Institut de Ci{\`e}ncies de l'Espai (IEEC/CSIC), the Institut de F{\'i}sica d'Altes Energies, Lawrence Berkeley National Laboratory, the Ludwig-Maximilians Universit{\"a}t M{\"u}nchen and the associated Excellence Cluster Universe, the University of Michigan, the National Optical Astronomy Observatory, the University of Nottingham, The Ohio State University, the OzDES Membership Consortium, the University of Pennsylvania, the University of Portsmouth, SLAC National Accelerator Laboratory, Stanford University, the University of Sussex, and Texas A\&M University.
Based in part on observations at Cerro Tololo Inter-American Observatory, National Optical Astronomy Observatory, which is operated by the Association of Universities for Research in Astronomy (AURA) under a cooperative agreement with the National Science Foundation.
\\
\\
The chains presented in this paper are plotted with {\sc GetDist} \citep{GetDist}, available at {\tt https://getdist.readthedocs.io}. 
\\
\\
For the purpose of open access, the author has applied a Creative Commons Attribution (CC BY) licence to any Author Accepted Manuscript version arising from this submission.
\\
\\
{\footnotesize All authors contributed to the development and writing of this paper. The authorship list is given in three groups: the lead authors (JHD, SH), followed by two alphabetical groups. The first alphabetical group includes those who are key contributors to both the scientific analysis and the data products (BG, NM, TT). The second group covers those who have either made a significant contribution to the data products, or to the scientific analysis.}


\section*{Data Availability}

The SLICS numerical simulations can be found at http://slics.roe.ac.uk/, while the SLICS-HR,  the cosmo-SLICS and the {\it Magneticum} can be made available upon request.
This work also uses public KiDS-1000 and DES-Y1 data, which can be found at  {\tt https://kids.strw.leidenuniv.nl/DR4/index.php} and {\tt https://des.ncsa.illinois.edu/releases/y1a1}, respectively.



\bibliographystyle{hapj}
\bibliography{peaks} 

\begin{thebibliography}{130}
\expandafter\ifx\csname natexlab\endcsname\relax\def\natexlab#1{#1}\fi

\bibitem[{{Abbott} {et~al.}(2018){Abbott}, {Abdalla}, {Allam}, {Amara},
  {Annis}, {Asorey}, {Avila}, {Ballester}, {Banerji}, {Barkhouse}, {Baruah},
  {Baumer}, {Bechtol}, {Becker}, {Benoit-L{\'e}vy}, {Bernstein}, {Bertin},
  {Blazek}, {Bocquet}, {Brooks}, {Brout}, {Buckley-Geer}, {Burke}, {Busti},
  {Campisano}, {Cardiel-Sas}, {Carnero Rosell}, {Carrasco Kind}, {Carretero},
  {Castander}, {Cawthon}, {Chang}, {Chen}, {Conselice}, {Costa}, {Crocce},
  {Cunha}, {D'Andrea}, {da Costa}, {Das}, {Daues}, {Davis}, {Davis}, {De
  Vicente}, {DePoy}, {DeRose}, {Desai}, {Diehl}, {Dietrich}, {Dodelson},
  {Doel}, {Drlica-Wagner}, {Eifler}, {Elliott}, {Evrard}, {Farahi}, {Fausti
  Neto}, {Fernandez}, {Finley}, {Flaugher}, {Foley}, {Fosalba}, {Friedel},
  {Frieman}, {Garc{\'\i}a-Bellido}, {Gaztanaga}, {Gerdes}, {Giannantonio},
  {Gill}, {Glazebrook}, {Goldstein}, {Gower}, {Gruen}, {Gruendl}, {Gschwend},
  {Gupta}, {Gutierrez}, {Hamilton}, {Hartley}, {Hinton}, {Hislop}, {Hollowood},
  {Honscheid}, {Hoyle}, {Huterer}, {Jain}, {James}, {Jeltema}, {Johnson},
  {Johnson}, {Kacprzak}, {Kent}, {Khullar}, {Klein}, {Kovacs}, {Koziol},
  {Krause}, {Kremin}, {Kron}, {Kuehn}, {Kuhlmann}, {Kuropatkin}, {Lahav},
  {Lasker}, {Li}, {Li}, {Liddle}, {Lima}, {Lin}, {L{\'o}pez-Reyes}, {MacCrann},
  {Maia}, {Maloney}, {Manera}, {March}, {Marriner}, {Marshall}, {Martini},
  {McClintock}, {McKay}, {McMahon}, {Melchior}, {Menanteau}, {Miller},
  {Miquel}, {Mohr}, {Morganson}, {Mould}, {Neilsen}, {Nichol}, {Nogueira},
  {Nord}, {Nugent}, {Nunes}, {Ogand o}, {Old}, {Pace}, {Palmese},
  {Paz-Chinch{\'o}n}, {Peiris}, {Percival}, {Petravick}, {Plazas}, {Poh},
  {Pond}, {Porredon}, {Pujol}, {Refregier}, {Reil}, {Ricker}, {Rollins},
  {Romer}, {Roodman}, {Rooney}, {Ross}, {Rykoff}, {Sako}, {Sanchez}, {Sanchez},
  {Santiago}, {Saro}, {Scarpine}, {Scolnic}, {Serrano}, {Sevilla-Noarbe},
  {Sheldon}, {Shipp}, {Silveira}, {Smith}, {Smith}, {Smith}, {Soares-Santos},
  {Sobreira}, {Song}, {Stebbins}, {Suchyta}, {Sullivan}, {Swanson}, {Tarle},
  {Thaler}, {Thomas}, {Thomas}, {Troxel}, {Tucker}, {Vikram}, {Vivas},
  {Walker}, {Wechsler}, {Weller}, {Wester}, {Wolf}, {Wu}, {Yanny}, {Zenteno},
  {Zhang}, {Zuntz}, {DES Collaboration}, {Juneau}, {Fitzpatrick}, {Nikutta},
  {Nidever}, {Olsen}, {Scott}, \& {NOAO Data Lab}}]{DESY1_data}
{Abbott}, T.~M.~C. {et~al.} 2018, \apjs, 239, 18, 1801.03181

\bibitem[{{Abbott} {et~al.}(2022){Abbott}, {Aguena}, {Alarcon}, {Allam},
  {Alves}, {Amon}, {Andrade-Oliveira}, {Annis}, {Avila}, {Bacon}, {Baxter},
  {Bechtol}, {Becker}, {Bernstein}, {Bhargava}, {Birrer}, {Blazek},
  {Brandao-Souza}, {Bridle}, {Brooks}, {Buckley-Geer}, {Burke}, {Camacho},
  {Campos}, {Carnero Rosell}, {Carrasco Kind}, {Carretero}, {Castander},
  {Cawthon}, {Chang}, {Chen}, {Chen}, {Choi}, {Conselice}, {Cordero},
  {Costanzi}, {Crocce}, {da Costa}, {da Silva Pereira}, {Davis}, {Davis}, {De
  Vicente}, {DeRose}, {Desai}, {Di Valentino}, {Diehl}, {Dietrich}, {Dodelson},
  {Doel}, {Doux}, {Drlica-Wagner}, {Eckert}, {Eifler}, {Elsner}, {Elvin-Poole},
  {Everett}, {Evrard}, {Fang}, {Farahi}, {Fernandez}, {Ferrero}, {Fert{\'e}},
  {Fosalba}, {Friedrich}, {Frieman}, {Garc{\'\i}a-Bellido}, {Gatti},
  {Gaztanaga}, {Gerdes}, {Giannantonio}, {Giannini}, {Gruen}, {Gruendl},
  {Gschwend}, {Gutierrez}, {Harrison}, {Hartley}, {Herner}, {Hinton},
  {Hollowood}, {Honscheid}, {Hoyle}, {Huff}, {Huterer}, {Jain}, {James},
  {Jarvis}, {Jeffrey}, {Jeltema}, {Kovacs}, {Krause}, {Kron}, {Kuehn},
  {Kuropatkin}, {Lahav}, {Leget}, {Lemos}, {Liddle}, {Lidman}, {Lima}, {Lin},
  {MacCrann}, {Maia}, {Marshall}, {Martini}, {McCullough}, {Melchior},
  {Mena-Fern{\'a}ndez}, {Menanteau}, {Miquel}, {Mohr}, {Morgan}, {Muir},
  {Myles}, {Nadathur}, {Navarro-Alsina}, {Nichol}, {Ogando}, {Omori},
  {Palmese}, {Pandey}, {Park}, {Paz-Chinch{\'o}n}, {Petravick}, {Pieres},
  {Plazas Malag{\'o}n}, {Porredon}, {Prat}, {Raveri}, {Rodriguez-Monroy},
  {Rollins}, {Romer}, {Roodman}, {Rosenfeld}, {Ross}, {Rykoff}, {Samuroff},
  {S{\'a}nchez}, {Sanchez}, {Sanchez}, {Sanchez Cid}, {Scarpine}, {Schubnell},
  {Scolnic}, {Secco}, {Serrano}, {Sevilla-Noarbe}, {Sheldon}, {Shin}, {Smith},
  {Soares-Santos}, {Suchyta}, {Swanson}, {Tabbutt}, {Tarle}, {Thomas}, {To},
  {Troja}, {Troxel}, {Tucker}, {Tutusaus}, {Varga}, {Walker}, {Weaverdyck},
  {Wechsler}, {Weller}, {Yanny}, {Yin}, {Zhang}, {Zuntz}, \& {DES
  Collaboration}}]{DESY3_3x2}
------. 2022, \prd, 105, 023520, 2105.13549

\bibitem[{{Abbott} {et~al.}(2023){Abbott}, {Aguena}, {Alarcon}, {Alves},
  {Amon}, {Andrade-Oliveira}, {Annis}, {Avila}, {Bacon}, {Baxter}, {Bechtol},
  {Becker}, {Bernstein}, {Birrer}, {Blazek}, {Bocquet}, {Brandao-Souza},
  {Bridle}, {Brooks}, {Burke}, {Camacho}, {Campos}, {Carnero Rosell}, {Carrasco
  Kind}, {Carretero}, {Castander}, {Cawthon}, {Chang}, {Chen}, {Chen}, {Choi},
  {Conselice}, {Cordero}, {Costanzi}, {Crocce}, {da Costa}, {Pereira}, {Davis},
  {Davis}, {DeRose}, {Desai}, {Di Valentino}, {Diehl}, {Dodelson}, {Doel},
  {Doux}, {Drlica-Wagner}, {Eckert}, {Eifler}, {Elsner}, {Elvin-Poole},
  {Everett}, {Fang}, {Farahi}, {Ferrero}, {Fert{\'e}}, {Flaugher}, {Fosalba},
  {Friedel}, {Friedrich}, {Frieman}, {Garc{\'\i}a-Bellido}, {Gatti}, {Giani},
  {Giannantonio}, {Giannini}, {Gruen}, {Gruendl}, {Gschwend}, {Gutierrez},
  {Hamaus}, {Harrison}, {Hartley}, {Herner}, {Hinton}, {Hollowood},
  {Honscheid}, {Huang}, {Huff}, {Huterer}, {Jain}, {James}, {Jarvis},
  {Jeffrey}, {Jeltema}, {Kovacs}, {Krause}, {Kuehn}, {Kuropatkin}, {Lahav},
  {Lee}, {Leget}, {Lemos}, {Leonard}, {Liddle}, {Lima}, {Lin}, {MacCrann},
  {Marshall}, {McCullough}, {Mena-Fern{\'a}ndez}, {Menanteau}, {Miquel},
  {Miranda}, {Mohr}, {Muir}, {Myles}, {Nadathur}, {Navarro-Alsina}, {Nichol},
  {Ogando}, {Omori}, {Palmese}, {Pandey}, {Park}, {Paterno},
  {Paz-Chinch{\'o}n}, {Percival}, {Pieres}, {Plazas Malag{\'o}n}, {Porredon},
  {Prat}, {Raveri}, {Rodriguez-Monroy}, {Rogozenski}, {Rollins}, {Romer},
  {Roodman}, {Rosenfeld}, {Ross}, {Rykoff}, {Samuroff}, {S{\'a}nchez},
  {Sanchez}, {Sanchez}, {Sanchez Cid}, {Scarpine}, {Scolnic}, {Secco},
  {Sevilla-Noarbe}, {Sheldon}, {Shin}, {Smith}, {Soares-Santos}, {Suchyta},
  {Tabbutt}, {Tarle}, {Thomas}, {To}, {Troja}, {Troxel}, {Tutusaus}, {Varga},
  {Vincenzi}, {Walker}, {Weaverdyck}, {Wechsler}, {Weller}, {Yanny}, {Yin},
  {Zhang}, {Zuntz}, \& {DES Collaboration}}]{DESY3_extensions}
------. 2023, \prd, 107, 083504, 2207.05766

\bibitem[{{Abdalla} {et~al.}(2022{\natexlab{a}}){Abdalla}, {Abell{\'a}n},
  {Aboubrahim}, {Agnello}, {Akarsu}, {Akrami}, {Alestas}, {Aloni}, {Amendola},
  {Anchordoqui}, {Anderson}, {Arendse}, {Asgari}, {Ballardini}, {Barger},
  {Basilakos}, {Batista}, {Battistelli}, {Battye}, {Benetti}, {Benisty},
  {Berlin}, {de Bernardis}, {Berti}, {Bidenko}, {Birrer}, {Blakeslee}, {Boddy},
  {Bom}, {Bonilla}, {Borghi}, {Bouchet}, {Braglia}, {Buchert}, {Buckley-Geer},
  {Calabrese}, {Caldwell}, {Camarena}, {Capozziello}, {Casertano}, {Chen},
  {Chluba}, {Chen}, {Chen}, {Chudaykin}, {Cicoli}, {Copi}, {Courbin},
  {Cyr-Racine}, {Czerny}, {Dainotti}, {D'Amico}, {Davis}, {de Cruz P{\'e}rez},
  {de Haro}, {Delabrouille}, {Denton}, {Dhawan}, {Dienes}, {Di Valentino},
  {Du}, {Eckert}, {Escamilla-Rivera}, {Fert{\'e}}, {Finelli}, {Fosalba},
  {Freedman}, {Frusciante}, {Gazta{\~n}aga}, {Giar{\`e}}, {Giusarma},
  {G{\'o}mez-Valent}, {Handley}, {Harrison}, {Hart}, {Hazra}, {Heavens},
  {Heinesen}, {Hildebrandt}, {Hill}, {Hogg}, {Holz}, {Hooper}, {Hosseininejad},
  {Huterer}, {Ishak}, {Ivanov}, {Jaffe}, {Jang}, {Jedamzik}, {Jimenez},
  {Joseph}, {Joudaki}, {Kamionkowski}, {Karwal}, {Kazantzidis}, {Keeley},
  {Klasen}, {Komatsu}, {Koopmans}, {Kumar}, {Lamagna}, {Lazkoz}, {Lee},
  {Lesgourgues}, {Levi Said}, {Lewis}, {L'Huillier}, {Lucca}, {Maartens},
  {Macri}, {Marfatia}, {Marra}, {Martins}, {Masi}, {Matarrese}, {Mazumdar},
  {Melchiorri}, {Mena}, {Mersini-Houghton}, {Mertens}, {Milakovi{\'c}},
  {Minami}, {Miranda}, {Moreno-Pulido}, {Moresco}, {Mota}, {Mottola}, {Mozzon},
  {Muir}, {Mukherjee}, {Mukherjee}, {Naselsky}, {Nath}, {Nesseris},
  {Niedermann}, {Notari}, {Nunes}, {{\'O} Colg{\'a}in}, {Owens},
  {{\"O}z{\"u}lker}, {Pace}, {Paliathanasis}, {Palmese}, {Pan}, {Paoletti},
  {Perez Bergliaffa}, {Perivolaropoulos}, {Pesce}, {Pettorino}, {Philcox},
  {Pogosian}, {Poulin}, {Poulot}, {Raveri}, {Reid}, {Renzi}, {Riess}, {Sabla},
  {Salucci}, {Salzano}, {Saridakis}, {Sathyaprakash}, {Schmaltz},
  {Sch{\"o}neberg}, {Scolnic}, {Sen}, {Sehgal}, {Shafieloo}, {Sheikh-Jabbari},
  {Silk}, {Silvestri}, {Skara}, {Sloth}, {Soares-Santos}, {Sol{\`a} Peracaula},
  {Songsheng}, {Soriano}, {Staicova}, {Starkman}, {Szapudi}, {Teixeira},
  {Thomas}, {Treu}, {Trott}, {van de Bruck}, {Vazquez}, {Verde}, {Visinelli},
  {Wang}, {Wang}, {Wang}, {Watkins}, {Watson}, {Webb}, {Weiner}, {Weltman},
  {Witte}, {Wojtak}, {Yadav}, {Yang}, {Zhao}, \&
  {Zumalac{\'a}rregui}}]{S8TensionReview}
{Abdalla}, E. {et~al.} 2022{\natexlab{a}}, Journal of High Energy Astrophysics,
  34, 49, 2203.06142

\bibitem[{{Abdalla} {et~al.}(2022{\natexlab{b}}){Abdalla}, {Abell{\'a}n},
  {Aboubrahim}, {Agnello}, {Akarsu}, {Akrami}, {Alestas}, {Aloni}, {Amendola},
  {Anchordoqui}, {Anderson}, {Arendse}, {Asgari}, {Ballardini}, {Barger},
  {Basilakos}, {Batista}, {Battistelli}, {Battye}, {Benetti}, {Benisty},
  {Berlin}, {de Bernardis}, {Berti}, {Bidenko}, {Birrer}, {Blakeslee}, {Boddy},
  {Bom}, {Bonilla}, {Borghi}, {Bouchet}, {Braglia}, {Buchert}, {Buckley-Geer},
  {Calabrese}, {Caldwell}, {Camarena}, {Capozziello}, {Casertano}, {Chen},
  {Chluba}, {Chen}, {Chen}, {Chudaykin}, {Cicoli}, {Copi}, {Courbin},
  {Cyr-Racine}, {Czerny}, {Dainotti}, {D'Amico}, {Davis}, {de Cruz P{\'e}rez},
  {de Haro}, {Delabrouille}, {Denton}, {Dhawan}, {Dienes}, {Di Valentino},
  {Du}, {Eckert}, {Escamilla-Rivera}, {Fert{\'e}}, {Finelli}, {Fosalba},
  {Freedman}, {Frusciante}, {Gazta{\~n}aga}, {Giar{\`e}}, {Giusarma},
  {G{\'o}mez-Valent}, {Handley}, {Harrison}, {Hart}, {Hazra}, {Heavens},
  {Heinesen}, {Hildebrandt}, {Hill}, {Hogg}, {Holz}, {Hooper}, {Hosseininejad},
  {Huterer}, {Ishak}, {Ivanov}, {Jaffe}, {Jang}, {Jedamzik}, {Jimenez},
  {Joseph}, {Joudaki}, {Kamionkowski}, {Karwal}, {Kazantzidis}, {Keeley},
  {Klasen}, {Komatsu}, {Koopmans}, {Kumar}, {Lamagna}, {Lazkoz}, {Lee},
  {Lesgourgues}, {Levi Said}, {Lewis}, {L'Huillier}, {Lucca}, {Maartens},
  {Macri}, {Marfatia}, {Marra}, {Martins}, {Masi}, {Matarrese}, {Mazumdar},
  {Melchiorri}, {Mena}, {Mersini-Houghton}, {Mertens}, {Milakovi{\'c}},
  {Minami}, {Miranda}, {Moreno-Pulido}, {Moresco}, {Mota}, {Mottola}, {Mozzon},
  {Muir}, {Mukherjee}, {Mukherjee}, {Naselsky}, {Nath}, {Nesseris},
  {Niedermann}, {Notari}, {Nunes}, {{\'O} Colg{\'a}in}, {Owens},
  {{\"O}z{\"u}lker}, {Pace}, {Paliathanasis}, {Palmese}, {Pan}, {Paoletti},
  {Perez Bergliaffa}, {Perivolaropoulos}, {Pesce}, {Pettorino}, {Philcox},
  {Pogosian}, {Poulin}, {Poulot}, {Raveri}, {Reid}, {Renzi}, {Riess}, {Sabla},
  {Salucci}, {Salzano}, {Saridakis}, {Sathyaprakash}, {Schmaltz},
  {Sch{\"o}neberg}, {Scolnic}, {Sen}, {Sehgal}, {Shafieloo}, {Sheikh-Jabbari},
  {Silk}, {Silvestri}, {Skara}, {Sloth}, {Soares-Santos}, {Sol{\`a} Peracaula},
  {Songsheng}, {Soriano}, {Staicova}, {Starkman}, {Szapudi}, {Teixeira},
  {Thomas}, {Treu}, {Trott}, {van de Bruck}, {Vazquez}, {Verde}, {Visinelli},
  {Wang}, {Wang}, {Wang}, {Watkins}, {Watson}, {Webb}, {Weiner}, {Weltman},
  {Witte}, {Wojtak}, {Yadav}, {Yang}, {Zhao}, \&
  {Zumalac{\'a}rregui}}]{CosmologyIntertwinedReview}
------. 2022{\natexlab{b}}, Journal of High Energy Astrophysics, 34, 49,
  2203.06142

\bibitem[{{Ajani} {et~al.}(2020){Ajani}, {Peel}, {Pettorino}, {Starck}, {Li},
  \& {Liu}}]{Ajani2020}
{Ajani}, V., {Peel}, A., {Pettorino}, V., {Starck}, J.-L., {Li}, Z., \& {Liu},
  J. 2020, \prd, 102, 103531

\bibitem[{{Akeson} {et~al.}(2019){Akeson}, {Armus}, {Bachelet}, {Bailey},
  {Bartusek}, {Bellini}, {Benford}, {Bennett}, {Bhattacharya}, {Bohlin},
  {Boyer}, {Bozza}, {Bryden}, {Calchi Novati}, {Carpenter}, {Casertano},
  {Choi}, {Content}, {Dayal}, {Dressler}, {Dor{\'e}}, {Fall}, {Fan}, {Fang},
  {Filippenko}, {Finkelstein}, {Foley}, {Furlanetto}, {Kalirai}, {Gaudi},
  {Gilbert}, {Girard}, {Grady}, {Greene}, {Guhathakurta}, {Heinrich},
  {Hemmati}, {Hendel}, {Henderson}, {Henning}, {Hirata}, {Ho}, {Huff},
  {Hutter}, {Jansen}, {Jha}, {Johnson}, {Jones}, {Kasdin}, {Kelly}, {Kirshner},
  {Koekemoer}, {Kruk}, {Lewis}, {Macintosh}, {Madau}, {Malhotra}, {Mandel},
  {Massara}, {Masters}, {McEnery}, {McQuinn}, {Melchior}, {Melton},
  {Mennesson}, {Peeples}, {Penny}, {Perlmutter}, {Pisani}, {Plazas}, {Poleski},
  {Postman}, {Ranc}, {Rauscher}, {Rest}, {Roberge}, {Robertson}, {Rodney},
  {Rhoads}, {Rhodes}, {Ryan}, {Sahu}, {Sand}, {Scolnic}, {Seth}, {Shvartzvald},
  {Siellez}, {Smith}, {Spergel}, {Stassun}, {Street}, {Strolger}, {Szalay},
  {Trauger}, {Troxel}, {Turnbull}, {van der Marel}, {von der Linden}, {Wang},
  {Weinberg}, {Williams}, {Windhorst}, {Wollack}, {Wu}, {Yee}, \&
  {Zimmerman}}]{Roman}
{Akeson}, R. {et~al.} 2019, arXiv e-prints, arXiv:1902.05569, 1902.05569

\bibitem[{{Alarcon} {et~al.}(2021){Alarcon}, {Gaztanaga}, {Eriksen}, {Baugh},
  {Cabayol}, {Casas}, {Carretero}, {Castander}, {De Vicente}, {Fernandez},
  {Garcia-Bellido}, {Hildebrandt}, {Hoekstra}, {Joachimi}, {Manzoni}, {Miquel},
  {Norberg}, {Padilla}, {Renard}, {Sanchez}, {Serrano}, {Sevilla-Noarbe},
  {Siudek}, \& {Tallada-Cresp{\'\i}}}]{Alarcon2021}
{Alarcon}, A. {et~al.} 2021, \mnras, 501, 6103, 2007.11132

\bibitem[{{Alonso} {et~al.}(2023){Alonso}, {Fabbian}, {Storey-Fisher},
  {Eilers}, {Garc{\'\i}a-Garc{\'\i}a}, {Hogg}, \& {Rix}}]{Quaia_Alonso2023}
{Alonso}, D., {Fabbian}, G., {Storey-Fisher}, K., {Eilers}, A.-C.,
  {Garc{\'\i}a-Garc{\'\i}a}, C., {Hogg}, D.~W., \& {Rix}, H.-W. 2023, arXiv
  e-prints, arXiv:2306.17748, 2306.17748

\bibitem[{{Alonso} {et~al.}(2019){Alonso}, {Sanchez}, {Slosar}, \& {LSST Dark
  Energy Science Collaboration}}]{NaMaster}
{Alonso}, D., {Sanchez}, J., {Slosar}, A., \& {LSST Dark Energy Science
  Collaboration}. 2019, \mnras, 484, 4127

\bibitem[{{Amon} {et~al.}(2022){Amon}, {Gruen}, {Troxel}, {MacCrann},
  {Dodelson}, {Choi}, {Doux}, {Secco}, {Samuroff}, {Krause}, {Cordero},
  {Myles}, {DeRose}, {Wechsler}, {Gatti}, {Navarro-Alsina}, {Bernstein},
  {Jain}, {Blazek}, {Alarcon}, {Fert{\'e}}, {Lemos}, {Raveri}, {Campos},
  {Prat}, {S{\'a}nchez}, {Jarvis}, {Alves}, {Andrade-Oliveira}, {Baxter},
  {Bechtol}, {Becker}, {Bridle}, {Camacho}, {Carnero Rosell}, {Carrasco Kind},
  {Cawthon}, {Chang}, {Chen}, {Chintalapati}, {Crocce}, {Davis}, {Diehl},
  {Drlica-Wagner}, {Eckert}, {Eifler}, {Elvin-Poole}, {Everett}, {Fang},
  {Fosalba}, {Friedrich}, {Gaztanaga}, {Giannini}, {Gruendl}, {Harrison},
  {Hartley}, {Herner}, {Huang}, {Huff}, {Huterer}, {Kuropatkin}, {Leget},
  {Liddle}, {McCullough}, {Muir}, {Pandey}, {Park}, {Porredon}, {Refregier},
  {Rollins}, {Roodman}, {Rosenfeld}, {Ross}, {Rykoff}, {Sanchez},
  {Sevilla-Noarbe}, {Sheldon}, {Shin}, {Troja}, {Tutusaus}, {Tutusaus},
  {Varga}, {Weaverdyck}, {Yanny}, {Yin}, {Zhang}, {Zuntz}, {Aguena}, {Allam},
  {Annis}, {Bacon}, {Bertin}, {Bhargava}, {Brooks}, {Buckley-Geer}, {Burke},
  {Carretero}, {Costanzi}, {da Costa}, {Pereira}, {De Vicente}, {Desai},
  {Dietrich}, {Doel}, {Ferrero}, {Flaugher}, {Frieman}, {Garc{\'\i}a-Bellido},
  {Gaztanaga}, {Gerdes}, {Giannantonio}, {Gschwend}, {Gutierrez}, {Hinton},
  {Hollowood}, {Honscheid}, {Hoyle}, {James}, {Kron}, {Kuehn}, {Lahav}, {Lima},
  {Lin}, {Maia}, {Marshall}, {Martini}, {Melchior}, {Menanteau}, {Miquel},
  {Mohr}, {Morgan}, {Ogando}, {Palmese}, {Paz-Chinch{\'o}n}, {Petravick},
  {Pieres}, {Romer}, {Sanchez}, {Scarpine}, {Schubnell}, {Serrano}, {Smith},
  {Soares-Santos}, {Tarle}, {Thomas}, {To}, {Weller}, \& {DES
  Collaboration}}]{DESY3_Amon}
{Amon}, A. {et~al.} 2022, \prd, 105, 023514

\bibitem[{{Amon} {et~al.}(2023){Amon}, {Robertson}, {Miyatake}, {Heymans},
  {White}, {DeRose}, {Yuan}, {Wechsler}, {Varga}, {Bocquet}, {Dvornik}, {More},
  {Ross}, {Hoekstra}, {Alarcon}, {Asgari}, {Blazek}, {Campos}, {Chen}, {Choi},
  {Crocce}, {Diehl}, {Doux}, {Eckert}, {Elvin-Poole}, {Everett}, {Fert{\'e}},
  {Gatti}, {Giannini}, {Gruen}, {Gruendl}, {Hartley}, {Herner}, {Hildebrandt},
  {Huang}, {Huff}, {Joachimi}, {Lee}, {MacCrann}, {Myles}, {Navarro-Alsina},
  {Nishimichi}, {Prat}, {Secco}, {Sevilla-Noarbe}, {Sheldon}, {Shin},
  {Tr{\"o}ster}, {Troxel}, {Tutusaus}, {Wright}, {Yin}, {Aguena}, {Allam},
  {Annis}, {Bacon}, {Bilicki}, {Brooks}, {Burke}, {Carnero Rosell},
  {Carretero}, {Castander}, {Cawthon}, {Costanzi}, {da Costa}, {Pereira}, {de
  Jong}, {De Vicente}, {Desai}, {Dietrich}, {Doel}, {Ferrero}, {Frieman},
  {Garc{\'\i}a-Bellido}, {Gerdes}, {Gschwend}, {Gutierrez}, {Hinton},
  {Hollowood}, {Honscheid}, {Huterer}, {Kannawadi}, {Kuehn}, {Kuropatkin},
  {Lahav}, {Lima}, {Maia}, {Marshall}, {Menanteau}, {Miquel}, {Mohr}, {Morgan},
  {Muir}, {Paz-Chinch{\'o}n}, {Pieres}, {Plazas Malag{\'o}n}, {Porredon},
  {Rodriguez-Monroy}, {Roodman}, {Sanchez}, {Serrano}, {Shan}, {Suchyta},
  {Swanson}, {Tarle}, {Thomas}, {To}, \& {Zhang}}]{LensingIsLow2023}
------. 2023, \mnras, 518, 477

\bibitem[{{Anbajagane} {et~al.}(2023){Anbajagane}, {Chang}, {Banerjee}, {Abel},
  {Gatti}, {Ajani}, {Alarcon}, {Amon}, {Baxter}, {Bechtol}, {Becker},
  {Bernstein}, {Campos}, {Carnero Rosell}, {Carrasco Kind}, {Chen}, {Choi},
  {Davis}, {DeRose}, {Diehl}, {Dodelson}, {Doux}, {Drlica-Wagner}, {Eckert},
  {Elvin-Poole}, {Everett}, {Fert'e}, {Gruen}, {Gruendl}, {Harrison},
  {Hartley}, {Huff}, {Jain}, {Jarvis}, {Jeffrey}, {Kacprzak}, {Kokron},
  {Kuropatkin}, {Leget}, {MacCrann}, {McCullough}, {Myles}, {Navarro-Alsina},
  {Pandey}, {Prat}, {Raveri}, {Rollins}, {Roodman}, {Rykoff}, {Sanchez},
  {Secco}, {Sevilla-Noarbe}, {Sheldon}, {Shin}, {Troxel}, {Tutusaus},
  {Whiteway}, {Yanny}, {Yin}, {Zhang}, {Abbott}, {Allam}, {Aguena}, {Alves},
  {Andrade-Oliveira}, {Annis}, {Bacon}, {Blazek}, {Brooks}, {Cawthon}, {da
  Costa}, {Pereira}, {Davis}, {Desai}, {Doel}, {Ferrero}, {Frieman},
  {Giannini}, {Gutierrez}, {Hinton}, {Hollowood}, {Honscheid}, {James},
  {Kuehn}, {Lahav}, {Marshall}, {Mena-Fernandez}, {Menanteau}, {Miquel},
  {Palmese}, {Pieres}, {Plazas Malag'on}, {Reil}, {Sanchez}, {Smith},
  {Swanson}, {Tarle}, \& {Wiseman}}]{DESY3_CDF}
{Anbajagane}, D. {et~al.} 2023, arXiv e-prints, arXiv:2308.03863, 2308.03863

\bibitem[{{Angulo} \& {Hahn}(2022)}]{AnguloHahn_Nbody}
{Angulo}, R.~E., \& {Hahn}, O. 2022, Living Reviews in Computational
  Astrophysics, 8, 1

\bibitem[{{Angulo} {et~al.}(2021){Angulo}, {Zennaro}, {Contreras}, {Aric{\`o}},
  {Pellejero-Iba{\~n}ez}, \& {St{\"u}cker}}]{BACCOEmu}
{Angulo}, R.~E., {Zennaro}, M., {Contreras}, S., {Aric{\`o}}, G.,
  {Pellejero-Iba{\~n}ez}, M., \& {St{\"u}cker}, J. 2021, \mnras, 507, 5869,
  2004.06245

\bibitem[{{Aric{\`o}} {et~al.}(2023){Aric{\`o}}, {Angulo}, {Zennaro},
  {Contreras}, {Chen}, \& {Hern{\'a}ndez-Monteagudo}}]{DESY3_BACCO}
{Aric{\`o}}, G., {Angulo}, R.~E., {Zennaro}, M., {Contreras}, S., {Chen}, A.,
  \& {Hern{\'a}ndez-Monteagudo}, C. 2023, \aap, 678, A109

\bibitem[{{Asgari} {et~al.}(2021){Asgari}, {Lin}, {Joachimi}, {Giblin},
  {Heymans}, {Hildebrandt}, {Kannawadi}, {St{\"o}lzner}, {Tr{\"o}ster}, {van
  den Busch}, {Wright}, {Bilicki}, {Blake}, {de Jong}, {Dvornik}, {Erben},
  {Getman}, {Hoekstra}, {K{\"o}hlinger}, {Kuijken}, {Miller}, {Radovich},
  {Schneider}, {Shan}, \& {Valentijn}}]{KiDS1000_Asgari}
{Asgari}, M. {et~al.} 2021, \aap, 645, A104

\bibitem[{{Asgari} {et~al.}(2020){Asgari}, {Tr{\"o}ster}, {Heymans},
  {Hildebrandt}, {van den Busch}, {Wright}, {Choi}, {Erben}, {Joachimi},
  {Joudaki}, {Kannawadi}, {Kuijken}, {Lin}, {Schneider}, \&
  {Zuntz}}]{Asgari_DES_KiDS_cosebi}
------. 2020, \aap, 634, A127

\bibitem[{{Begeman} {et~al.}(2013){Begeman}, {Belikov}, {Boxhoorn}, \&
  {Valentijn}}]{astroWISE}
{Begeman}, K., {Belikov}, A.~N., {Boxhoorn}, D.~R., \& {Valentijn}, E.~A. 2013,
  Experimental Astronomy, 35, 1

\bibitem[{{Ben{\'{\i}}tez}(2000)}]{BPZ}
{Ben{\'{\i}}tez}, N. 2000, \apj, 536, 571

\bibitem[{{Bilicki} {et~al.}(2021){Bilicki}, {Dvornik}, {Hoekstra}, {Wright},
  {Chisari}, {Vakili}, {Asgari}, {Giblin}, {Heymans}, {Hildebrandt},
  {Holwerda}, {Hopkins}, {Johnston}, {Kannawadi}, {Kuijken}, {Nakoneczny},
  {Shan}, {Sonnenfeld}, \& {Valentijn}}]{KiDS1000_bright}
{Bilicki}, M. {et~al.} 2021, \aap, 653, A82

\bibitem[{{Blazek} {et~al.}(2019){Blazek}, {MacCrann}, {Troxel}, \&
  {Fang}}]{TATT}
{Blazek}, J.~A., {MacCrann}, N., {Troxel}, M.~A., \& {Fang}, X. 2019, \prd,
  100, 103506

\bibitem[{{Bridle} \& {King}(2007)}]{2007NJPh....9..444B}
{Bridle}, S., \& {King}, L. 2007, New Journal of Physics, 9, 444

\bibitem[{{Brouwer} {et~al.}(2018){Brouwer}, {Demchenko}, {Harnois-D{\'e}raps},
  {Bilicki}, {Heymans}, {Hoekstra}, {Kuijken}, {Alpaslan}, {Brough}, {Cai},
  {Costa-Duarte}, {Dvornik}, {Erben}, {Hildebrandt}, {Holwerda}, {Schneider},
  {Sif{\'o}n}, \& {van Uitert}}]{Brouwer2018}
{Brouwer}, M.~M. {et~al.} 2018, \mnras, 481, 5189

\bibitem[{{Burger} {et~al.}(2022){Burger}, {Friedrich}, {Harnois-D{\'e}raps},
  \& {Schneider}}]{KiDS1000_BurgerDSS}
{Burger}, P., {Friedrich}, O., {Harnois-D{\'e}raps}, J., \& {Schneider}, P.
  2022, \aap, 661, A137

\bibitem[{{Burger} {et~al.}(2023){Burger}, {Porth}, {Heydenreich}, {Linke},
  {Wielders}, {Schneider}, {Asgari}, {Castro}, {Dolag}, {Harnois-Deraps},
  {Kuijken}, \& {Martinet}}]{KiDS1000_Map3}
{Burger}, P.~A. {et~al.} 2023, arXiv e-prints, 2309.08602

\bibitem[{{Castro} {et~al.}(2018){Castro}, {Quartin}, {Giocoli}, {Borgani}, \&
  {Dolag}}]{LensingPDF_baryons}
{Castro}, T., {Quartin}, M., {Giocoli}, C., {Borgani}, S., \& {Dolag}, K. 2018,
  \mnras, 478, 1305

\bibitem[{{Chintalapati} {et~al.}(2022){Chintalapati}, {Gutierrez}, \&
  {Wang}}]{LensingProjection}
{Chintalapati}, P.~R.~V., {Gutierrez}, G., \& {Wang}, M.~H.~L.~S. 2022, \prd,
  105, 043515

\bibitem[{{Dalal} {et~al.}(2023){Dalal}, {Li}, {Nicola}, {Zuntz}, {Strauss},
  {Sugiyama}, {Zhang}, {Rau}, {Mandelbaum}, {Takada}, {More}, {Miyatake},
  {Kannawadi}, {Shirasaki}, {Taniguchi}, {Takahashi}, {Osato}, {Hamana},
  {Oguri}, {Nishizawa}, {Plazas Malag{\'o}n}, {Sunayama}, {Alonso}, {Slosar},
  {Armstrong}, {Bosch}, {Komiyama}, {Lupton}, {Lust}, {MacArthur}, {Miyazaki},
  {Murayama}, {Nishimichi}, {Okura}, {Price}, {Tait}, {Tanaka}, \&
  {Wang}}]{HSCY3_Cl}
{Dalal}, R. {et~al.} 2023, arXiv e-prints, arXiv:2304.00701, 2304.00701

\bibitem[{{Dark Energy Survey and Kilo-Degree Survey
  Collaboration}(2023)}]{DESY3-KiDS1000}
{Dark Energy Survey and Kilo-Degree Survey Collaboration}. 2023, The Open
  Journal of Astrophysics, 6, 36

\bibitem[{{DESI Collaboration} {et~al.}(2024){DESI Collaboration}, {Adame},
  {Aguilar}, {Ahlen}, {Alam}, {Alexander}, {Alvarez}, {Alves}, {Anand},
  {Andrade}, {Armengaud}, {Avila}, {Aviles}, {Awan}, {Bahr-Kalus}, {Bailey},
  {Baltay}, {Bault}, {Behera}, {BenZvi}, {Bera}, {Beutler}, {Bianchi}, {Blake},
  {Blum}, {Brieden}, {Brodzeller}, {Brooks}, {Buckley-Geer}, {Burtin},
  {Calderon}, {Canning}, {Carnero Rosell}, {Cereskaite}, {Cervantes-Cota},
  {Chabanier}, {Chaussidon}, {Chaves-Montero}, {Chen}, {Chen}, {Claybaugh},
  {Cole}, {Cuceu}, {Davis}, {Dawson}, {de la Macorra}, {de Mattia}, {Deiosso},
  {Dey}, {Dey}, {Ding}, {Doel}, {Edelstein}, {Eftekharzadeh}, {Eisenstein},
  {Elliott}, {Fagrelius}, {Fanning}, {Ferraro}, {Ereza}, {Findlay}, {Flaugher},
  {Font-Ribera}, {Forero-S{\'a}nchez}, {Forero-Romero}, {Frenk},
  {Garcia-Quintero}, {Gazta{\~n}aga}, {Gil-Mar{\'\i}n}, {Gontcho},
  {Gonzalez-Morales}, {Gonzalez-Perez}, {Gordon}, {Green}, {Gruen}, {Gsponer},
  {Gutierrez}, {Guy}, {Hadzhiyska}, {Hahn}, {Hanif}, {Herrera-Alcantar},
  {Honscheid}, {Howlett}, {Huterer}, {Ir{\v{s}}i{\v{c}}}, {Ishak}, {Juneau},
  {Kara{\c{c}}ayl{\i}}, {Kehoe}, {Kent}, {Kirkby}, {Kremin}, {Krolewski},
  {Lai}, {Lan}, {Landriau}, {Lang}, {Lasker}, {Le Goff}, {Le Guillou},
  {Leauthaud}, {Levi}, {Li}, {Linder}, {Lodha}, {Magneville}, {Manera},
  {Margala}, {Martini}, {Maus}, {McDonald}, {Medina-Varela}, {Meisner},
  {Mena-Fern{\'a}ndez}, {Miquel}, {Moon}, {Moore}, {Moustakas}, {Mudur},
  {Mueller}, {Mu{\~n}oz-Guti{\'e}rrez}, {Myers}, {Nadathur}, {Napolitano},
  {Neveux}, {Newman}, {Nguyen}, {Nie}, {Niz}, {Noriega}, {Padmanabhan},
  {Paillas}, {Palanque-Delabrouille}, {Pan}, {Penmetsa}, {Percival}, {Pieri},
  {Pinon}, {Poppett}, {Porredon}, {Prada}, {P{\'e}rez-Fern{\'a}ndez},
  {P{\'e}rez-R{\`a}fols}, {Rabinowitz}, {Raichoor}, {Ram{\'\i}rez-P{\'e}rez},
  {Ramirez-Solano}, {Ravoux}, {Rashkovetskyi}, {Rezaie}, {Rich}, {Rocher},
  {Rockosi}, {Roe}, {Rosado-Marin}, {Ross}, {Rossi}, {Ruggeri},
  {Ruhlmann-Kleider}, {Samushia}, {Sanchez}, {Saulder}, {Schlafly}, {Schlegel},
  {Schubnell}, {Seo}, {Shafieloo}, {Sharples}, {Silber}, {Slosar}, {Smith},
  {Sprayberry}, {Tan}, {Tarl{\'e}}, {Taylor}, {Trusov}, {Ure{\~n}a-L{\'o}pez},
  {Vaisakh}, {Valcin}, {Valdes}, {Vargas-Maga{\~n}a}, {Verde}, {Walther},
  {Wang}, {Wang}, {Weaver}, {Weaverdyck}, {Wechsler}, {Weinberg}, {White},
  {Yu}, {Yu}, {Yuan}, {Y{\`e}che}, {Zaborowski}, {Zarrouk}, {Zhang}, {Zhao},
  {Zhao}, {Zhou}, {Zhuang}, \& {Zou}}]{DESI_cosmo}
{DESI Collaboration} {et~al.} 2024, arXiv e-prints, arXiv:2404.03002,
  2404.03002

\bibitem[{{Duncan} {et~al.}(2022){Duncan}, {Harnois-D{\'e}raps}, {Miller}, \&
  {Langedijk}}]{Duncan2022}
{Duncan}, C. A.~J., {Harnois-D{\'e}raps}, J., {Miller}, L., \& {Langedijk}, A.
  2022, \mnras, 515, 1130

\bibitem[{{Edge} {et~al.}(2013){Edge}, {Sutherland}, {Kuijken}, {Driver},
  {McMahon}, {Eales}, \& {Emerson}}]{VIKING}
{Edge}, A., {Sutherland}, W., {Kuijken}, K., {Driver}, S., {McMahon}, R.,
  {Eales}, S., \& {Emerson}, J.~P. 2013, The Messenger, 154, 32

\bibitem[{{Erben} {et~al.}(2013){Erben}, {Hildebrandt}, {Miller}, {van
  Waerbeke}, {Heymans}, {Hoekstra}, {Kitching}, {Mellier}, {Benjamin}, {Blake},
  {Bonnett}, {Cordes}, {Coupon}, {Fu}, {Gavazzi}, {Gillis}, {Grocutt}, {Gwyn},
  {Holhjem}, {Hudson}, {Kilbinger}, {Kuijken}, {Milkeraitis}, {Rowe},
  {Schrabback}, {Semboloni}, {Simon}, {Smit}, {Toader}, {Vafaei}, {van Uitert},
  \& {Velander}}]{2013MNRAS.433.2545E}
{Erben}, T. {et~al.} 2013, \mnras, 433, 2545

\bibitem[{{Euclid Collaboration: Ajani} {et~al.}(2023){Euclid Collaboration:
  Ajani}, {Baldi}, {Barthelemy}, {Boyle}, {Burger}, {Cardone}, {Cheng},
  {Codis}, {Giocoli}, {Harnois-D{\'e}raps}, {Heydenreich}, {Kansal},
  {Kilbinger}, {Linke}, {Llinares}, {Martinet}, {Parroni}, {Peel}, {Pires},
  {Porth}, {Tereno}, {Uhlemann}, {Vicinanza}, {Vinciguerra}, {Aghanim},
  {Auricchio}, {Bonino}, {Branchini}, {Brescia}, {Brinchmann}, {Camera},
  {Capobianco}, {Carbone}, {Carretero}, {Castander}, {Castellano}, {Cavuoti},
  {Cimatti}, {Cledassou}, {Congedo}, {Conselice}, {Conversi}, {Corcione},
  {Courbin}, {Cropper}, {Da Silva}, {Degaudenzi}, {Di Giorgio}, {Dinis},
  {Douspis}, {Dubath}, {Dupac}, {Farrens}, {Ferriol}, {Fosalba}, {Frailis},
  {Franceschi}, {Galeotta}, {Garilli}, {Gillis}, {Grazian}, {Grupp},
  {Hoekstra}, {Holmes}, {Hornstrup}, {Hudelot}, {Jahnke}, {Jhabvala},
  {K{\"u}mmel}, {Kitching}, {Kunz}, {Kurki-Suonio}, {Lilje}, {Lloro},
  {Maiorano}, {Mansutti}, {Marggraf}, {Markovic}, {Marulli}, {Massey}, {Mei},
  {Mellier}, {Meneghetti}, {Moresco}, {Moscardini}, {Niemi}, {Nightingale},
  {Nutma}, {Padilla}, {Paltani}, {Pedersen}, {Pettorino}, {Polenta}, {Poncet},
  {Popa}, {Raison}, {Renzi}, {Rhodes}, {Riccio}, {Romelli}, {Roncarelli},
  {Rossetti}, {Saglia}, {Sapone}, {Sartoris}, {Schneider}, {Schrabback},
  {Secroun}, {Seidel}, {Serrano}, {Sirignano}, {Stanco}, {Starck},
  {Tallada-Cresp{\'\i}}, {Taylor}, {Toledo-Moreo}, {Torradeflot}, {Tutusaus},
  {Valentijn}, {Valenziano}, {Vassallo}, {Wang}, {Weller}, {Zamorani},
  {Zoubian}, {Andreon}, {Bardelli}, {Boucaud}, {Bozzo}, {Colodro-Conde}, {Di
  Ferdinando}, {Fabbian}, {Farina}, {Graci{\'a}-Carpio}, {Keih{\"a}nen},
  {Lindholm}, {Maino}, {Mauri}, {Neissner}, {Schirmer}, {Scottez}, {Zucca},
  {Akrami}, {Baccigalupi}, {Balaguera-Antol{\'\i}nez}, {Ballardini},
  {Bernardeau}, {Biviano}, {Blanchard}, {Borgani}, {Borlaff}, {Burigana},
  {Cabanac}, {Cappi}, {Carvalho}, {Casas}, {Castignani}, {Castro}, {Chambers},
  {Cooray}, {Coupon}, {Courtois}, {Davini}, {de la Torre}, {De Lucia},
  {Desprez}, {Dole}, {Escartin}, {Escoffier}, {Ferrero}, {Finelli}, {Ganga},
  {Garcia-Bellido}, {George}, {Giacomini}, {Gozaliasl}, {Hildebrandt}, {Jimenez
  Mu{\~n}oz}, {Joachimi}, {Kajava}, {Kirkpatrick}, {Legrand}, {Loureiro},
  {Magliocchetti}, {Maoli}, {Marcin}, {Martinelli}, {Martins}, {Matthew},
  {Maurin}, {Metcalf}, {Monaco}, {Morgante}, {Nadathur}, {Nucita}, {Popa},
  {Potter}, {Pourtsidou}, {P{\"o}ntinen}, {Reimberg}, {S{\'a}nchez}, {Sakr},
  {Schneider}, {Sefusatti}, {Sereno}, {Shulevski}, {Spurio Mancini},
  {Steinwagner}, {Teyssier}, {Valiviita}, {Veropalumbo}, {Viel}, \&
  {Zinchenko}}]{HOWLS_paper1}
{Euclid Collaboration: Ajani}, V. {et~al.} 2023, \aap, 675, A120

\bibitem[{{Euclid Collaboration: Knabenhans} {et~al.}(2019){Euclid
  Collaboration: Knabenhans}, {Stadel}, {Marelli}, {Potter}, {Teyssier},
  {Legrand }, {Schneider}, {Sudret}, {Blot}, {Awan}, {Burigana}, {Carvalho},
  {Kurki-Suonio}, \& {Sirri}}]{EuclidEmulator}
{Euclid Collaboration: Knabenhans}, M. {et~al.} 2019, \mnras, 484, 5509

\bibitem[{{Fenech Conti} {et~al.}(2017){Fenech Conti}, {Herbonnet}, {Hoekstra},
  {Merten}, {Miller}, \& {Viola}}]{FenechConti17}
{Fenech Conti}, I., {Herbonnet}, R., {Hoekstra}, H., {Merten}, J., {Miller},
  L., \& {Viola}, M. 2017, \mnras, 467, 1627

\bibitem[{{Feroz} {et~al.}(2009){Feroz}, {Hobson}, \& {Bridges}}]{Multinest}
{Feroz}, F., {Hobson}, M.~P., \& {Bridges}, M. 2009, \mnras, 398, 1601

\bibitem[{{Fluri} {et~al.}(2019){Fluri}, {Kacprzak}, {Lucchi}, {Refregier},
  {Amara}, {Hofmann}, \& {Schneider}}]{Fluri2019}
{Fluri}, J., {Kacprzak}, T., {Lucchi}, A., {Refregier}, A., {Amara}, A.,
  {Hofmann}, T., \& {Schneider}, A. 2019, \prd, 100, 063514

\bibitem[{{Fluri} {et~al.}(2022){Fluri}, {Kacprzak}, {Lucchi}, {Schneider},
  {Refregier}, \& {Hofmann}}]{KiDS1000_Fluri}
{Fluri}, J., {Kacprzak}, T., {Lucchi}, A., {Schneider}, A., {Refregier}, A., \&
  {Hofmann}, T. 2022, \prd, 105, 083518

\bibitem[{{Fortuna} {et~al.}(2021){Fortuna}, {Hoekstra}, {Joachimi},
  {Johnston}, {Chisari}, {Georgiou}, \& {Mahony}}]{Fortuna2021}
{Fortuna}, M.~C., {Hoekstra}, H., {Joachimi}, B., {Johnston}, H., {Chisari},
  N.~E., {Georgiou}, C., \& {Mahony}, C. 2021, \mnras, 501, 2983

\bibitem[{{Fosalba} {et~al.}(2015){Fosalba}, {Gazta{\~n}aga}, {Castander}, \&
  {Crocce}}]{Fosalba2013}
{Fosalba}, P., {Gazta{\~n}aga}, E., {Castander}, F.~J., \& {Crocce}, M. 2015,
  \mnras, 447, 1319

\bibitem[{{Fosalba} {et~al.}(2008){Fosalba}, {Gazta{\~n}aga}, {Castander}, \&
  {Manera}}]{Onion}
{Fosalba}, P., {Gazta{\~n}aga}, E., {Castander}, F.~J., \& {Manera}, M. 2008,
  \mnras, 391, 435

\bibitem[{{Fu} {et~al.}(2014){Fu}, {Kilbinger}, {Erben}, {Heymans},
  {Hildebrandt}, {Hoekstra}, {Kitching}, {Mellier}, {Miller}, {Semboloni},
  {Simon}, {Van Waerbeke}, {Coupon}, {Harnois-D{\'e}raps}, {Hudson}, {Kuijken},
  {Rowe}, {Schrabback}, {Vafaei}, \& {Velander}}]{Fu2014}
{Fu}, L. {et~al.} 2014, \mnras, 441, 2725

\bibitem[{{Gatti} {et~al.}(2020){Gatti}, {Chang}, {Friedrich}, {Jain}, {Bacon},
  {Crocce}, {DeRose}, {Ferrero}, {Fosalba}, {Gaztanaga}, {Gruen}, {Harrison},
  {Jeffrey}, {MacCrann}, {McClintock}, {Secco}, {Whiteway}, {Abbott}, {Allam},
  {Annis}, {Avila}, {Brooks}, {Buckley-Geer}, {Burke}, {Carnero Rosell},
  {Carrasco Kind}, {Carretero}, {Cawthon}, {da Costa}, {De Vicente}, {Desai},
  {Diehl}, {Doel}, {Eifler}, {Estrada}, {Everett}, {Evrard}, {Frieman},
  {Garc{\'\i}a-Bellido}, {Gerdes}, {Gruendl}, {Gschwend}, {Gutierrez}, {James},
  {Johnson}, {Krause}, {Kuehn}, {Lima}, {Maia}, {March}, {Marshall},
  {Melchior}, {Menanteau}, {Miquel}, {Palmese}, {Paz-Chinch{\'o}n}, {Plazas},
  {S{\'a}nchez}, {Sanchez}, {Scarpine}, {Schubnell}, {Santiago},
  {Sevilla-Noarbe}, {Smith}, {Soares-Santos}, {Suchyta}, {Swanson}, {Tarle},
  {Thomas}, {Troxel}, {Zuntz}, \& {DES Collaboration}}]{Gatti20}
{Gatti}, M. {et~al.} 2020, \mnras, 498, 4060

\bibitem[{{Gatti} {et~al.}(2023{\natexlab{a}}){Gatti}, {Jeffrey}, {Whiteway},
  {Ajani}, {Kacprzak}, {Z{\"u}rcher}, {Chang}, {Jain}, {Blazek}, {Krause},
  {Alarcon}, {Amon}, {Bechtol}, {Becker}, {Bernstein}, {Campos}, {Chen},
  {Choi}, {Davis}, {Derose}, {Diehl}, {Dodelson}, {Doux}, {Eckert},
  {Elvin-Poole}, {Everett}, {Ferte}, {Gruen}, {Gruendl}, {Harrison}, {Hartley},
  {Herner}, {Huff}, {Jarvis}, {Kuropatkin}, {Leget}, {MacCrann}, {McCullough},
  {Myles}, {Navarro-Alsina}, {Pandey}, {Prat}, {Raveri}, {Rollins}, {Roodman},
  {Sanchez}, {Secco}, {Sevilla-Noarbe}, {Sheldon}, {Shin}, {Troxel},
  {Tutusaus}, {Varga}, {Yanny}, {Yin}, {Zhang}, {Zuntz}, {Allam}, {Alves},
  {Aguena}, {Bacon}, {Bertin}, {Brooks}, {Burke}, {Carnero Rosell},
  {Carretero}, {Cawthon}, {da Costa}, {Davis}, {De Vicente}, {Desai}, {Doel},
  {Garc{\'\i}a-Bellido}, {Giannini}, {Gutierrez}, {Ferrero}, {Frieman},
  {Hinton}, {Hollowood}, {Honscheid}, {James}, {Kuehn}, {Lahav}, {Marshall},
  {Mena-Fern{\'a}ndez}, {Miquel}, {Ogando}, {Palmese}, {Pereira}, {Plazas
  Malag{\'o}n}, {Rodriguez-Monroy}, {Samuroff}, {Sanchez}, {Schubnell},
  {Smith}, {Sobreira}, {Suchyta}, {Swanson}, {Tarle}, {Weaverdyck}, \&
  {Wiseman}}]{DESY3_Gatti_source_clust}
------. 2023{\natexlab{a}}, arXiv e-prints, 2307.13860

\bibitem[{{Gatti} {et~al.}(2023{\natexlab{b}}){Gatti}, {Jeffrey}, {Whiteway},
  {Williamson}, {Jain}, {Ajani}, {Anbajagane}, {Giannini}, {Zhou}, {Porredon},
  {Prat}, {Yamamoto}, {Blazek}, {Kacprzak}, {Samuroff}, {Alarcon}, {Amon},
  {Bechtol}, {Becker}, {Bernstein}, {Campos}, {Chang}, {Chen}, {Choi}, {Davis},
  {Derose}, {Diehl}, {Dodelson}, {Doux}, {Eckert}, {Elvin-Poole}, {Everett},
  {Ferte}, {Gruen}, {Gruendl}, {Harrison}, {Hartley}, {Herner}, {Huff},
  {Jarvis}, {Kuropatkin}, {Leget}, {MacCrann}, {McCullough}, {Myles},
  {Navarro-Alsina}, {Pandey}, {Raveri}, {Rollins}, {Roodman}, {Sanchez},
  {Secco}, {Sevilla-Noarbe}, {Sheldon}, {Shin}, {Troxel}, {Tutusaus}, {Varga},
  {Yanny}, {Yin}, {Zhang}, {Zuntz}, {Aguena}, {Alves}, {Annis}, {Brooks},
  {Carretero}, {Castander}, {Cawthon}, {da Costa}, {De Vicente}, {Desai},
  {Ferrero}, {Flaugher}, {Friedel}, {Frieman}, {Garc{\'\i}a-Bellido},
  {Gutierrez}, {Hinton}, {Hollowood}, {Honscheid}, {James}, {Kuehn}, {Lahav},
  {Lee}, {Marshall}, {Mena-Fern{\'a}ndez}, {Menanteau}, {Miquel}, {Ogando},
  {Pereira}, {Pieres}, {Plazas Malag{\'o}n}, {Sanchez}, {Smith}, {Suchyta},
  {Swanson}, {Tarle}, {Weaverdyck}, {Weller}, \& {Wiseman}}]{Gatti2023}
------. 2023{\natexlab{b}}, arXiv e-prints, 2310.17557

\bibitem[{{Giblin} {et~al.}(2023){Giblin}, {Cai}, \&
  {Harnois-D{\'e}raps}}]{Giblin_PDF}
{Giblin}, B., {Cai}, Y.-C., \& {Harnois-D{\'e}raps}, J. 2023, \mnras, 520, 1721

\bibitem[{{Giblin} {et~al.}(2021){Giblin}, {Heymans}, {Asgari}, {Hildebrandt},
  {Hoekstra}, {Joachimi}, {Kannawadi}, {Kuijken}, {Lin}, {Miller},
  {Tr{\"o}ster}, {van den Busch}, {Wright}, {Bilicki}, {Blake}, {de Jong},
  {Dvornik}, {Erben}, {Getman}, {Napolitano}, {Schneider}, {Shan}, \&
  {Valentijn}}]{KiDS1000_Giblin}
{Giblin}, B. {et~al.} 2021, \aap, 645, 105

\bibitem[{{Giblin} {et~al.}(2018){Giblin}, {Heymans}, {Harnois-D{\'e}raps},
  {Simpson}, {Dietrich}, {Van Waerbeke}, {Amon}, {Asgari}, {Erben},
  {Hildebrandt}, {Joachimi}, {Kuijken}, {Martinet}, {Schneider}, \&
  {Tr{\"o}ster}}]{Giblin18}
------. 2018, \mnras, 480, 5529

\bibitem[{{Gruen} \& {Brimioulle}(2017)}]{GruenBrimioulle}
{Gruen}, D., \& {Brimioulle}, F. 2017, \mnras, 468, 769

\bibitem[{{Gruen} {et~al.}(2018){Gruen}, {Friedrich}, {Krause}, {DeRose},
  {Cawthon}, {Davis}, {Elvin-Poole}, {Rykoff}, {Wechsler}, {Alarcon},
  {Bernstein}, {Blazek}, {Chang}, {Clampitt}, {Crocce}, {De Vicente}, {Gatti},
  {Gill}, {Hartley}, {Hilbert}, {Hoyle}, {Jain}, {Jarvis}, {Lahav}, {MacCrann},
  {McClintock}, {Prat}, {Rollins}, {Ross}, {Rozo}, {Samuroff}, {S{\'a}nchez},
  {Sheldon}, {Troxel}, {Zuntz}, {Abbott}, {Abdalla}, {Allam}, {Annis},
  {Bechtol}, {Benoit-L{\'e}vy}, {Bertin}, {Bridle}, {Brooks}, {Buckley-Geer},
  {Carnero Rosell}, {Carrasco Kind}, {Carretero}, {Cunha}, {D'Andrea}, {da
  Costa}, {Desai}, {Diehl}, {Dietrich}, {Doel}, {Drlica-Wagner}, {Fernandez},
  {Flaugher}, {Fosalba}, {Frieman}, {Garc{\'\i}a-Bellido}, {Gaztanaga},
  {Giannantonio}, {Gruendl}, {Gschwend}, {Gutierrez}, {Honscheid}, {James},
  {Jeltema}, {Kuehn}, {Kuropatkin}, {Lima}, {March}, {Marshall}, {Martini},
  {Melchior}, {Menanteau}, {Miquel}, {Mohr}, {Plazas}, {Roodman}, {Sanchez},
  {Scarpine}, {Schubnell}, {Sevilla-Noarbe}, {Smith}, {Smith}, {Soares-Santos},
  {Sobreira}, {Swanson}, {Tarle}, {Thomas}, {Vikram}, {Walker}, {Weller},
  {Zhang}, \& {DES Collaboration}}]{Gruen2017}
{Gruen}, D. {et~al.} 2018, \prd, 98, 023507

\bibitem[{Gupta \& Nagar(1999)}]{Gupta1999-matrix}
Gupta, A., \& Nagar, D. 1999, Matrix Variate Distributions, Monographs and
  Surveys in Pure and Applied Mathematics (Taylor \& Francis)

\bibitem[{{Hamana} {et~al.}(2020){Hamana}, {Shirasaki}, {Miyazaki}, {Hikage},
  {Oguri}, {More}, {Armstrong}, {Leauthaud}, {Mandelbaum}, {Miyatake},
  {Nishizawa}, {Simet}, {Takada}, {Aihara}, {Bosch}, {Komiyama}, {Lupton},
  {Murayama}, {Strauss}, \& {Tanaka}}]{HSCY1_2PCF}
{Hamana}, T. {et~al.} 2020, \pasj, 72, 16

\bibitem[{{Harnois-D{\'e}raps} {et~al.}(2019){Harnois-D{\'e}raps}, {Giblin}, \&
  {Joachimi}}]{cosmoSLICS}
{Harnois-D{\'e}raps}, J., {Giblin}, B., \& {Joachimi}, B. 2019, \aap, 631, A160

\bibitem[{{Harnois-D{\'e}raps}
  {et~al.}(2021{\natexlab{a}}){Harnois-D{\'e}raps}, {Martinet}, {Castro},
  {Dolag}, {Giblin}, {Heymans}, {Hildebrandt}, \& {Xia}}]{HD21}
{Harnois-D{\'e}raps}, J., {Martinet}, N., {Castro}, T., {Dolag}, K., {Giblin},
  B., {Heymans}, C., {Hildebrandt}, H., \& {Xia}, Q. 2021{\natexlab{a}},
  \mnras, 506, 1623

\bibitem[{{Harnois-D{\'e}raps}
  {et~al.}(2021{\natexlab{b}}){Harnois-D{\'e}raps}, {Martinet}, \&
  {Reischke}}]{Tidalator}
{Harnois-D{\'e}raps}, J., {Martinet}, N., \& {Reischke}, R. 2021{\natexlab{b}},
  \mnras

\bibitem[{{Harnois-D{\'e}raps} \& {van Waerbeke}(2015)}]{SLICS_1}
{Harnois-D{\'e}raps}, J., \& {van Waerbeke}, L. 2015, \mnras, 450, 2857,
  1406.0543

\bibitem[{{Hartlap} {et~al.}(2007){Hartlap}, {Simon}, \&
  {Schneider}}]{Hartlap2007}
{Hartlap}, J., {Simon}, P., \& {Schneider}, P. 2007, \aap, 464, 399

\bibitem[{{Heitmann} {et~al.}(2014){Heitmann}, {Lawrence}, {Kwan}, {Habib}, \&
  {Higdon}}]{Coyote3}
{Heitmann}, K., {Lawrence}, E., {Kwan}, J., {Habib}, S., \& {Higdon}, D. 2014,
  \apj, 780, 111

\bibitem[{{Heydenreich} {et~al.}(2022){Heydenreich}, {Br{\"u}ck}, {Burger},
  {Harnois-D{\'e}raps}, {Unruh}, {Castro}, {Dolag}, \&
  {Martinet}}]{DESY1_Heydenreich}
{Heydenreich}, S., {Br{\"u}ck}, B., {Burger}, P., {Harnois-D{\'e}raps}, J.,
  {Unruh}, S., {Castro}, T., {Dolag}, K., \& {Martinet}, N. 2022, arXiv
  e-prints, arXiv:2204.11831, 2204.11831

\bibitem[{{Heydenreich} {et~al.}(2021){Heydenreich}, {Br{\"u}ck}, \&
  {Harnois-D{\'e}raps}}]{Homology}
{Heydenreich}, S., {Br{\"u}ck}, B., \& {Harnois-D{\'e}raps}, J. 2021, \aap,
  648, A74, 2007.13724

\bibitem[{{Hilbert} {et~al.}(2020){Hilbert}, {Barreira}, {Fabbian}, {Fosalba},
  {Giocoli}, {Bose}, {Calabrese}, {Carbone}, {Davies}, {Li}, {Llinares}, \&
  {Monaco}}]{Hilbert_LensSimAccuracy}
{Hilbert}, S. {et~al.} 2020, \mnras, 493, 305, 1910.10625

\bibitem[{{Hildebrandt} {et~al.}(2021){Hildebrandt}, {van den Busch}, {Wright},
  {Blake}, {Joachimi}, {Kuijken}, {Tr{\"o}ster}, {Asgari}, {Bilicki}, {de
  Jong}, {Dvornik}, {Erben}, {Getman}, {Giblin}, {Heymans}, {Kannawadi}, {Lin},
  \& {Shan}}]{KiDS1000_redshifts}
{Hildebrandt}, H. {et~al.} 2021, \aap, 647, A124

\bibitem[{{Ho} {et~al.}(2022){Ho}, {Bird}, \& {Shelton}}]{Bird_Multifidelity}
{Ho}, M.-F., {Bird}, S., \& {Shelton}, C.~R. 2022, \mnras, 509, 2551,
  2105.01081

\bibitem[{{Hoekstra} {et~al.}(2015){Hoekstra}, {Herbonnet}, {Muzzin}, {Babul},
  {Mahdavi}, {Viola}, \& {Cacciato}}]{2015MNRAS.449..685H}
{Hoekstra}, H., {Herbonnet}, R., {Muzzin}, A., {Babul}, A., {Mahdavi}, A.,
  {Viola}, M., \& {Cacciato}, M. 2015, \mnras, 449, 685, 1502.01883

\bibitem[{Hotelling(1931)}]{Hotelling_paper}
Hotelling, H. 1931, The Annals of Mathematical Statistics, 2, 360

\bibitem[{{Hoyle} {et~al.}(2018){Hoyle}, {Gruen}, {Bernstein}, {Rau}, {De
  Vicente}, {Hartley}, {Gaztanaga}, {DeRose}, {Troxel}, {Davis}, {Alarcon},
  {MacCrann}, {Prat}, {S{\'a}nchez}, {Sheldon}, {Wechsler}, {Asorey}, {Becker},
  {Bonnett}, {Carnero Rosell}, {Carollo}, {Carrasco Kind}, {Castander},
  {Cawthon}, {Chang}, {Childress}, {Davis}, {Drlica-Wagner}, {Gatti},
  {Glazebrook}, {Gschwend}, {Hinton}, {Hoormann}, {Kim}, {King}, {Kuehn},
  {Lewis}, {Lidman}, {Lin}, {Macaulay}, {Maia}, {Martini}, {Mudd},
  {M{\"o}ller}, {Nichol}, {Ogando}, {Rollins}, {Roodman}, {Ross}, {Rozo},
  {Rykoff}, {Samuroff}, {Sevilla-Noarbe}, {Sharp}, {Sommer}, {Tucker}, {Uddin},
  {Varga}, {Vielzeuf}, {Yuan}, {Zhang}, {Abbott}, {Abdalla}, {Allam}, {Annis},
  {Bechtol}, {Benoit-L{\'e}vy}, {Bertin}, {Brooks}, {Buckley-Geer}, {Burke},
  {Busha}, {Capozzi}, {Carretero}, {Crocce}, {D'Andrea}, {da Costa}, {DePoy},
  {Desai}, {Diehl}, {Doel}, {Eifler}, {Estrada}, {Evrard}, {Fernandez},
  {Flaugher}, {Fosalba}, {Frieman}, {Garc{\'\i}a-Bellido}, {Gerdes},
  {Giannantonio}, {Goldstein}, {Gruendl}, {Gutierrez}, {Honscheid}, {James},
  {Jarvis}, {Jeltema}, {Johnson}, {Johnson}, {Kirk}, {Krause}, {Kuhlmann},
  {Kuropatkin}, {Lahav}, {Li}, {Lima}, {March}, {Marshall}, {Melchior},
  {Menanteau}, {Miquel}, {Nord}, {O'Neill}, {Plazas}, {Romer}, {Sako},
  {Sanchez}, {Santiago}, {Scarpine}, {Schindler}, {Schubnell}, {Smith},
  {Smith}, {Soares-Santos}, {Sobreira}, {Suchyta}, {Swanson}, {Tarle},
  {Thomas}, {Tucker}, {Vikram}, {Walker}, {Weller}, {Wester}, {Wolf}, {Yanny},
  {Zuntz}, \& {DES Collaboration}}]{DESY1_redshifts}
{Hoyle}, B. {et~al.} 2018, \mnras, 478, 592, 1708.01532

\bibitem[{{Ivezi{\'c}} {et~al.}(2019){Ivezi{\'c}}, {Kahn}, {Tyson}, {Abel},
  {Acosta}, {Allsman}, {Alonso}, {AlSayyad}, {Anderson}, {Andrew}, {Angel},
  {Angeli}, {Ansari}, {Antilogus}, {Araujo}, {Armstrong}, {Arndt}, {Astier},
  {Aubourg}, {Auza}, {Axelrod}, {Bard}, {Barr}, {Barrau}, {Bartlett}, {Bauer},
  {Bauman}, {Baumont}, {Bechtol}, {Bechtol}, {Becker}, {Becla}, {Beldica},
  {Bellavia}, {Bianco}, {Biswas}, {Blanc}, {Blazek}, {Blandford}, {Bloom},
  {Bogart}, {Bond}, {Booth}, {Borgland}, {Borne}, {Bosch}, {Boutigny},
  {Brackett}, {Bradshaw}, {Brandt}, {Brown}, {Bullock}, {Burchat}, {Burke},
  {Cagnoli}, {Calabrese}, {Callahan}, {Callen}, {Carlin}, {Carlson},
  {Chandrasekharan}, {Charles-Emerson}, {Chesley}, {Cheu}, {Chiang}, {Chiang},
  {Chirino}, {Chow}, {Ciardi}, {Claver}, {Cohen-Tanugi}, {Cockrum}, {Coles},
  {Connolly}, {Cook}, {Cooray}, {Covey}, {Cribbs}, {Cui}, {Cutri}, {Daly},
  {Daniel}, {Daruich}, {Daubard}, {Daues}, {Dawson}, {Delgado}, {Dellapenna},
  {de Peyster}, {de Val-Borro}, {Digel}, {Doherty}, {Dubois},
  {Dubois-Felsmann}, {Durech}, {Economou}, {Eifler}, {Eracleous}, {Emmons},
  {Fausti Neto}, {Ferguson}, {Figueroa}, {Fisher-Levine}, {Focke}, {Foss},
  {Frank}, {Freemon}, {Gangler}, {Gawiser}, {Geary}, {Gee}, {Geha}, {Gessner},
  {Gibson}, {Gilmore}, {Glanzman}, {Glick}, {Goldina}, {Goldstein}, {Goodenow},
  {Graham}, {Gressler}, {Gris}, {Guy}, {Guyonnet}, {Haller}, {Harris},
  {Hascall}, {Haupt}, {Hernandez}, {Herrmann}, {Hileman}, {Hoblitt}, {Hodgson},
  {Hogan}, {Howard}, {Huang}, {Huffer}, {Ingraham}, {Innes}, {Jacoby}, {Jain},
  {Jammes}, {Jee}, {Jenness}, {Jernigan}, {Jevremovi{\'c}}, {Johns}, {Johnson},
  {Johnson}, {Jones}, {Juramy-Gilles}, {Juri{\'c}}, {Kalirai}, {Kallivayalil},
  {Kalmbach}, {Kantor}, {Karst}, {Kasliwal}, {Kelly}, {Kessler}, {Kinnison},
  {Kirkby}, {Knox}, {Kotov}, {Krabbendam}, {Krughoff}, {Kub{\'a}nek},
  {Kuczewski}, {Kulkarni}, {Ku}, {Kurita}, {Lage}, {Lambert}, {Lange},
  {Langton}, {Le Guillou}, {Levine}, {Liang}, {Lim}, {Lintott}, {Long},
  {Lopez}, {Lotz}, {Lupton}, {Lust}, {MacArthur}, {Mahabal}, {Mandelbaum},
  {Markiewicz}, {Marsh}, {Marshall}, {Marshall}, {May}, {McKercher}, {McQueen},
  {Meyers}, {Migliore}, {Miller}, {Mills}, {Miraval}, {Moeyens}, {Moolekamp},
  {Monet}, {Moniez}, {Monkewitz}, {Montgomery}, {Morrison}, {Mueller},
  {Muller}, {Mu{\~n}oz Arancibia}, {Neill}, {Newbry}, {Nief}, {Nomerotski},
  {Nordby}, {O'Connor}, {Oliver}, {Olivier}, {Olsen}, {O'Mullane}, {Ortiz},
  {Osier}, {Owen}, {Pain}, {Palecek}, {Parejko}, {Parsons}, {Pease},
  {Peterson}, {Peterson}, {Petravick}, {Libby Petrick}, {Petry},
  {Pierfederici}, {Pietrowicz}, {Pike}, {Pinto}, {Plante}, {Plate}, {Plutchak},
  {Price}, {Prouza}, {Radeka}, {Rajagopal}, {Rasmussen}, {Regnault}, {Reil},
  {Reiss}, {Reuter}, {Ridgway}, {Riot}, {Ritz}, {Robinson}, {Roby}, {Roodman},
  {Rosing}, {Roucelle}, {Rumore}, {Russo}, {Saha}, {Sassolas}, {Schalk},
  {Schellart}, {Schindler}, {Schmidt}, {Schneider}, {Schneider}, {Schoening},
  {Schumacher}, {Schwamb}, {Sebag}, {Selvy}, {Sembroski}, {Seppala}, {Serio},
  {Serrano}, {Shaw}, {Shipsey}, {Sick}, {Silvestri}, {Slater}, {Smith},
  {Smith}, {Sobhani}, {Soldahl}, {Storrie-Lombardi}, {Stover}, {Strauss},
  {Street}, {Stubbs}, {Sullivan}, {Sweeney}, {Swinbank}, {Szalay}, {Takacs},
  {Tether}, {Thaler}, {Thayer}, {Thomas}, {Thornton}, {Thukral}, {Tice},
  {Trilling}, {Turri}, {Van Berg}, {Vanden Berk}, {Vetter}, {Virieux},
  {Vucina}, {Wahl}, {Walkowicz}, {Walsh}, {Walter}, {Wang}, {Wang}, {Warner},
  {Wiecha}, {Willman}, {Winters}, {Wittman}, {Wolff}, {Wood-Vasey}, {Wu},
  {Xin}, {Yoachim}, \& {Zhan}}]{LSST-Design}
{Ivezi{\'c}}, {\v{Z}}. {et~al.} 2019, \apj, 873, 111, 0805.2366

\bibitem[{{Izard} {et~al.}(2018){Izard}, {Fosalba}, \& {Crocce}}]{ICE-COLA}
{Izard}, A., {Fosalba}, P., \& {Crocce}, M. 2018, \mnras, 473, 3051

\bibitem[{{Jarvis} {et~al.}(2004){Jarvis}, {Bernstein}, \& {Jain}}]{TreeCorr}
{Jarvis}, M., {Bernstein}, G., \& {Jain}, B. 2004, \mnras, 352, 338

\bibitem[{{Jeffrey} {et~al.}(2021){Jeffrey}, {Alsing}, \&
  {Lanusse}}]{Jeffrey2021}
{Jeffrey}, N., {Alsing}, J., \& {Lanusse}, F. 2021, \mnras, 501, 954,
  2009.08459

\bibitem[{{Joachimi} {et~al.}(2021){Joachimi}, {Lin}, {Asgari}, {Tr{\"o}ster},
  {Heymans}, {Hildebrandt}, {K{\"o}hlinger}, {S{\'a}nchez}, {Wright},
  {Bilicki}, {Blake}, {van den Busch}, {Crocce}, {Dvornik}, {Erben}, {Getman},
  {Giblin}, {Hoekstra}, {Kannawadi}, {Kuijken}, {Napolitano}, {Schneider},
  {Scoccimarro}, {Sellentin}, {Shan}, {von Wietersheim-Kramsta}, \&
  {Zuntz}}]{KiDS1000_Joachimi}
{Joachimi}, B. {et~al.} 2021, \aap, 646, A129, 2007.01844

\bibitem[{{Joudaki} {et~al.}(2020){Joudaki}, {Hildebrandt}, {Traykova},
  {Chisari}, {Heymans}, {Kannawadi}, {Kuijken}, {Wright}, {Asgari}, {Erben},
  {Hoekstra}, {Joachimi}, {Miller}, {Tr{\"o}ster}, \& {van den
  Busch}}]{KiDS_DES_Joudaki}
{Joudaki}, S. {et~al.} 2020, \aap, 638, L1, 1906.09262

\bibitem[{{Kacprzak} {et~al.}(2016){Kacprzak}, {Kirk}, {Friedrich}, {Amara},
  {Refregier}, {Marian}, {Dietrich}, {Suchyta}, {Aleksi{\'c}}, {Bacon},
  {Becker}, {Bonnett}, {Bridle}, {Chang}, {Eifler}, {Hartley}, {Huff},
  {Krause}, {MacCrann}, {Melchior}, {Nicola}, {Samuroff}, {Sheldon}, {Troxel},
  {Weller}, {Zuntz}, {Abbott}, {Abdalla}, {Armstrong}, {Benoit-L{\'e}vy},
  {Bernstein}, {Bernstein}, {Bertin}, {Brooks}, {Burke}, {Carnero Rosell},
  {Carrasco Kind}, {Carretero}, {Castander}, {Crocce}, {D'Andrea}, {da Costa},
  {Desai}, {Diehl}, {Evrard}, {Neto}, {Flaugher}, {Fosalba}, {Frieman},
  {Gerdes}, {Goldstein}, {Gruen}, {Gruendl}, {Gutierrez}, {Honscheid}, {Jain},
  {James}, {Jarvis}, {Kuehn}, {Kuropatkin}, {Lahav}, {Lima}, {March},
  {Marshall}, {Martini}, {Miller}, {Miquel}, {Mohr}, {Nichol}, {Nord},
  {Plazas}, {Romer}, {Roodman}, {Rykoff}, {Sanchez}, {Scarpine}, {Schubnell},
  {Sevilla-Noarbe}, {Smith}, {Soares-Santos}, {Sobreira}, {Swanson}, {Tarle},
  {Thomas}, {Vikram}, {Walker}, {Zhang}, \& {DES Collaboration}}]{Kacprzak2016}
{Kacprzak}, T. {et~al.} 2016, \mnras, 463, 3653

\bibitem[{{Kannawadi} {et~al.}(2019){Kannawadi}, {Hoekstra}, {Miller}, {Viola},
  {Fenech Conti}, {Herbonnet}, {Erben}, {Heymans}, {Hildebrandt}, {Kuijken},
  {Vakili}, \& {Wright}}]{2019A&A...624A..92K}
{Kannawadi}, A. {et~al.} 2019, \aap, 624, A92, 1812.03983

\bibitem[{{Kilbinger}(2015)}]{2015RPPh...78h6901K}
{Kilbinger}, M. 2015, Reports on Progress in Physics, 78, 086901

\bibitem[{{Kilbinger} {et~al.}(2017){Kilbinger}, {Heymans}, {Asgari},
  {Joudaki}, {Schneider}, {Simon}, {Van Waerbeke}, {Harnois-D{\'e}raps},
  {Hildebrand t}, {K{\"o}hlinger}, {Kuijken}, \& {Viola}}]{Kilbinger17}
{Kilbinger}, M. {et~al.} 2017, \mnras, 472, 2126, 1702.05301

\bibitem[{{Kilbinger} {et~al.}(2006){Kilbinger}, {Schneider}, \&
  {Eifler}}]{2006A&A...457...15K}
{Kilbinger}, M., {Schneider}, P., \& {Eifler}, T. 2006, \aap, 457, 15,
  astro-ph/0604520

\bibitem[{{Kuijken} {et~al.}(2019){Kuijken}, {Heymans}, {Dvornik},
  {Hildebrandt}, {de Jong}, {Wright}, {Erben}, {Bilicki}, {Giblin}, {Shan},
  {Getman}, {Grado}, {Hoekstra}, {Miller}, {Napolitano}, {Paolilo}, {Radovich},
  {Schneider}, {Sutherland}, {Tewes}, {Tortora}, {Valentijn}, \& {Verdoes
  Kleijn}}]{KiDS1000_Kuijken}
{Kuijken}, K. {et~al.} 2019, \aap, 625, A2

\bibitem[{{Kuijken} {et~al.}(2015){Kuijken}, {Heymans}, {Hildebrandt},
  {Nakajima}, {Erben}, {de Jong}, {Viola}, {Choi}, {Hoekstra}, {Miller}, {van
  Uitert}, {Amon}, {Blake}, {Brouwer}, {Buddendiek}, {Conti}, {Eriksen},
  {Grado}, {Harnois-D{\'e}raps}, {Helmich}, {Herbonnet}, {Irisarri},
  {Kitching}, {Klaes}, {La Barbera}, {Napolitano}, {Radovich}, {Schneider},
  {Sif{\'o}n}, {Sikkema}, {Simon}, {Tudorica}, {Valentijn}, {Verdoes Kleijn},
  \& {van Waerbeke}}]{2015MNRAS.454.3500K}
------. 2015, \mnras, 454, 3500, 1507.00738

\bibitem[{{Laigle} {et~al.}(2016){Laigle}, {McCracken}, {Ilbert}, {Hsieh},
  {Davidzon}, {Capak}, {Hasinger}, {Silverman}, {Pichon}, {Coupon}, {Aussel},
  {Le Borgne}, {Caputi}, {Cassata}, {Chang}, {Civano}, {Dunlop}, {Fynbo},
  {Kartaltepe}, {Koekemoer}, {Le F{\`e}vre}, {Le Floc'h}, {Leauthaud}, {Lilly},
  {Lin}, {Marchesi}, {Milvang-Jensen}, {Salvato}, {Sanders}, {Scoville},
  {Smolcic}, {Stockmann}, {Taniguchi}, {Tasca}, {Toft}, {Vaccari}, \&
  {Zabl}}]{COSMOS15}
{Laigle}, C. {et~al.} 2016, \apjs, 224, 24, 1604.02350

\bibitem[{{Lange}(2023)}]{Nautilus}
{Lange}, J.~U. 2023, \mnras, 525, 3181, 2306.16923

\bibitem[{{Laureijs} {et~al.}(2011){Laureijs}, {Amiaux}, {Arduini},
  {Augu{\`e}res}, {Brinchmann}, {Cole}, {Cropper}, {Dabin}, {Duvet}, {Ealet},
  {Garilli}, {Gondoin}, {Guzzo}, {Hoar}, {Hoekstra}, {Holmes}, {Kitching},
  {Maciaszek}, {Mellier}, {Pasian}, {Percival}, {Rhodes}, {Saavedra Criado},
  {Sauvage}, {Scaramella}, {Valenziano}, {Warren}, {Bender}, {Castander},
  {Cimatti}, {Le F{\`e}vre}, {Kurki-Suonio}, {Levi}, {Lilje}, {Meylan},
  {Nichol}, {Pedersen}, {Popa}, {Rebolo Lopez}, {Rix}, {Rottgering},
  {Zeilinger}, {Grupp}, {Hudelot}, {Massey}, {Meneghetti}, {Miller}, {Paltani},
  {Paulin-Henriksson}, {Pires}, {Saxton}, {Schrabback}, {Seidel}, {Walsh},
  {Aghanim}, {Amendola}, {Bartlett}, {Baccigalupi}, {Beaulieu}, {Benabed},
  {Cuby}, {Elbaz}, {Fosalba}, {Gavazzi}, {Helmi}, {Hook}, {Irwin}, {Kneib},
  {Kunz}, {Mannucci}, {Moscardini}, {Tao}, {Teyssier}, {Weller}, {Zamorani},
  {Zapatero Osorio}, {Boulade}, {Foumond}, {Di Giorgio}, {Guttridge}, {James},
  {Kemp}, {Martignac}, {Spencer}, {Walton}, {Bl{\"u}mchen}, {Bonoli},
  {Bortoletto}, {Cerna}, {Corcione}, {Fabron}, {Jahnke}, {Ligori}, {Madrid},
  {Martin}, {Morgante}, {Pamplona}, {Prieto}, {Riva}, {Toledo}, {Trifoglio},
  {Zerbi}, {Abdalla}, {Douspis}, {Grenet}, {Borgani}, {Bouwens}, {Courbin},
  {Delouis}, {Dubath}, {Fontana}, {Frailis}, {Grazian}, {Koppenh{\"o}fer},
  {Mansutti}, {Melchior}, {Mignoli}, {Mohr}, {Neissner}, {Noddle}, {Poncet},
  {Scodeggio}, {Serrano}, {Shane}, {Starck}, {Surace}, {Taylor},
  {Verdoes-Kleijn}, {Vuerli}, {Williams}, {Zacchei}, {Altieri}, {Escudero
  Sanz}, {Kohley}, {Oosterbroek}, {Astier}, {Bacon}, {Bardelli}, {Baugh},
  {Bellagamba}, {Benoist}, {Bianchi}, {Biviano}, {Branchini}, {Carbone},
  {Cardone}, {Clements}, {Colombi}, {Conselice}, {Cresci}, {Deacon}, {Dunlop},
  {Fedeli}, {Fontanot}, {Franzetti}, {Giocoli}, {Garcia-Bellido}, {Gow},
  {Heavens}, {Hewett}, {Heymans}, {Holland}, {Huang}, {Ilbert}, {Joachimi},
  {Jennins}, {Kerins}, {Kiessling}, {Kirk}, {Kotak}, {Krause}, {Lahav}, {van
  Leeuwen}, {Lesgourgues}, {Lombardi}, {Magliocchetti}, {Maguire}, {Majerotto},
  {Maoli}, {Marulli}, {Maurogordato}, {McCracken}, {McLure}, {Melchiorri},
  {Merson}, {Moresco}, {Nonino}, {Norberg}, {Peacock}, {Pello}, {Penny},
  {Pettorino}, {Di Porto}, {Pozzetti}, {Quercellini}, {Radovich}, {Rassat},
  {Roche}, {Ronayette}, {Rossetti}, {Sartoris}, {Schneider}, {Semboloni},
  {Serjeant}, {Simpson}, {Skordis}, {Smadja}, {Smartt}, {Spano}, {Spiro},
  {Sullivan}, {Tilquin}, {Trotta}, {Verde}, {Wang}, {Williger}, {Zhao},
  {Zoubian}, \& {Zucca}}]{RedBook}
{Laureijs}, R. {et~al.} 2011, arXiv e-prints, arXiv:1110.3193, 1110.3193

\bibitem[{{Lemos} {et~al.}(2022){Lemos}, {Weaverdyck}, {Rollins}, {Muir},
  {Fert{\'e}}, {Liddle}, {Campos}, {Huterer}, {Raveri}, {Zuntz}, {Di
  Valentino}, {Fang}, {Hartley}, {Aguena}, {Allam}, {Annis}, {Bertin},
  {Bocquet}, {Brooks}, {Burke}, {Carnero Rosell}, {Carrasco Kind}, {Carretero},
  {Castander}, {Choi}, {Costanzi}, {Crocce}, {da Costa}, {Pereira}, {Dietrich},
  {Everett}, {Ferrero}, {Frieman}, {Garc{\'\i}a-Bellido}, {Gatti}, {Gaztanaga},
  {Gerdes}, {Gruen}, {Gruendl}, {Gschwend}, {Gutierrez}, {Hinton}, {Hollowood},
  {Honscheid}, {James}, {Kuehn}, {Kuropatkin}, {Lima}, {March}, {Melchior},
  {Menanteau}, {Miquel}, {Morgan}, {Palmese}, {Paz-Chinch{\'o}n}, {Pieres},
  {Plazas Malag{\'o}n}, {Porredon}, {Sanchez}, {Scarpine}, {Schubnell},
  {Serrano}, {Sevilla-Noarbe}, {Smith}, {Suchyta}, {Swanson}, {Tarle},
  {Thomas}, {To}, {Varga}, \& {Weller}}]{Lemos}
{Lemos}, P. {et~al.} 2022, arXiv e-prints, arXiv:2202.08233, 2202.08233

\bibitem[{{Lewis}(2019)}]{GetDist}
{Lewis}, A. 2019, arXiv e-prints, arXiv:1910.13970, 1910.13970

\bibitem[{{Li} {et~al.}(2023{\natexlab{a}}){Li}, {Hoekstra}, {Kuijken},
  {Asgari}, {Bilicki}, {Giblin}, {Heymans}, {Hildebrandt}, {Joachimi},
  {Miller}, {van den Busch}, {Wright}, {Kannawadi}, {Reischke}, \&
  {Shan}}]{KiDS1000_Li}
{Li}, S.-S. {et~al.} 2023{\natexlab{a}}, arXiv e-prints, arXiv:2306.11124,
  2306.11124

\bibitem[{{Li} {et~al.}(2023{\natexlab{b}}){Li}, {Zhang}, {Sugiyama}, {Dalal},
  {Rau}, {Mandelbaum}, {Takada}, {More}, {Strauss}, {Miyatake}, {Shirasaki},
  {Hamana}, {Oguri}, {Luo}, {Nishizawa}, {Takahashi}, {Nicola}, {Osato},
  {Kannawadi}, {Sunayama}, {Armstrong}, {Komiyama}, {Lupton}, {Lust},
  {Miyazaki}, {Murayama}, {Nishimichi}, {Okura}, {Price}, {Tait}, {Tanaka}, \&
  {Wang}}]{HSCY3_2PCF}
{Li}, X. {et~al.} 2023{\natexlab{b}}, arXiv e-prints, arXiv:2304.00702,
  2304.00702

\bibitem[{{Li} {et~al.}(2019){Li}, {Liu}, {Zorrilla Matilla}, \&
  {Coulton}}]{MassiveNu1}
{Li}, Z., {Liu}, J., {Zorrilla Matilla}, J.~M., \& {Coulton}, W.~R. 2019, \prd,
  99, 063527, 1810.01781

\bibitem[{{Lima} {et~al.}(2008){Lima}, {Cunha}, {Oyaizu}, {Frieman}, {Lin}, \&
  {Sheldon}}]{DIR}
{Lima}, M., {Cunha}, C.~E., {Oyaizu}, H., {Frieman}, J., {Lin}, H., \&
  {Sheldon}, E.~S. 2008, \mnras, 390, 118, 0801.3822

\bibitem[{{Lin} {et~al.}(2023){Lin}, {von wietersheim-Kramsta}, {Joachimi}, \&
  {Feeney}}]{KiDS1000_SBI}
{Lin}, K., {von wietersheim-Kramsta}, M., {Joachimi}, B., \& {Feeney}, S. 2023,
  \mnras, 524, 6167, 2212.04521

\bibitem[{{Linke} {et~al.}(2023){Linke}, {Burger}, {Heydenreich}, {Porth}, \&
  {Schneider}}]{SSC_Linke}
{Linke}, L., {Burger}, P.~A., {Heydenreich}, S., {Porth}, L., \& {Schneider},
  P. 2023, arXiv e-prints, arXiv:2302.12277, 2302.12277

\bibitem[{{Liu} {et~al.}(2023){Liu}, {Yuan}, {Pan}, {Zhang}, {Wang}, \&
  {Fan}}]{HSCY1_peaks_th}
{Liu}, X., {Yuan}, S., {Pan}, C., {Zhang}, T., {Wang}, Q., \& {Fan}, Z. 2023,
  \mnras, 519, 594, 2210.07853

\bibitem[{{Longley} {et~al.}(2023){Longley}, {Chang}, {Walter}, {Zuntz},
  {Ishak}, {Mandelbaum}, {Miyatake}, {Nicola}, {Pedersen}, {Pereira}, {Prat},
  {S{\'a}nchez}, {Secco}, {Tr{\"o}ster}, {Troxel}, {Wright}, \& {LSST Dark
  Energy Science Collaboration}}]{Longley2023}
{Longley}, E.~P. {et~al.} 2023, \mnras, 520, 5016, 2208.07179

\bibitem[{{Loureiro} {et~al.}(2022){Loureiro}, {Whittaker}, {Spurio Mancini},
  {Joachimi}, {Cuceu}, {Asgari}, {St{\"o}lzner}, {Tr{\"o}ster}, {Wright},
  {Bilicki}, {Dvornik}, {Giblin}, {Heymans}, {Hildebrandt}, {Shan}, {Amara},
  {Auricchio}, {Bodendorf}, {Bonino}, {Branchini}, {Brescia}, {Capobianco},
  {Carbone}, {Carretero}, {Castellano}, {Cavuoti}, {Cimatti}, {Cledassou},
  {Congedo}, {Conversi}, {Copin}, {Corcione}, {Cropper}, {Da Silva}, {Douspis},
  {Dubath}, {Duncan}, {Dupac}, {Dusini}, {Farrens}, {Ferriol}, {Fosalba},
  {Frailis}, {Franceschi}, {Fumana}, {Garilli}, {Gillis}, {Giocoli}, {Grazian},
  {Grupp}, {Haugan}, {Holmes}, {Hormuth}, {Jahnke}, {K{\"u}mmel}, {Kermiche},
  {Kiessling}, {Kilbinger}, {Kitching}, {Kuijken}, {Kunz}, {Kurki-Suonio},
  {Ligori}, {Lilje}, {Lloro}, {Mansutti}, {Marggraf}, {Markovic}, {Marulli},
  {Massey}, {Meneghetti}, {Meylan}, {Moresco}, {Morin}, {Moscardini}, {Munari},
  {Niemi}, {Padilla}, {Paltani}, {Pasian}, {Pedersen}, {Pettorino}, {Pires},
  {Poncet}, {Popa}, {Raison}, {Rhodes}, {Rix}, {Roncarelli}, {Saglia},
  {Schneider}, {Secroun}, {Serrano}, {Sirignano}, {Sirri}, {Stanco}, {Starck},
  {Tallada-Cresp{\'\i}}, {Taylor}, {Tereno}, {Toledo-Moreo}, {Torradeflot},
  {Valentijn}, {Wang}, {Welikala}, {Weller}, {Zamorani}, {Zoubian}, {Andreon},
  {Baldi}, {Camera}, {Farinelli}, {Polenta}, \& {Tessore}}]{KiDS1000_Loureiro}
{Loureiro}, A. {et~al.} 2022, \aap, 665, A56

\bibitem[{{MacCrann} {et~al.}(2022){MacCrann}, {Becker}, {McCullough}, {Amon},
  {Gruen}, {Jarvis}, {Choi}, {Troxel}, {Sheldon}, {Yanny}, {Herner},
  {Dodelson}, {Zuntz}, {Eckert}, {Rollins}, {Varga}, {Bernstein}, {Gruendl},
  {Harrison}, {Hartley}, {Sevilla-Noarbe}, {Pieres}, {Bridle}, {Myles},
  {Alarcon}, {Everett}, {S{\'a}nchez}, {Huff}, {Tarsitano}, {Gatti}, {Secco},
  {Abbott}, {Aguena}, {Allam}, {Annis}, {Bacon}, {Bertin}, {Brooks}, {Burke},
  {Carnero Rosell}, {Carrasco Kind}, {Carretero}, {Costanzi}, {Crocce},
  {Pereira}, {De Vicente}, {Desai}, {Diehl}, {Dietrich}, {Doel}, {Eifler},
  {Ferrero}, {Fert{\'e}}, {Flaugher}, {Fosalba}, {Frieman},
  {Garc{\'\i}a-Bellido}, {Gaztanaga}, {Gerdes}, {Giannantonio}, {Gschwend},
  {Gutierrez}, {Hinton}, {Hollowood}, {Honscheid}, {James}, {Lahav}, {Lima},
  {Maia}, {March}, {Marshall}, {Martini}, {Melchior}, {Menanteau}, {Miquel},
  {Mohr}, {Morgan}, {Muir}, {Ogando}, {Palmese}, {Paz-Chinch{\'o}n}, {Plazas},
  {Rodriguez-Monroy}, {Roodman}, {Samuroff}, {Sanchez}, {Scarpine}, {Serrano},
  {Smith}, {Soares-Santos}, {Suchyta}, {Swanson}, {Tarle}, {Thomas}, {To},
  {Wilkinson}, {Wilkinson}, \& {DES Collaboration}}]{DESY3_MacCrann}
{MacCrann}, N. {et~al.} 2022, \mnras, 509, 3371, 2012.08567

\bibitem[{{Mandelbaum}(2018)}]{Mandelbaum18}
{Mandelbaum}, R. 2018, \araa, 56, 393, 1710.03235

\bibitem[{{Marques} {et~al.}(2023){Marques}, {Liu}, {Shirasaki}, {Thiele},
  {Grand{\'o}n}, {Huffenberger}, {Cheng}, {Harnois-D{\'e}raps}, {Osato}, \&
  {Coulton}}]{HSCY1_peaks_sims}
{Marques}, G.~A. {et~al.} 2023, arXiv e-prints, arXiv:2308.10866, 2308.10866

\bibitem[{{Martinet} {et~al.}(2021){Martinet}, {Castro}, {Harnois-D{\'e}raps},
  {Jullo}, {Giocoli}, \& {Dolag}}]{Martinet21}
{Martinet}, N., {Castro}, T., {Harnois-D{\'e}raps}, J., {Jullo}, E., {Giocoli},
  C., \& {Dolag}, K. 2021, \aap, 648, A115, 2012.09614

\bibitem[{{Martinet} {et~al.}(2020){Martinet}, {Harnois-D{\'e}raps}, {Jullo},
  \& {Schneider}}]{Martinet20}
{Martinet}, N., {Harnois-D{\'e}raps}, J., {Jullo}, E., \& {Schneider}, P. 2020,
  arXiv e-prints, arXiv:2010.07376, 2010.07376

\bibitem[{{Martinet} {et~al.}(2018){Martinet}, {Schneider}, {Hildebrandt},
  {Shan}, {Asgari}, {Dietrich}, {Harnois-D{\'e}raps}, {Erben}, {Grado},
  {Heymans}, {Hoekstra}, {Klaes}, {Kuijken}, {Merten}, \&
  {Nakajima}}]{Martinet18}
{Martinet}, N. {et~al.} 2018, \mnras, 474, 712, 1709.07678

\bibitem[{{McCarthy} {et~al.}(2017){McCarthy}, {Schaye}, {Bird}, \& {Le
  Brun}}]{BAHAMAS}
{McCarthy}, I.~G., {Schaye}, J., {Bird}, S., \& {Le Brun}, A.~M.~C. 2017,
  \mnras, 465, 2936, 1603.02702

\bibitem[{{Mead} {et~al.}(2016){Mead}, {Heymans}, {Lombriser}, {Peacock},
  {Steele}, \& {Winther}}]{2016MNRAS.459.1468M}
{Mead}, A.~J., {Heymans}, C., {Lombriser}, L., {Peacock}, J.~A., {Steele},
  O.~I., \& {Winther}, H.~A. 2016, \mnras, 459, 1468

\bibitem[{{Miller} {et~al.}(2013){Miller}, {Heymans}, {Kitching}, {van
  Waerbeke}, {Erben}, {Hildebrandt}, {Hoekstra}, {Mellier}, {Rowe}, {Coupon},
  {Dietrich}, {Fu}, {Harnois-D{\'e}raps}, {Hudson}, {Kilbinger}, {Kuijken},
  {Schrabback}, {Semboloni}, {Vafaei}, \& {Velander}}]{2013MNRAS.429.2858M}
{Miller}, L. {et~al.} 2013, \mnras, 429, 2858

\bibitem[{Pedregosa {et~al.}(2011)Pedregosa, Varoquaux, Gramfort, Michel,
  Thirion, Grisel, Blondel, Prettenhofer, Weiss, Dubourg, Vanderplas, Passos,
  Cournapeau, Brucher, Perrot, \& Duchesnay}]{scikit}
Pedregosa, F. {et~al.} 2011, Journal of Machine Learning Research, 12, 2825

\bibitem[{{Percival} {et~al.}(2022){Percival}, {Friedrich}, {Sellentin}, \&
  {Heavens}}]{PercivalLike}
{Percival}, W.~J., {Friedrich}, O., {Sellentin}, E., \& {Heavens}, A. 2022,
  \mnras, 510, 3207, 2108.10402

\bibitem[{{Planck Collaboration} {et~al.}(2020){Planck Collaboration},
  {Aghanim}, {Akrami}, {Ashdown}, {Aumont}, {Baccigalupi}, {Ballardini},
  {Banday}, {Barreiro}, {Bartolo}, {Basak}, {Battye}, {Benabed}, {Bernard},
  {Bersanelli}, {Bielewicz}, {Bock}, {Bond}, {Borrill}, {Bouchet}, {Boulanger},
  {Bucher}, {Burigana}, {Butler}, {Calabrese}, {Cardoso}, {Carron},
  {Challinor}, {Chiang}, {Chluba}, {Colombo}, {Combet}, {Contreras}, {Crill},
  {Cuttaia}, {de Bernardis}, {de Zotti}, {Delabrouille}, {Delouis}, {Di
  Valentino}, {Diego}, {Dor{\'e}}, {Douspis}, {Ducout}, {Dupac}, {Dusini},
  {Efstathiou}, {Elsner}, {En{\ss}lin}, {Eriksen}, {Fantaye}, {Farhang},
  {Fergusson}, {Fernandez-Cobos}, {Finelli}, {Forastieri}, {Frailis},
  {Fraisse}, {Franceschi}, {Frolov}, {Galeotta}, {Galli}, {Ganga},
  {G{\'e}nova-Santos}, {Gerbino}, {Ghosh}, {Gonz{\'a}lez-Nuevo}, {G{\'o}rski},
  {Gratton}, {Gruppuso}, {Gudmundsson}, {Hamann}, {Handley}, {Hansen},
  {Herranz}, {Hildebrandt}, {Hivon}, {Huang}, {Jaffe}, {Jones}, {Karakci},
  {Keih{\"a}nen}, {Keskitalo}, {Kiiveri}, {Kim}, {Kisner}, {Knox},
  {Krachmalnicoff}, {Kunz}, {Kurki-Suonio}, {Lagache}, {Lamarre}, {Lasenby},
  {Lattanzi}, {Lawrence}, {Le Jeune}, {Lemos}, {Lesgourgues}, {Levrier},
  {Lewis}, {Liguori}, {Lilje}, {Lilley}, {Lindholm}, {L{\'o}pez-Caniego},
  {Lubin}, {Ma}, {Mac{\'\i}as-P{\'e}rez}, {Maggio}, {Maino}, {Mandolesi},
  {Mangilli}, {Marcos-Caballero}, {Maris}, {Martin}, {Martinelli},
  {Mart{\'\i}nez-Gonz{\'a}lez}, {Matarrese}, {Mauri}, {McEwen}, {Meinhold},
  {Melchiorri}, {Mennella}, {Migliaccio}, {Millea}, {Mitra},
  {Miville-Desch{\^e}nes}, {Molinari}, {Montier}, {Morgante}, {Moss}, {Natoli},
  {N{\o}rgaard-Nielsen}, {Pagano}, {Paoletti}, {Partridge}, {Patanchon},
  {Peiris}, {Perrotta}, {Pettorino}, {Piacentini}, {Polastri}, {Polenta},
  {Puget}, {Rachen}, {Reinecke}, {Remazeilles}, {Renzi}, {Rocha}, {Rosset},
  {Roudier}, {Rubi{\~n}o-Mart{\'\i}n}, {Ruiz-Granados}, {Salvati}, {Sandri},
  {Savelainen}, {Scott}, {Shellard}, {Sirignano}, {Sirri}, {Spencer},
  {Sunyaev}, {Suur-Uski}, {Tauber}, {Tavagnacco}, {Tenti}, {Toffolatti},
  {Tomasi}, {Trombetti}, {Valenziano}, {Valiviita}, {Van Tent}, {Vibert},
  {Vielva}, {Villa}, {Vittorio}, {Wandelt}, {Wehus}, {White}, {White},
  {Zacchei}, \& {Zonca}}]{PlanckLegacyCosmo}
{Planck Collaboration} {et~al.} 2020, \aap, 641, A6, 1807.06209

\bibitem[{{Pyne} \& {Joachimi}(2021)}]{PyneJoachimi}
{Pyne}, S., \& {Joachimi}, B. 2021, \mnras, 503, 2300, 2010.00614

\bibitem[{{Schirmer} {et~al.}(2007){Schirmer}, {Erben}, {Hetterscheidt}, \&
  {Schneider}}]{Schirmer2007}
{Schirmer}, M., {Erben}, T., {Hetterscheidt}, M., \& {Schneider}, P. 2007,
  \aap, 462, 875, astro-ph/0607022

\bibitem[{{Schneider} {et~al.}(2019){Schneider}, {Teyssier}, {Stadel},
  {Chisari}, {Le Brun}, {Amara}, \& {Refregier}}]{Baryonification2}
{Schneider}, A., {Teyssier}, R., {Stadel}, J., {Chisari}, N.~E., {Le Brun}, A.
  M.~C., {Amara}, A., \& {Refregier}, A. 2019, \jcap, 2019, 020, 1810.08629

\bibitem[{{Schneider}(1996)}]{Schneider1996}
{Schneider}, P. 1996, \mnras, 283, 837, astro-ph/9601039

\bibitem[{{Secco} {et~al.}(2022{\natexlab{a}}){Secco}, {Jarvis}, {Jain},
  {Chang}, {Gatti}, {Frieman}, {Adhikari}, {Alarcon}, {Amon}, {Bechtol},
  {Becker}, {Bernstein}, {Blazek}, {Campos}, {Carnero Rosell}, {Carrasco Kind},
  {Choi}, {Cordero}, {DeRose}, {Dodelson}, {Doux}, {Drlica-Wagner}, {Everett},
  {Giannini}, {Gruen}, {Gruendl}, {Harrison}, {Hartley}, {Herner}, {Krause},
  {MacCrann}, {McCullough}, {Myles}, {Navarro-Alsina}, {Prat}, {Rollins},
  {Samuroff}, {S{\'a}nchez}, {Sevilla-Noarbe}, {Sheldon}, {Troxel}, {Zeurcher},
  {Aguena}, {Andrade-Oliveira}, {Annis}, {Bacon}, {Bertin}, {Bocquet},
  {Brooks}, {Burke}, {Carretero}, {Castander}, {Crocce}, {da Costa}, {Pereira},
  {De Vicente}, {Diehl}, {Doel}, {Eckert}, {Ferrero}, {Flaugher}, {Friedel},
  {Garc{\'\i}a-Bellido}, {Gutierrez}, {Hinton}, {Hollowood}, {Honscheid},
  {Huterer}, {Kuehn}, {Kuropatkin}, {Maia}, {Marshall}, {Menanteau}, {Miquel},
  {Mohr}, {Morgan}, {Muir}, {Paz-Chinch{\'o}n}, {Pieres}, {Plazas Malag{\'o}n},
  {Rodriguez-Monroy}, {Roodman}, {Sanchez}, {Serrano}, {Suchyta}, {Swanson},
  {Tarle}, {Thomas}, {To}, {Weller}, \& {DES Collaboration}}]{DESY3_3pt}
{Secco}, L.~F. {et~al.} 2022{\natexlab{a}}, \prd, 105, 103537, 2201.05227

\bibitem[{{Secco} {et~al.}(2022{\natexlab{b}}){Secco}, {Samuroff}, {Krause},
  {Jain}, {Blazek}, {Raveri}, {Campos}, {Amon}, {Chen}, {Doux}, {Choi},
  {Gruen}, {Bernstein}, {Chang}, {DeRose}, {Myles}, {Fert{\'e}}, {Lemos},
  {Huterer}, {Prat}, {Troxel}, {MacCrann}, {Liddle}, {Kacprzak}, {Fang},
  {S{\'a}nchez}, {Pandey}, {Dodelson}, {Chintalapati}, {Hoffmann}, {Alarcon},
  {Alves}, {Andrade-Oliveira}, {Baxter}, {Bechtol}, {Becker}, {Brandao-Souza},
  {Camacho}, {Carnero Rosell}, {Carrasco Kind}, {Cawthon}, {Cordero}, {Crocce},
  {Davis}, {Di Valentino}, {Drlica-Wagner}, {Eckert}, {Eifler}, {Elidaiana},
  {Elsner}, {Elvin-Poole}, {Everett}, {Fosalba}, {Friedrich}, {Gatti},
  {Giannini}, {Gruendl}, {Harrison}, {Hartley}, {Herner}, {Huang}, {Huff},
  {Jarvis}, {Jeffrey}, {Kuropatkin}, {Leget}, {Muir}, {Mccullough}, {Navarro
  Alsina}, {Omori}, {Park}, {Porredon}, {Rollins}, {Roodman}, {Rosenfeld},
  {Ross}, {Rykoff}, {Sanchez}, {Sevilla-Noarbe}, {Sheldon}, {Shin}, {Troja},
  {Tutusaus}, {Varga}, {Weaverdyck}, {Wechsler}, {Yanny}, {Yin}, {Zhang},
  {Zuntz}, {Abbott}, {Aguena}, {Allam}, {Annis}, {Bacon}, {Bertin}, {Bhargava},
  {Bridle}, {Brooks}, {Buckley-Geer}, {Burke}, {Carretero}, {Costanzi}, {da
  Costa}, {De Vicente}, {Diehl}, {Dietrich}, {Doel}, {Ferrero}, {Flaugher},
  {Frieman}, {Garc{\'\i}a-Bellido}, {Gaztanaga}, {Gerdes}, {Giannantonio},
  {Gschwend}, {Gutierrez}, {Hinton}, {Hollowood}, {Honscheid}, {Hoyle},
  {James}, {Jeltema}, {Kuehn}, {Lahav}, {Lima}, {Lin}, {Maia}, {Marshall},
  {Martini}, {Melchior}, {Menanteau}, {Miquel}, {Mohr}, {Morgan}, {Ogando},
  {Palmese}, {Paz-Chinch{\'o}n}, {Petravick}, {Pieres}, {Plazas Malag{\'o}n},
  {Rodriguez-Monroy}, {Romer}, {Sanchez}, {Scarpine}, {Schubnell}, {Scolnic},
  {Serrano}, {Smith}, {Soares-Santos}, {Suchyta}, {Swanson}, {Tarle}, {Thomas},
  {To}, \& {DES Collaboration}}]{DESY3_Secco}
------. 2022{\natexlab{b}}, \prd, 105, 023515, 2105.13544

\bibitem[{{Sellentin} \& {Heavens}(2016)}]{SellentinHeavens}
{Sellentin}, E., \& {Heavens}, A.~F. 2016, \mnras, 456, L132

\bibitem[{{Shan} {et~al.}(2018){Shan}, {Liu}, {Hildebrandt}, {Pan}, {Martinet},
  {Fan}, {Schneider}, {Asgari}, {Harnois-D{\'e}raps}, {Hoekstra}, {Wright},
  {Dietrich}, {Erben}, {Getman}, {Grado}, {Heymans}, {Klaes}, {Kuijken},
  {Merten}, {Puddu}, {Radovich}, \& {Wang}}]{Shan18}
{Shan}, H. {et~al.} 2018, \mnras, 474, 1116, 1709.07651

\bibitem[{{Sheldon} \& {Huff}(2017)}]{Metacal}
{Sheldon}, E.~S., \& {Huff}, E.~M. 2017, \apj, 841, 24

\bibitem[{{Takahashi} {et~al.}(2017){Takahashi}, {Hamana}, {Shirasaki},
  {Namikawa}, {Nishimichi}, {Osato}, \& {Shiroyama}}]{HSCmocks}
{Takahashi}, R., {Hamana}, T., {Shirasaki}, M., {Namikawa}, T., {Nishimichi},
  T., {Osato}, K., \& {Shiroyama}, K. 2017, \apj, 850, 24

\bibitem[{{Takahashi} {et~al.}(2012){Takahashi}, {Sato}, {Nishimichi},
  {Taruya}, \& {Oguri}}]{Takahashi2012}
{Takahashi}, R., {Sato}, M., {Nishimichi}, T., {Taruya}, A., \& {Oguri}, M.
  2012, \apj, 761, 152

\bibitem[{{Tr{\"o}ster} {et~al.}(2021){Tr{\"o}ster}, {Asgari}, {Blake},
  {Cataneo}, {Heymans}, {Hildebrandt}, {Joachimi}, {Lin}, {S{\'a}nchez},
  {Wright}, {Bilicki}, {Bose}, {Crocce}, {Dvornik}, {Erben}, {Giblin},
  {Glazebrook}, {Hoekstra}, {Joudaki}, {Kannawadi}, {K{\"o}hlinger}, {Kuijken},
  {Lidman}, {Lombriser}, {Mead}, {Parkinson}, {Shan}, {Wolf}, \&
  {Xia}}]{KiDS1000_Troester}
{Tr{\"o}ster}, T. {et~al.} 2021, \aap, 649, A88

\bibitem[{{Troxel} {et~al.}(2018){Troxel}, {MacCrann}, {Zuntz}, {Eifler},
  {Krause}, {Dodelson}, {Gruen}, {Blazek}, {Friedrich}, {Samuroff}, {Prat},
  {Secco}, {Davis}, {Fert{\'e}}, {DeRose}, {Alarcon}, {Amara}, {Baxter},
  {Becker}, {Bernstein}, {Bridle}, {Cawthon}, {Chang}, {Choi}, {De Vicente},
  {Drlica-Wagner}, {Elvin-Poole}, {Frieman}, {Gatti}, {Hartley}, {Honscheid},
  {Hoyle}, {Huff}, {Huterer}, {Jain}, {Jarvis}, {Kacprzak}, {Kirk}, {Kokron},
  {Krawiec}, {Lahav}, {Liddle}, {Peacock}, {Rau}, {Refregier}, {Rollins},
  {Rozo}, {Rykoff}, {S{\'a}nchez}, {Sevilla-Noarbe}, {Sheldon}, {Stebbins},
  {Varga}, {Vielzeuf}, {Wang}, {Wechsler}, {Yanny}, {Abbott}, {Abdalla},
  {Allam}, {Annis}, {Bechtol}, {Benoit-L{\'e}vy}, {Bertin}, {Brooks},
  {Buckley-Geer}, {Burke}, {Carnero Rosell}, {Carrasco Kind}, {Carretero},
  {Castander}, {Crocce}, {Cunha}, {D'Andrea}, {da Costa}, {DePoy}, {Desai},
  {Diehl}, {Dietrich}, {Doel}, {Fernandez}, {Flaugher}, {Fosalba},
  {Garc{\'\i}a-Bellido}, {Gaztanaga}, {Gerdes}, {Giannantonio}, {Goldstein},
  {Gruendl}, {Gschwend}, {Gutierrez}, {James}, {Jeltema}, {Johnson}, {Johnson},
  {Kent}, {Kuehn}, {Kuhlmann}, {Kuropatkin}, {Li}, {Lima}, {Lin}, {Maia},
  {March}, {Marshall}, {Martini}, {Melchior}, {Menanteau}, {Miquel}, {Mohr},
  {Neilsen}, {Nichol}, {Nord}, {Petravick}, {Plazas}, {Romer}, {Roodman},
  {Sako}, {Sanchez}, {Scarpine}, {Schindler}, {Schubnell}, {Smith}, {Smith},
  {Soares-Santos}, {Sobreira}, {Suchyta}, {Swanson}, {Tarle}, {Thomas},
  {Tucker}, {Vikram}, {Walker}, {Weller}, {Zhang}, \& {DES
  Collaboration}}]{DESY1_Troxel}
{Troxel}, M.~A. {et~al.} 2018, \prd, 98, 043528, 1708.01538

\bibitem[{{Vakili} {et~al.}(2020){Vakili}, {Hoekstra}, {Bilicki}, {Fortuna},
  {Kuijken}, {Wright}, {Asgari}, {Brown}, {Dombrovskij}, {Erben}, {Giblin},
  {Heymans}, {Hildebrandt}, {Johnston}, {Joudaki}, \&
  {Kannawadi}}]{KiDS1000_LRGs}
{Vakili}, M. {et~al.} 2020, arXiv e-prints, arXiv:2008.13154, 2008.13154

\bibitem[{{van den Busch} {et~al.}(2020){van den Busch}, {Hildebrandt},
  {Wright}, {Morrison}, {Blake}, {Joachimi}, {Erben}, {Heymans}, {Kuijken}, \&
  {Taylor}}]{KiDS1000_MICE}
{van den Busch}, J.~L. {et~al.} 2020, \aap, 642, A200, 2007.01846

\bibitem[{{van den Busch} {et~al.}(2022){van den Busch}, {Wright},
  {Hildebrandt}, {Bilicki}, {Asgari}, {Joudaki}, {Blake}, {Heymans},
  {Kannawadi}, {Shan}, \& {Tr{\"o}ster}}]{KiDS1000_vdB}
------. 2022, \aap, 664, A170, 2204.02396

\bibitem[{{van Waerbeke} {et~al.}(2013)}]{VanWaerbeke2013}
{van Waerbeke}, L., {et~al.} 2013, \mnras, 433, 3373

\bibitem[{{Wright} {et~al.}(2020){Wright}, {Hildebrandt}, {van den Busch},
  {Heymans}, {Joachimi}, {Kannawadi}, \& {Kuijken}}]{Wright_KV450_SOM}
{Wright}, A.~H., {Hildebrandt}, H., {van den Busch}, J.~L., {Heymans}, C.,
  {Joachimi}, B., {Kannawadi}, A., \& {Kuijken}, K. 2020, \aap, 640, L14,
  2005.04207

\bibitem[{{Xavier} {et~al.}(2016){Xavier}, {Abdalla}, \& {Joachimi}}]{FLASK}
{Xavier}, H.~S., {Abdalla}, F.~B., \& {Joachimi}, B. 2016, {FLASK: Full-sky
  Lognormal Astro-fields Simulation Kit}, Astrophysics Source Code Library,
  record ascl:1606.015, 1606.015

\bibitem[{{Zorrilla Matilla} {et~al.}(2020){Zorrilla Matilla}, {Waterval}, \&
  {Haiman}}]{LensingHyperparameters}
{Zorrilla Matilla}, J.~M., {Waterval}, S., \& {Haiman}, Z. 2020, \aj, 159, 284,
  1909.12345

\bibitem[{{Zuntz} {et~al.}(2015){Zuntz}, {Paterno}, {Jennings}, {Rudd},
  {Manzotti}, {Dodelson}, {Bridle}, {Sehrish}, \& {Kowalkowski}}]{cosmoSIS}
{Zuntz}, J. {et~al.} 2015, Astronomy and Computing, 12, 45

\bibitem[{{Z{\"u}rcher} {et~al.}(2022){Z{\"u}rcher}, {Fluri}, {Sgier},
  {Kacprzak}, {Gatti}, {Doux}, {Whiteway}, {R{\'e}fr{\'e}gier}, {Chang},
  {Jeffrey}, {Jain}, {Lemos}, {Bacon}, {Alarcon}, {Amon}, {Bechtol}, {Becker},
  {Bernstein}, {Campos}, {Chen}, {Choi}, {Davis}, {Derose}, {Dodelson},
  {Elsner}, {Elvin-Poole}, {Everett}, {Ferte}, {Gruen}, {Harrison}, {Huterer},
  {Jarvis}, {Leget}, {Maccrann}, {Mccullough}, {Muir}, {Myles}, {Navarro
  Alsina}, {Pandey}, {Prat}, {Raveri}, {Rollins}, {Roodman}, {Sanchez},
  {Secco}, {Sheldon}, {Shin}, {Troxel}, {Tutusaus}, {Yin}, {Aguena}, {Allam},
  {Andrade-Oliveira}, {Annis}, {Bertin}, {Brooks}, {Burke}, {Carnero Rosell},
  {Carrasco Kind}, {Carretero}, {Castander}, {Cawthon}, {Conselice},
  {Costanzi}, {da Costa}, {da Silva Pereira}, {Davis}, {De Vicente}, {Desai},
  {Diehl}, {Dietrich}, {Doel}, {Eckert}, {Evrard}, {Ferrero}, {Flaugher},
  {Fosalba}, {Friedel}, {Frieman}, {Garcia-Bellido}, {Gaztanaga}, {Gerdes},
  {Giannantonio}, {Gruendl}, {Gschwend}, {Gutierrez}, {Hinton}, {Hollowood},
  {Honscheid}, {Hoyle}, {James}, {Kuehn}, {Kuropatkin}, {Lahav}, {Lidman},
  {Lima}, {Maia}, {Marshall}, {Melchior}, {Menanteau}, {Miquel}, {Morgan},
  {Palmese}, {Paz-Chinchon}, {Pieres}, {Plazas Malag{\'o}n}, {Reil}, {Rodriguez
  Monroy}, {Romer}, {Sanchez}, {Scarpine}, {Schubnell}, {Serrano}, {Sevilla},
  {Smith}, {Suchyta}, {Tarle}, {Thomas}, {To}, {Varga}, {Weller}, {Wilkinson},
  \& {DES Collaboration}}]{DESY3_Zuercher}
{Z{\"u}rcher}, D. {et~al.} 2022, \mnras, 511, 2075

\bibitem[{{Z{\"u}rcher} {et~al.}(2020){Z{\"u}rcher}, {Fluri}, {Sgier},
  {Kacprzak}, \& {Refregier}}]{Zuercher2020a}
{Z{\"u}rcher}, D., {Fluri}, J., {Sgier}, R., {Kacprzak}, T., \& {Refregier}, A.
  2020, arXiv e-prints, arXiv:2006.12506, 2006.12506

\end{thebibliography}




\appendix

\section{$B$-modes}
\label{sec:Bmodes}

\begin{figure}
\begin{center}
\includegraphics[width=3.2in]{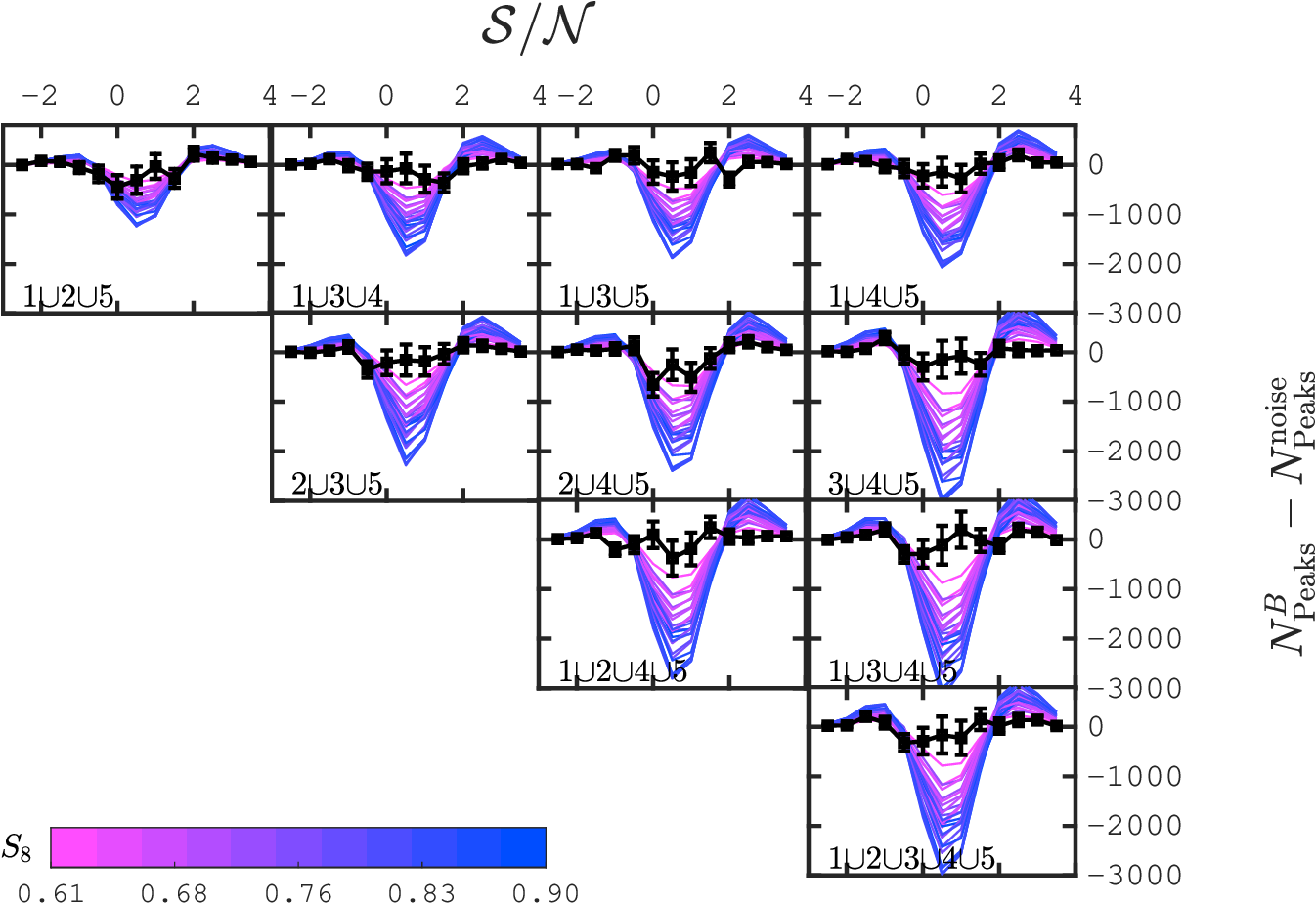}
\caption{Noise-subtracted peak count statistics measured from $B$-modes data (black squares) for a representative subset of the 30 tomographic bins, compared with the $E$-modes cosmological predictions. }
\label{fig:SNR_Bmodes}
\end{center}
\end{figure}

To leading order, $B$-modes are not produced by gravitational lensing, hence their detection in cosmic shear data is generally regarded as an indication of residual systematics. As mentioned in Sec. \ref{subsec:null}, the aperture mass map statistics constructed on a grid inevitably induces $B$-modes from the missing sub-pixel contributions, resulting in a non-zero $M_{\times}$ signal. This section presents a careful investigation of the amplitude, origins and consequences of these induced $B$-modes. In particular,  and we find that  finite sampling of the shear field itself is also a source of $B$-modes in aperture mass maps, on top of pixelisation.

We first quantify here the strength of these effects by measuring the peak function from $M_{\times}(\boldsymbol{\theta})$ in our data, i.e. aperture mass maps in which the galaxies  are rotated by 45 deg. 
The (noise-subtracted)  signal $N_{\rm peaks}^{B}$  is shown in Fig. \ref{fig:SNR_Bmodes} for a representative subset of the tomographic bins. We observe that the residual signal is much flatter than what we would expect from a cosmological signal consistent with pure noise with a $p$-value of $p=0.12$,  above the threshold of $p=0.01$ \citep[the same threshold is used in the main text, and in the DES-Y3 results for this type of hypothesis testing, see Appendices G and D of][respectively]{DESY3_Zuercher, DESY3_3x2}. 
This agrees with A21, namely that there is no evidence of residual $B$-modes in the KiDS-1000 data. It is therefore safe to keep all data entries in our inference, but investigate further the source of the $M_{\times}(\boldsymbol{\theta})$ signal seen by eye in Fig. \ref{fig:SNR_Bmodes}  to confirm it is not problematic.

We carried out peak count measurements of $M_{\times}(\boldsymbol{\theta})$ on 20 full survey realisations from the {\it Covariance Training Set} (again, these are pure $E$-mode mocks rotated by 45 degrees for this exercise), expecting to find large $p$-values in all of these trials. Instead, this test revealed that $p$-values range from $10^{-10}$ to 0.1. Some of these trials seem to rule out completely the null hypothesis (that the $B$-modes are consistent with pure noise), even though no $B$-mode exists at the catalogue level. The observed $M_{\times}$ signal must therefore come from the aperture map method itself, and is consequently a poor test for residual observational systematics.

Interestingly, 
the measured $N_{\rm peaks}^{B}$  averaged over 20 noise realisations has a $p$-value of 1.0, namely $\langle N_{\rm peaks}^{B} \rangle$ = $N_{\rm peaks}^{\rm noise}$, suggesting that these $B$-modes contain mostly noise, even though on a case-by-case some realisations see strong deviations. We hypothesise that this stems from $E$-modes leaking into $B$-modes due to an incomplete knowledge of the shear field: assuming a noiseless, pure $E$-mode shear field, the average cross-shear $\gamma_\times$ on a circle around every point in the field vanishes by definition, and thus $M_\times(\boldsymbol{\theta})\equiv 0$ holds everywhere. However, that is no longer guaranteed once the shear field is only known at a discrete set of positions, as the average $\gamma_\times$ on a circle no longer necessarily vanishes.
We investigate this by varying the number of source galaxies in our simulations. We achieve this by running our measurements on Stage-IV mocks created with a number density of 30.0 \mbox{arcmin}, introduced in \citet{Homology}, without any tomographic split. We measure on these the  $M_{\times}$ signal from maps sampled a) at every pixel location, b) at all galaxy positions, and c) at galaxy positions downsampled to match the KiDS-1000 number density. In the first case, we find that the $B$-mode field $M_\times(\boldsymbol{\theta})$ vanishes completely (up to numerical precision). The second case induces $B$-modes of approximately 0.5\% of the $E$-mode signal, whereas the third case (KiDS-1000-like number density) yields $B$-modes of approximately 4\% of the $E$-mode signal. We note that this is likely to be exacerbated by the splitting the galaxies into different tomographic bins, which further decreases the number density per aperture. We further note that these tests were performed in the absence of shape noise to better isolate this effect. 

More importantly, since these non-zero $N_{\rm peaks}^{B}$ are caused by finite sampling of the shear field, and that this sampling is exactly the same for the data and the {\it Cosmology Training Set}, the same amount of leakage should occur on average. In particular this should be fully converged in simulation-based model once averaged over the 50 mock survey $\times$ 10 noise realisations per cosmology (20 was shown to be enough in the discussion above). Therefore our inference must be immune to these by construction.

\section{Validation of the cosmology inference pipeline}
\label{sec:syst_pipeline}


\begin{figure}
\begin{center}
\includegraphics[width=3.5in]{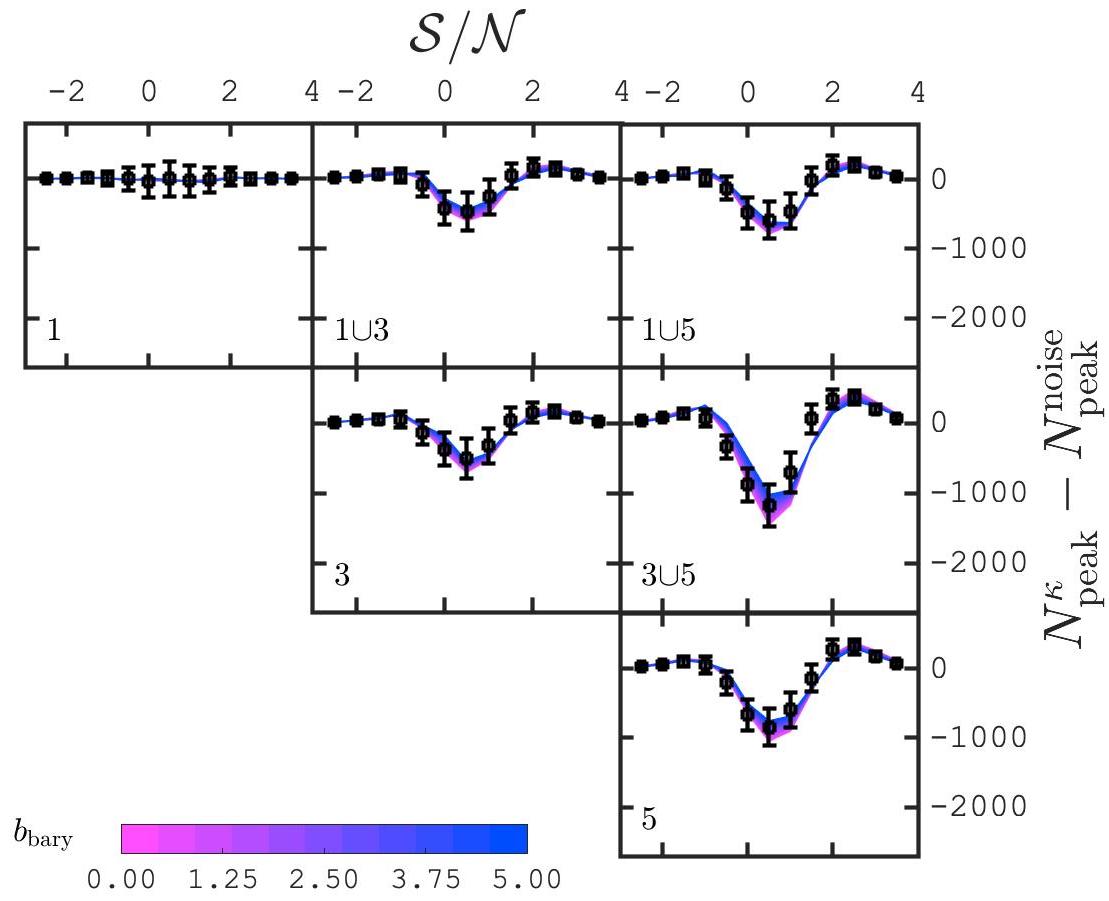}
\caption{Tomographic weak lensing peak function in the {\it Baryons Training Set}.  The coloured lines are obtained by scaling the GPR predictions (at the {\it Magneticum} cosmology) by the $b_{\rm bary}$ parameter, over the full prior range, demonstrating that peak statistics are fairly insensitive to changes in baryon feedback.}
\label{fig:N_peaks_Magneticum}
\end{center}
\end{figure}

\begin{figure}
\begin{center}
\includegraphics[width=3.0in]{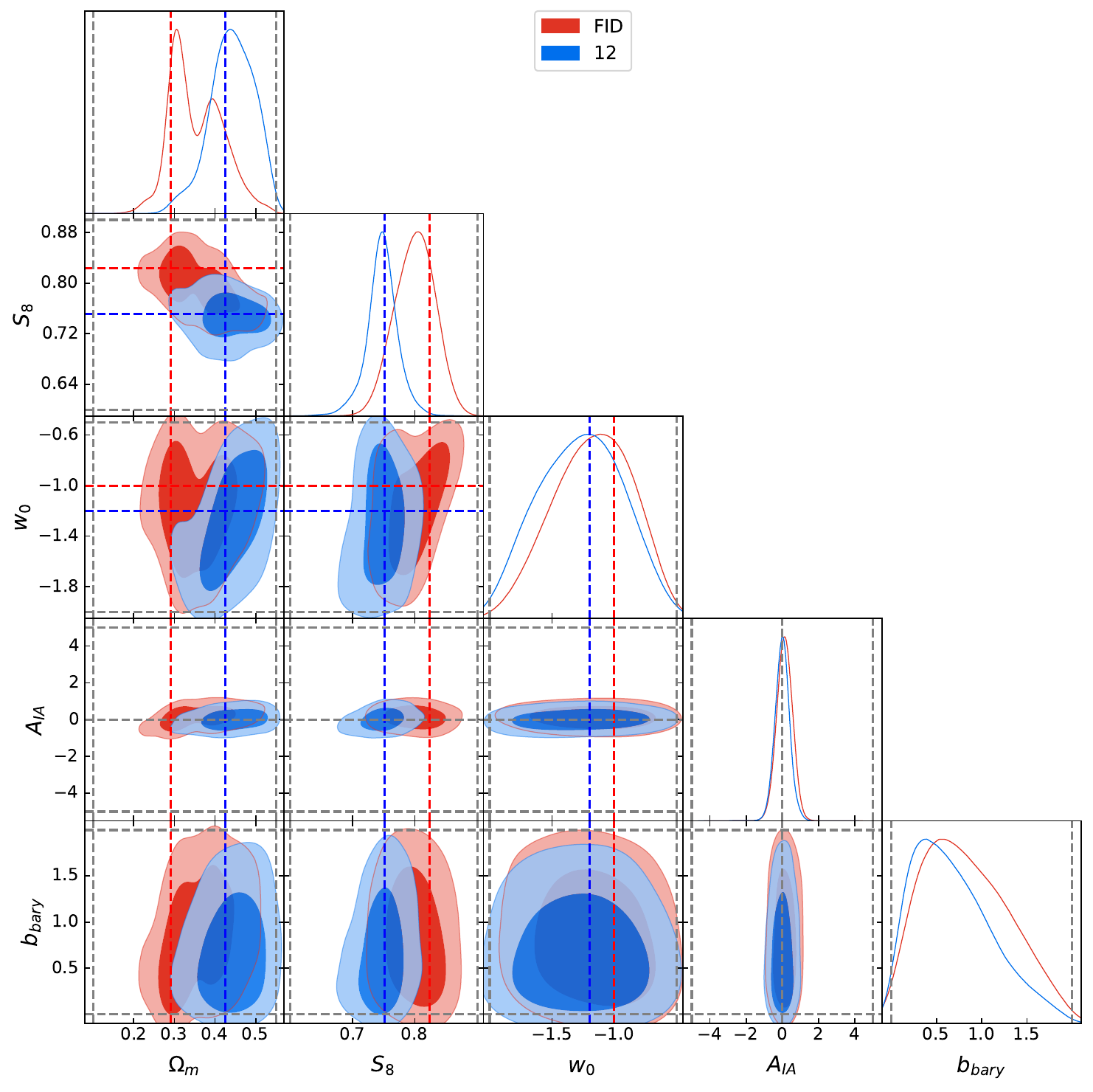}
\caption{Cosmological constraints inferred from mock data vectors extracted from two of our {\it Cosmology Training Set} models.}
\label{fig:cosmoSLICS}
\end{center}
\end{figure}

\begin{figure}
\begin{center}
\includegraphics[width=3.0in]{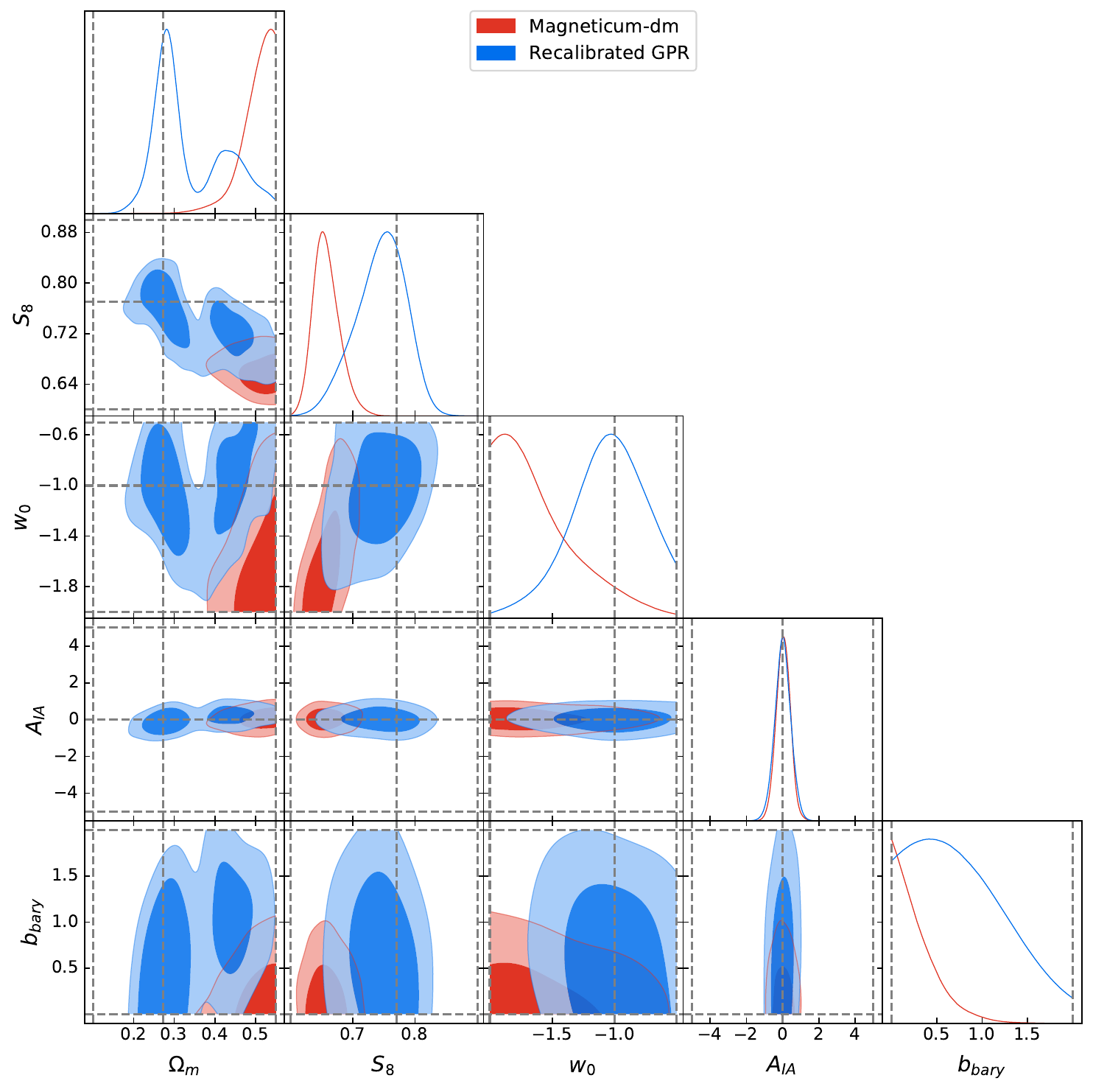}
\caption{ Cosmological constraints inferred from mock data vectors extracted from the {\it Magneticum} dark matter-only model, with (blue) and without (red) recalibrating the emulator (see main text). Similar results are obtained with the T17 simulations.}
\label{fig:MagRecal}
\end{center}
\end{figure}

\begin{figure}
\begin{center}
\includegraphics[width=3.0in]{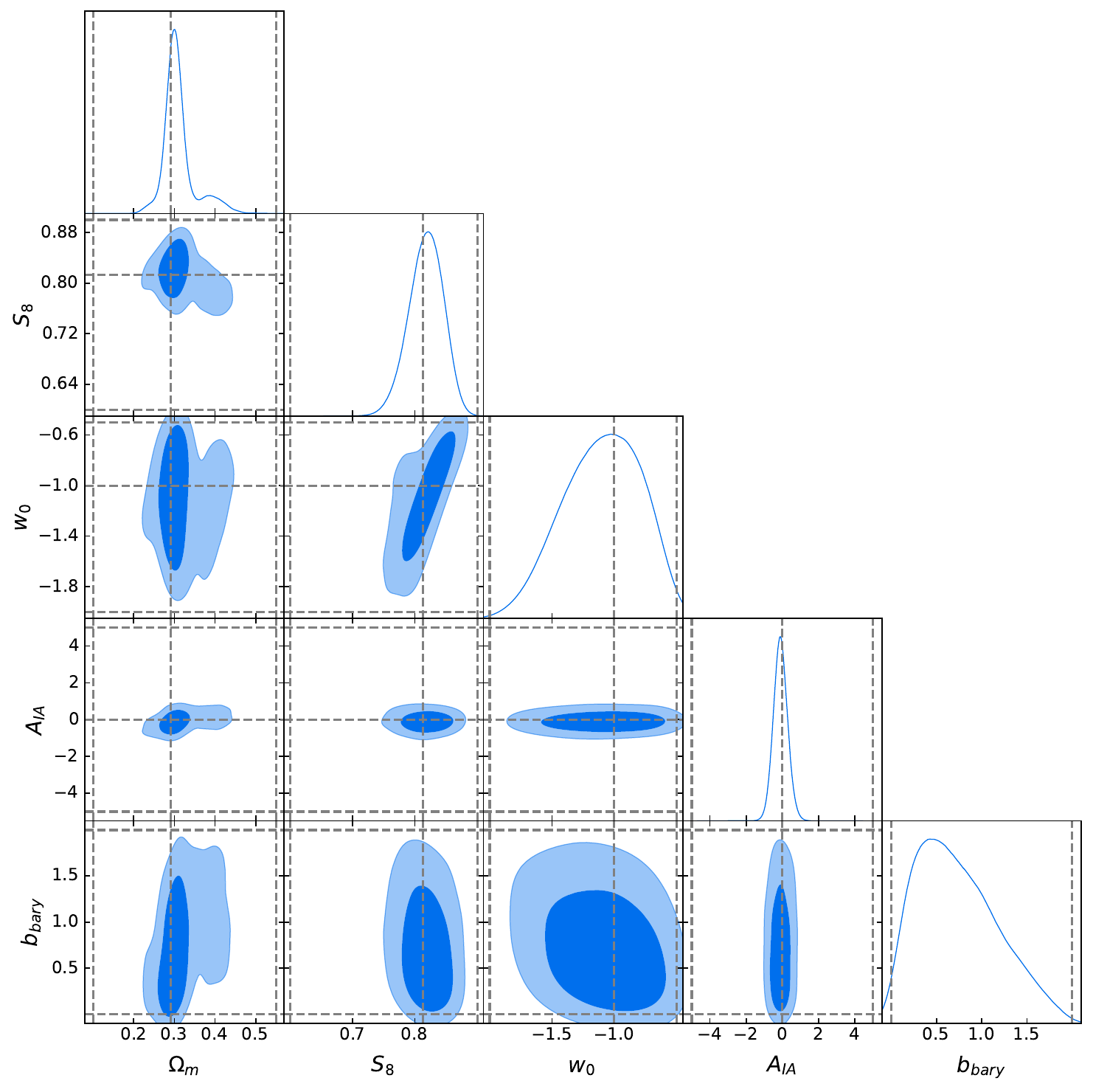}
\caption{Joint-survey mock analysis of the {\it Validation Training Set}. 
}
\label{fig:Joint_inference_SLICS}
\end{center}
\end{figure}

In this section we present a series of validation tests we performed on our cosmology inference pipeline.
First, we verified that the derivatives $\partial N_{\rm peaks} / \partial \Delta m_a$, $\partial N_{\rm peaks} / \partial \Delta z_a$, $\partial N_{\rm peaks} / \partial b_{\rm bary}$ and $\partial N_{\rm peaks} / \partial A_{\rm IA}$ are consistent with the results found in HD21 and \citet{Tidalator}. The element-by-element values are different since these are survey-specific, but they agree qualitatively. 
We next verified that the impact of increased $N$-body force is of no consequence, consistent with HD21. This is achieved by carrying the inference with the {\it Validation Set} (high-resolution) instead of the mean of the {\it Covariance Training Set}, as done in HD21.

We also verified that the peak function measured from the {\it Baryons Training Set} is consistent with the {\it Cosmology Training Set} in Fig. \ref{fig:N_peaks_Magneticum}. This is an important test, as the baryon mocks based on a completely independent $N$-body code.  This figures also shows the impact of varying $b_{\rm bary}$ on the data vector. Stronger feedback models (purple) tend to have fewer large  peaks ($\mathcal{S}/\mathcal{N} > 2$), and more   in the range $-1 < \mathcal{S}/\mathcal{N} < 1.5$. This is best seen in panel 3$\cup$5 but a common feature to most panels.  A few points lie outside the GPR predictions, which suggest that differences between $N$-body solvers / ray-tracing codes have a non-negligible impact on the data vector. We investigate this below, but first we  examine in  Fig. \ref{fig:cosmoSLICS} the accuracy of our KiDS-1000 cosmology inference pipeline by showing the recovery of input parameters for two different cosmologies selected from the {\it Cosmology Training Set}: the fiducial $\Lambda$CDM model as well as $w$CDM model 12.  We find again an excellent agreement, except that the double peak solution in $\Omega_{\rm m}$ when analysing the former model. This was first seen with the {\it Validation Training Set} in Fig. \ref{fig:cosmo_syst_infusion}, but is absent from model 12; it therefore seems to be a cosmology-dependent feature, most likely associated with limits in our GPR emulation.

 No parts of the data vector can easily explain the double peak solution in $\Omega_{\rm m}$. We have examined the posterior distributions resulting from our likelihood sampling  and identified three tomographic bins (bins 1, 1$\cup$2 and 1$\cup$2$\cup$3) where the data points were scattering outside the posteriors. Removing these from the analysis made minor differences, slightly broadening the contours;  we therefore do not deem justified to remove them from the main analysis.

 When inferring the cosmology from the {\it Magneticum} directly or the T17 simulations,  the results on all cosmological parameters are severely biased, as seen by the red contours  in Fig. \ref{fig:MagRecal}. As discussed earlier, this is likely caused by differences in the resolution of the $N$-body calculations and/or the ray-tracing algorithm being used in the creation of these mocks. In particular, the {\it Magneticum} and T17 mocks both have lower mass resolution than the main {\it cosmo}-SLICS training set, which inevitably affects the accuracy of their measured peak statistics at scales as non-linear as those targeted by this analysis. One way to avoid these biases is to calibrate our emulator and explicitly enforce the desired data vector ({\it Magneticum or T17}) at some point in parameter space. This can be achieved for instance by multiplying our theory by a calibration factor computed from the target data vector and the GPR prediction at the target cosmology. This is shown as the blue contours in Fig. \ref{fig:MagRecal}, where the input cosmology is now correctly recovered, but the double $\Omega_{\rm m}$ solution persists, even if varying only that single parameter in the MCMC.  We note that the  {\it Magneticum} and T17 mocks require distinct calibrations, and that swapping them yields results almost as biased as the original case, due to  their differences in small scales resolution. Because of this non-universality we do not calibrate the prediction in our main analysis, but optionally include the spread in these correction factors in the covariance matrix, accounting for added uncertainty about small scales physics.

 The important conclusions drawn from these tests are  that 1) small-scales structures that are not fully resolved or converged in $N$-body simulations greatly affect the non-Gaussian statistics we are investigating here, hence future analyses with increased precision will need to pay particular attention to such considerations, and  2) the sparsity of our cosmological  training nodes impacts the GPR emulator mostly on the $\Omega_{\rm m}$ dimension, while all other cosmological parameters are well recovered. This means that the current analysis is robust in its measurements of $S_8$, $w_0$, $A_{\rm IA}$ and $b_{\rm bary}$, however our constraints on the matter density are unstable and hence we do not report on them. Since we use the same training nodes for the KiDS and DES analyses, this applies also to the joint-survey constraints.

\begin{figure*}
\begin{center}
\includegraphics[width=1.81in]{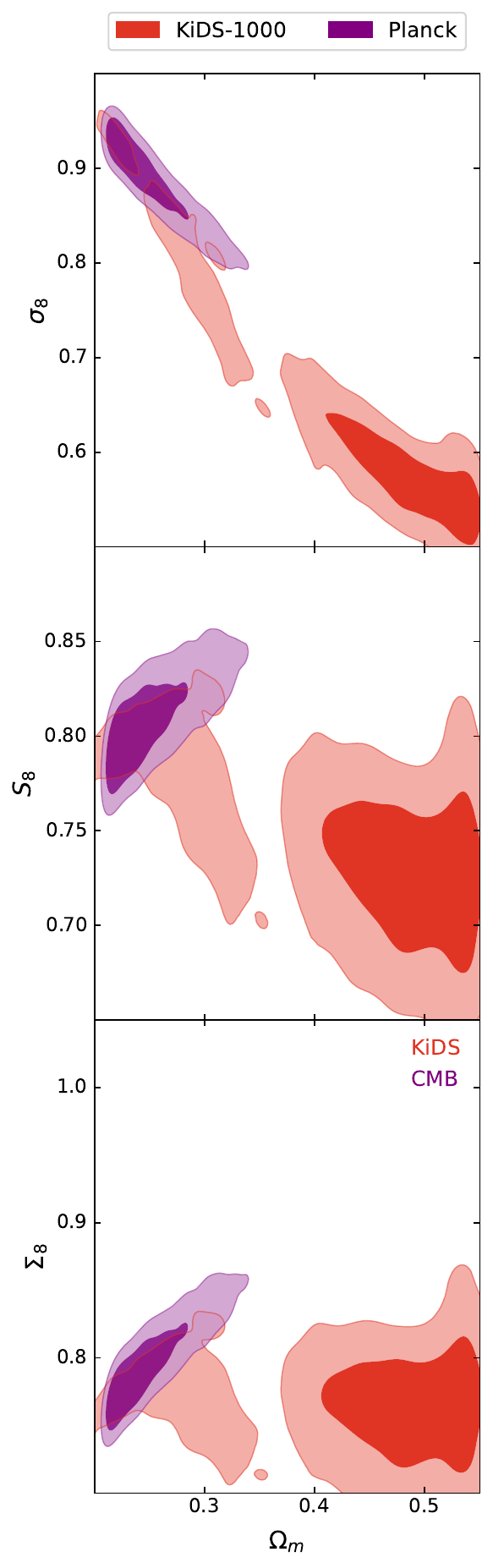}
\includegraphics[width=1.5in]{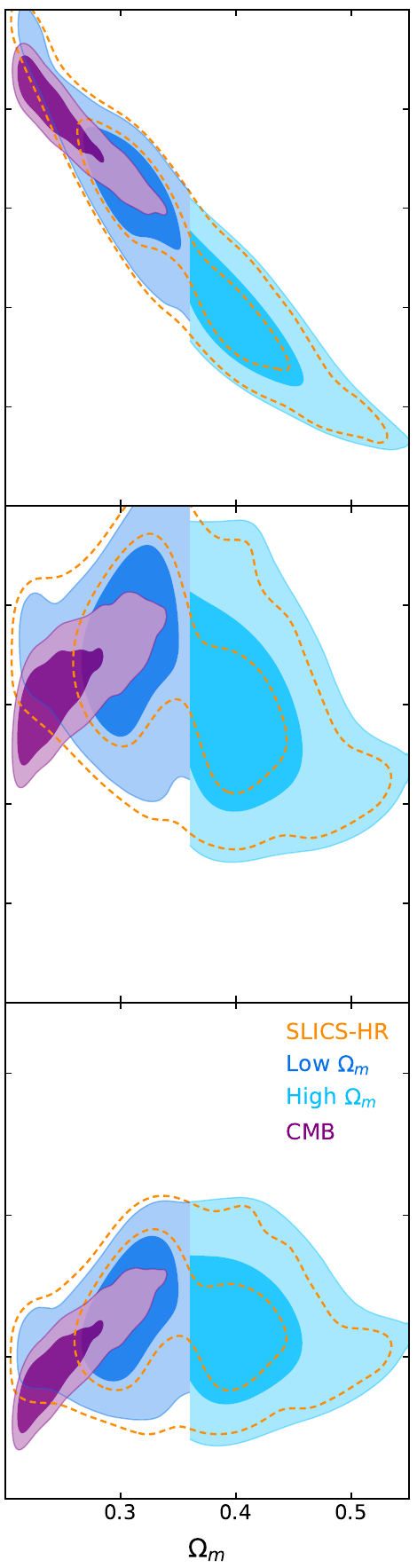}
\includegraphics[width=1.497in]{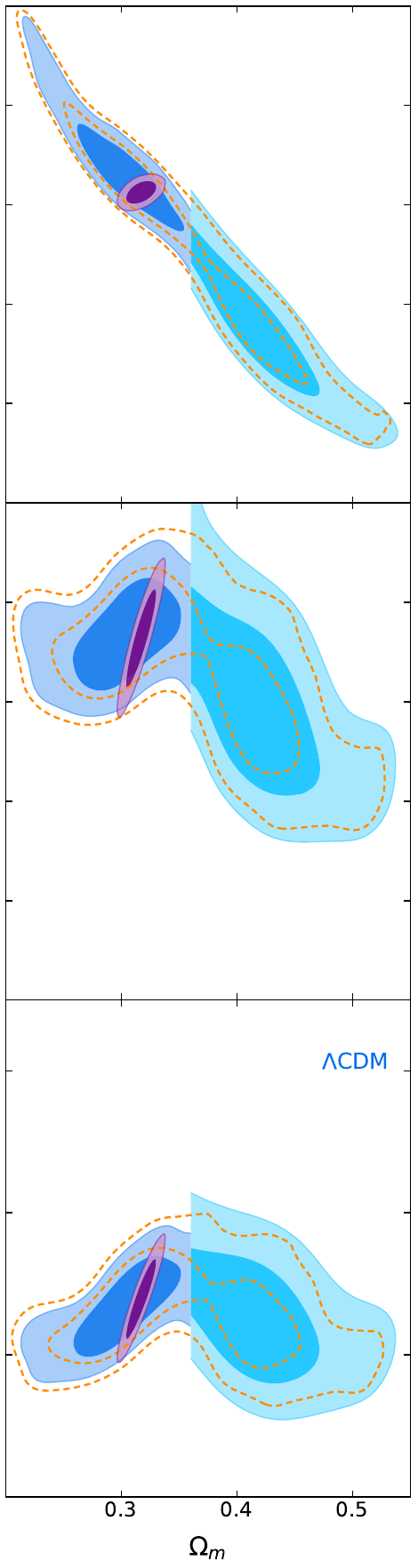}
\caption{({\it left panel:}) KiDS-1000 analysis: posterior distributions on the matter density and on the three clustering parameters ($\sigma_8, S_8, \Sigma_8$). The inferred value of $\Omega_{\rm m}$ from the data is higher than in previous KiDS-1000 analyses, consistent from being drawn from the biased secondary solution discussed in the main text, caused by our emulator. Over-plotted in purple are the $\it Planck$ results, which are in mild tension with KiDS for the first  two clustering parameters, but in full agreement with $\Sigma_8$. ({\it middle:}) Mock analysis, here carried out on the  {\it Validation Training Set} (orange, dashed). Our sample is further split into low- and high-$\Omega_{\rm m}$ parts (dark and pale blue, respectively), highlighting the residual $[S_8 - \Omega_{\rm m}]$ degeneracy, which vanishes for $\Sigma_8$. ({\it right:}) Same as middle panel, but here showing $\Lambda$CDM analyses. }
\label{fig:Sigma8}
\end{center}
\end{figure*}

We finally tested the joint-survey pipeline with the {\it Validation Set} defined for both KiDS-1000 and DES-Y1 analyses, and report our results in Fig. \ref{fig:Joint_inference_SLICS}. 
We observe that it recovers very well  the input truth:  the best-fit value is $S_8=0.818^{+0.030}_{-0.025}$, the maximum likelihood value is 0.831, while the truth is 0.813. 
These results are obtained from $w$CDM pipeline assuming the ``clean'' selection of $\mathcal{S}/\mathcal{N}$ bin, marginalising over all nuisance parameters. All input parameters are accurately recovered,  and we see here again the double $\Omega_{\rm m}$ solution, demonstrating that this parameter is subject to artificial degeneracies caused by our emulator, supporting our choice to not trust nor report its value in the main analysis. We also verified that the chain elements that fall in the secondary solution also tend to have a lower $S_8$ by about 0.03, which is larger than our statistical precision, however these are highly suppressed compared to the KiDS-1000 only pipeline, justifying our choice not to include this as a stand-alone systematic error. Again, the secondary solution yields unbiased $\Sigma_8$ inference, making the latter a more robust statistics. 

We illustrate this last point  in Fig. \ref{fig:Sigma8} where we  present the projected posteriors on $\sigma_8$, $S_8$ and $\Sigma_8$ versus $\Omega_{\rm m}$ for our analyses of the KiDS-1000 data and of the {\it Validation Training Set}. The KiDS-1000 inference prefers large $\Omega_{\rm m}$ values, consistent with being drawn from the secondary solution discussed earlier. If that is the case, the inferred value of $S_8$ might be biased low, but $\Sigma_8$ is robust. To illustrate this, we split the fiducial MCMC chain of the simulation analysis into low- and high-$\Omega_{\rm m}$ regions, and recover that both yield the similar $\Sigma_8$ posteriors, while their $S_8$ values differ by up to 0.03. We also show in this figure how the tension with {\it Planck} evolves under these change of variables.

\section{Goodness-of-fit for Student-$t$ likelihoods}
\label{sec:pvalues}

Noisy numerical covariance matrices need to be treated carefully in likelihood analyses to avoid biases incurred during the inversion. A commonly used approach is to debias the inverse matrix with the Hartlap-Anderson coefficient \citep{Hartlap2007}, however this  often leads to over-estimates in the contours. Instead, \citet{SellentinHeavens} suggested to replace the Hartlap-corrected multivariate Gaussian likelihood by a Student-$t$ distribution, which better accounts for the noise present when  estimating the matrix from $N_{\rm sims}$ realisations of the data.

Once the likelihood has been sampled and the best-fit parameters found, one of the key subsequent steps is to estimate the goodness-of-fit. This is usually achieved by means of the $p$-value, which determines how likely it is that the difference between the best-fit model and the measured data is due to a random noise fluctuation. Given the number of degrees of freedom $\nu$, best-fit $\chi^2$ measurements from data that is well described by a multi-Gaussian likelihood will be sampled from a $\chi^2_\nu$ distribution. Using this metric with noisy numerical covariance matrices will yield $p$-values that are biased towards low values if the inverse matrix is not Hartlap-corrected. Conversely, if corrected, the $p$-values are at risk to be on the high-side \citep{SellentinHeavens}. 

This is demonstrated by a toy model, which is created to follow our analysis: we generate a matrix $A$ with $210^2$ Gaussian random numbers (the same dimension as our KiDS-1000 analysis) and define a `true' covariance matrix $\Sigma = A^T A$. We also define the `true' data vector as the zero-vector. 

Afterwards, the following procedure is repeated $10\,000$ times:
we generate 1240 realizations of a multivariate normal distribution with mean 0 and covariance $\Sigma$, from which we estimate our sample covariance matrix $C$, mimicking the {\it Covariance Training Set}. We then also draw one additional realisation $\mathcal{X}$ of the same multivariate normal distribution, which constitutes our measurement. Finally, we calculate the $p$-value given $\mathcal{X}$ and $C$ and a chosen $p$-value test, assuming that the degrees of freedom equal the number of elements in the data vector (since our toy model contains no free parameters).

\begin{figure}
\begin{center}
\includegraphics[width=2.7in]{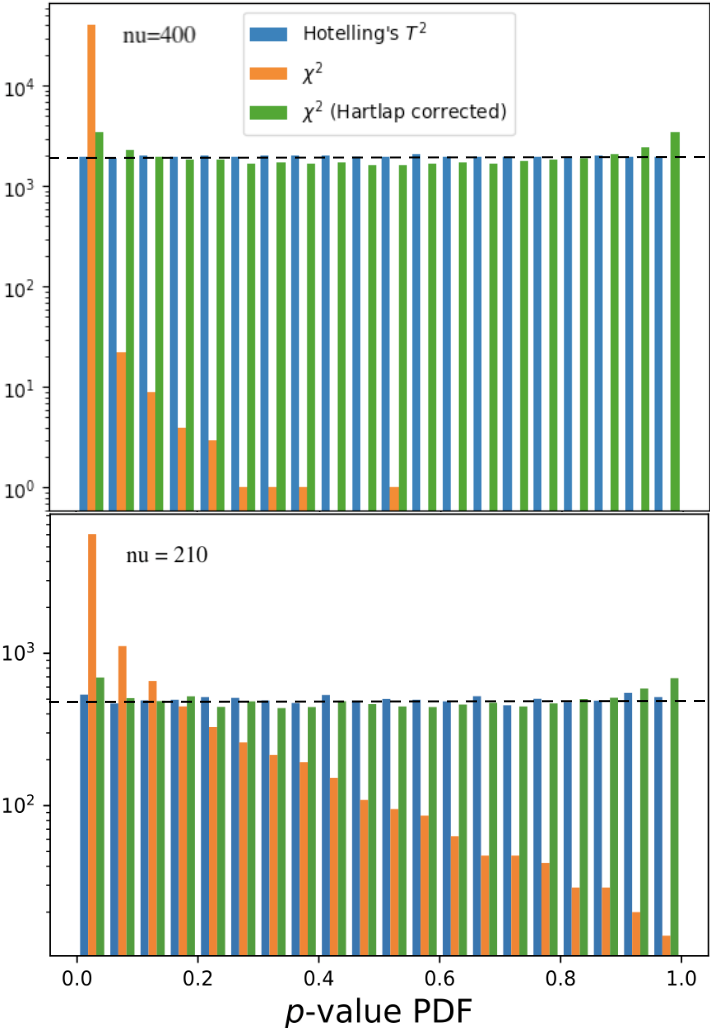}
\caption{Distribution of $p$-values extracted from our toy examples based on three commonly used goodness-of-fit statistics, for $\nu=210$ (lower) and $\nu=400$ (upper). Given a noisy numerical covariance matrix, only the Hotelling's $T^2$ distribution returns $p$-values evenly sampling the range [0, 1]; the $\chi^2$ distribution (orange) is heavily skewed towards low $p$-values, while the Hartlap-corrected $\chi^2$ (green) is slightly skewed towards extrema $p$-values. }
\label{fig:hotellings}
\end{center}
\end{figure}

This procedure yields $10\,000$ $p$-values which, if the chosen test is appropriate for our analysis, form a uniform distribution between 0 and 1. As can be seen in the lower panel of Fig.~\ref{fig:hotellings}, the $\chi^2$-based $p$-value tests are heavily biased towards 0, as is expected. The Hartlap-corrected $p$-values are more uniformly distributed, but still show slight biases towards 0 and 1, and consequently a reduced probability towards central values. Although this effect is relatively weak for our setup, it becomes more prominent if the degrees of freedom increase. Nevertheless, this means that a Hartlap-corrected $p$-value test is more likely to favour extreme values, but it appears to be relatively robust.

An unbiased solution to this problem can be achieved by deriving the sampling distribution of our quadratic statistics\footnote{The quadratic statistics described by Eq.~\ref{eq:T2def} should not be labelled `$\chi^2$' unless it is sampling a $\chi^2_\nu$ distribution. We used the $\chi^2$ notation in the main text only to align with the notation in the weak lensing literature. } defined in Eq.~\eqref{eq:T2def}, specifically:
\begin{equation}
\label{eq:T2bestfit}
    T^2_\mathrm{best-fit} = \left(\boldsymbol{d} - \boldsymbol{x}(\boldsymbol{\pi}_\mathrm{best-fit})\right)^TC^{-1}\left(\boldsymbol{d} - \boldsymbol{x}(\boldsymbol{\pi}_\mathrm{best-fit})\right)\,,
\end{equation}
where the data $\boldsymbol{d}$ has dimension $p$ and is drawn from a normal distribution $\boldsymbol{d}\sim N(\boldsymbol{\mu}, \Sigma)$, for unknown mean $\boldsymbol{\mu}$ and unknown covariance $\Sigma$, and the $C$ covariance is drawn from a Wishart distribution with $N_\mathrm{sims}-1$ degrees of freedom $(N_\mathrm{sims}-1)C\sim W_p(\Sigma, N_\mathrm{sims}-1)$.

We now define $LL^T = \Sigma^{-1}$ and $\boldsymbol{w} = L\left( \boldsymbol{d} - \boldsymbol{x}(\boldsymbol{\pi}_\mathrm{best-fit})\right)$, such that $\mathrm{Cov}\left[\boldsymbol{w}, \boldsymbol{w}\right] = \mathds{1}_p$, the $p\times p$ identity matrix. 
With this, we can express Eq.~\eqref{eq:T2bestfit} as
\begin{equation}
\label{eq:T2wV}
    T^2_\mathrm{best-fit} 
    = (N_\mathrm{sims}-1)\boldsymbol{w}^T V^{-1}\boldsymbol{w}\,,
\end{equation}
where we have defined $V = (N_\mathrm{sims}-1) L C L^T$ and note that $V\sim W_p(\mathds{1}_p, N_\mathrm{sims}-1)$.
We can now introduce an orthogonal matrix $M^TM=\mathds{1}_p$ with the first row being $\frac{\boldsymbol{w}}{\lVert\boldsymbol{w}\rVert}$ and the others orthogonal to it, such that
\begin{equation}
    M\boldsymbol{w} = \begin{pmatrix}
        \lVert\boldsymbol{w}\rVert \\
        0 \\
        \vdots \\
        0
    \end{pmatrix}\,.
\end{equation}
Conditional on $M$, we have that $Q = M V M^T\sim W_p(\mathds{1}_p, N_\mathrm{sims}-1)$. 
Since this does not depend on $M$, $Q\sim W_p(\mathds{1}_p, N_\mathrm{sims}-1)$ also holds in the unconditional case.
With this transformation, Eq.~\eqref{eq:T2wV} can be written as
\begin{equation}
\label{eq:T2wQ}
    T^2_\mathrm{best-fit} = (N_\mathrm{sims}-1)\lVert\boldsymbol{w}\rVert^2 \left(Q^{-1}\right)^2_{11}\,,
\end{equation}
with $\left(Q^{-1}\right)^2_{11}$ being the 1-1 entry of $Q^{-1}$.
Writing out $\lVert\boldsymbol{w}\rVert^2$ as 
\begin{equation}
    \begin{split}        
    \boldsymbol{w}^T\boldsymbol{w} &= \left(\boldsymbol{d} -  \boldsymbol{x}(\boldsymbol{\pi}_\mathrm{best-fit})\right)^TL^TL\left(\boldsymbol{d} - \boldsymbol{x}(\boldsymbol{\pi}_\mathrm{best-fit})\right) \\
    &= \left(\boldsymbol{d} - \boldsymbol{x}(\boldsymbol{\pi}_\mathrm{best-fit})\right)^T\Sigma^{-1}\left(\boldsymbol{d} - \boldsymbol{x}(\boldsymbol{\pi}_\mathrm{best-fit})\right)\,,
    \end{split}
\end{equation}
we recognise this as the usual $\chi^2$ quantity where the true covariance $\Sigma$ is assumed to be known. 
In other words, $\lVert\boldsymbol{w}\rVert^2\sim\chi^2_\nu$, with $\nu = p - n_\mathrm{eff}$, where $n_\mathrm{eff}$ is the effective number of free parameters that are being varied when finding $\boldsymbol{\pi}_\mathrm{best-fit}$,  which accounts for the fact that the model may contain both constrained and unconstrained parameters.

Returning to the last term in Eq.~\eqref{eq:T2wQ}, we have 
\begin{equation}
    \left(Q^{-1}\right)_{11}^{-1} = \frac{1}{\left(Q^{-1}\right)_{11}} = Q_{11} - Q_{12}Q_{22}^{-1}Q_{21} \,,
\end{equation}
where $Q_{12}$ and $Q_{22}$ are the $1\times (p-1)$ and $(p-1)\times (p-1)$ sub-matrices of $Q$. 
From this follows \citep[e.g.][]{Gupta1999-matrix} that $\frac{1}{\left(Q^{-1}\right)_{11}}\sim W_1(I_1, N_\mathrm{sims} - p) = \chi^2_{N_\mathrm{sims} - p}$. 
Putting things together, we therefore have that
\begin{equation}
    T^2_\mathrm{best-fit}\sim (N_\mathrm{sims}-1)\frac{\chi^2_{p-n_\mathrm{eff}}}{\chi^2_{N_\mathrm{sims} - p}} = \frac{(N_\mathrm{sims}-1)(p-n_\mathrm{eff})}{(N_\mathrm{sims} - p)} F_{p-n_\mathrm{eff}, N_\mathrm{sims} - p}\,,
\end{equation}
where $F_{p-n_\mathrm{eff}, N_\mathrm{sims} - p}$ is the $F$-distribution. 
For the case of no free parameters, $n_\mathrm{eff} = 0$, this reduces to Hotelling's $T^2$ distribution \citep{Hotelling_paper}.



To calculate a $p$-value, one just has to evaluate the cumulative distribution function of the $F$-statistics at $\frac{N_\mathrm{sims}-p}{(p-n_\mathrm{eff})(N_\mathrm{sims}-1)}T^2$, replacing the $\chi^2(p-n_\mathrm{eff})$ distribution. Clearly seen in Fig.~\ref{fig:hotellings}, this constitutes an ideal solution for our toy model, 
 so we use this statistics to estimate the $p$-values of our measurements.

When applied to our fiducial KiDS-1000 peak count data-vector, along with the best-fit model and our numerical covariance matrix, we obtain $p$-values of 0.43 with the (unbiased) Hotelling's statistics and for the (slightly biased) Hartlap-corrected $\chi^2$ approach, and 0.02 for the (heavily biased) normal $\chi^2$ statistics.

In this toy example, the difference between the Hotelling's and the Hartlap-corrected $p$-values distributions is quite small, however this is not always the case. The upper panel of Fig.~\ref{fig:hotellings} shows a second case where now the number of degrees of freedom has been increased to $\nu=400$, overshooting our joint-survey setup, but close to typical sizes of  data vectors used in 2pt statistics. Keeping $N_{\rm sims}$ unchanged, in this case the Hartlap-corrected distribution shows a clear excess towards low and high $p$-values. The Hotelling's distribution is still flat however, showcasing the advantage of  the $F$-statistics.



\bsp	
\label{lastpage}
\end{document}